# Valid auto-models for spatially autocorrelated occupancy and abundance data

**David C Bardos[1*], Gurutzeta Guillera-Arroita[1] and Brendan A Wintle[1]**

[1]*ARC Centre of Excellence for Environmental Decisions, School of Botany, The University of Melbourne, Parkville, Victoria 3010, Australia*

**Abstract.** Spatially autocorrelated species abundance or distribution datasets typically generate spatially autocorrelated residuals in generalized linear models; a broader modelling framework is therefore required. Auto-logistic and related auto-models, implemented approximately as autocovariate regression, provide simple and direct modelling of spatial dependence. The autologistic model has been widely applied in ecology since Augustin, Mugglestone and Buckland (*Journal of Applied Ecology,* 1996, **33,** 339) analysed red deer census data using a hybrid estimation approach, combining maximum pseudo-likelihood estimation with Gibbs sampling of missing data. However Dormann (*Ecological Modelling*, 2007, **207**, 234) questioned the validity of auto-logistic regression even for fully-observed data, giving examples of apparent underestimation of covariate parameters in analysis of simulated 'snouter' data. Extending this critique to include auto-Poisson and uncentered auto-normal models, Dormann et al. (*Ecography*, 2007, **30**, 609) found likewise that covariate parameters estimated via autocovariate regression bore little resemblance to values used to generate 'snouter' data. We note that all the above studies employ neighbourhood weighting schemes inconsistent with auto-model definitions; in the auto-Poisson case, a further inconsistency was the failure to exclude cooperative interactions. We investigate the impact of these implementation errors on auto-model estimation using both empirical and simulated datasets. We show that when 'snouter' data is reanalysed using valid weightings, very different estimates are obtained for covariate parameters. For auto-logistic and auto-normal models, the new estimates agree closely with values used to generate the 'snouter' simulations. Re-analysis of the red deer data shows that invalid neighbourhood weightings generate only small estimation errors for the full dataset, but larger errors occur on geographic subsamples. A substantial fraction of papers employing auto-logistic regression use these invalid neighbourhood weightings, which are embedded as default options in the widely used 'spdep' spatial dependence package for R. Auto-logistic analyses conducted using invalid neighbourhood weightings will be erroneous to an extent that can vary widely. These analyses can easily be corrected by using valid neighbourhood weightings available in 'spdep'. The hybrid estimation approach for missing data is readily adapted for valid neighbourhood weighting schemes and is implemented here in R for application to sparse presence-absence data.

Key words: abundance, autologistic, autocovariate, conditional auto regression, occupancy, species distribution model, SDM, spatial autocorrelation, spdep

*Corresponding author. E-mail: dcbardos@unimelb.edu.au

# Introduction

Spatial autocorrelation (SAC) of species distribution data can arise from intrinsic population processes, such as dispersal or biotic interactions, or in response to spatially autocorrelated environmental variables (Legendre 1993; Elith & Leathwick 2009). Intrinsic processes introduce statistical dependence between observations, violating a key assumption (McCullagh & Nelder 1989) of generalized linear models (GLMs) such as the logistic model. Environmental covariates with strong SAC induce correlations between nearby observations, but these covariates will be missing when data is unavailable or if key environmental drivers for a species are unknown. Both intrinsic processes and missing covariates can result in strong SAC in GLM residuals, indicating the presence of structure in the data that cannot be modelled within the GLM framework. Species distribution modelling therefore requires some manner of directly incorporating SAC (Legendre 1993; Lichstein *et al.* 2002).

The autologistic model (Besag 1974) achieves this by extending the logistic model for presence-absence data to allow dependence between nearby observations. It is of fundamental importance in ecology where many datasets are in the form of presence-absence maps and has been widely used since its application by Augustin, Mugglestone and Buckland (1996) to spatial grids with sparse observations. The broader family of "auto-models" introduced by Besag (1974) includes auto-binomial and auto-normal models that can be applied to spatially-dependent count data and real-valued data respectively. The auto-Poisson model can be applied to count data but permits only competitive interactions in which abundance at a given site tends to be reduced by high abundance at nearby sites. The Conditional Autoregressive (CAR) family of models are the most widely applied form of auto-normal model. Autologistic and related auto-models offer distinct advantages for analysing spatially autocorrelated ecological data, including simple and direct spatial modelling of dependence between nearby observations and straightforward implementation using autocovariate regression to obtain approximate parameter estimates.

However, Dormann (2007) took issue with the validity of autologistic regression, reporting apparent underestimation of environmental covariate parameters when strong SAC was injected into simulated 'snouter' data. Dormann *et al.* (2007) extended this analysis to auto-normal and auto-Poisson models, likewise reporting underestimation relative to GLM estimates and to values used to generate the simulations. These papers, hereafter 'Dormann et al.', are heavily cited and their negative conclusions concerning autocovariate regression appear to be widely accepted in the literature, with some authors relying on these conclusions to choose against using auto-models. This has significant implications for ecologists dealing with spatial autocorrelation in the residuals of occupancy and abundance models such as species distribution models.

Yet no explanation has been established for the apparent estimation anomalies, nor for the conflict with previous studies (Huffer & Wu 1998; Hoeting, Leecaster & Bowden 2000; He, Zhou & Zhu 2003) in which autologistic regression estimates were close to values used to generate simulated data. In particular, Hoeting, Leecaster and Bowden (2000) obtained good performance in a challenging problem where most of the observations were missing, in contrast to Dormann et al.'s fully-observed simulated datasets. Dormann et al. (2007) suggested that this contradiction might be due to (a) more advanced estimation methods employed in the better performing studies, (b) reduced importance of the autocovariate in Hoeting, Leecaster and Bowden (2000) due to the high proportion of missing-data, or (c) different methods of generating SAC in simulated data.

The first two suggestions are problematic, however, because (a) the maximum pseudo-likelihood estimation (MPLE) method evaluated in Huffer and Wu (1998), Hoeting, Leecaster and Bowden (2000) and He, Zhou and Zhu (2003) is equivalent (Possolo 1986) to the autocovariate regression used in Dormann et al., while (b) Hoeting, Leecaster and Bowden (2000) obtained strong autocovariate parameter estimates. Suggestion (c) must be considered, but turns out to be inessential for the 'snouter' examples.

We note that in common with Augustin, Mugglestone and Buckland (1996), Dormann et al. used weighted means to construct their autocovariates. In contrast, Huffer and Wu (1998), Hoeting, Leecaster and Bowden (2000) and He, Zhou and Zhu (2003) used weighted sums. The widely used weighted-mean construction (the default option in the 'spdep' spatial dependence package (Bivand 2014) for R) is mathematically inconsistent with auto-model definitions, while the weighted sum conforms to the definitions. Estimation errors will consequently arise when the weighted mean is used. However, since the two constructions differ only on boundary sites, it might reasonably be expected that errors would tend to be small in most ecological applications.

We provide a detailed investigation into estimation errors arising from using weighted-mean autovariates, as follows:

We present essential theoretical background on auto-models and the conditions required for their valid estimation, including the requirement for *valid neighbourhood weightings*, which is fulfilled by the weighted-sum autocovariate construction. We then investigate estimation errors arising from weighted-mean autocovariates for the "snouter" data of Dormann et al. and use new simulations to investigate errors as a function of lattice size and covariate SAC and to test suggestion (c) above concerning the effect of different methods of generating intrinsic SAC. We reanalyse the red deer census dataset (Buckland & Elston 1993; Augustin, Mugglestone & Buckland 1996) to explore errors due to weighted-mean autocovariates in estimates from empirical data and briefly discuss the usage of weighted-mean autocovariates in previous studies.

We provide R code implementing valid autologistic regression, using the hybrid approach of Augustin, Mugglestone and Buckland (1996) to iterate MPLE (Besag 1975) for the model parameters and Gibbs sampling of missing observations. The code implements weighted-sum autocovariates without making use of `spdep', allowing us to conduct numerical testing of a non-default scheme in 'spdep' expected to generate weighted-sum autocovariates. As an illustration we apply this code for a 20% random sample of the abovementioned red deer data.

# Valid neighbourhood weightings and parameter regions

## Theoretical background

We consider a collection of $S$ sites labelled $n = 1, 2, \ldots S$, each of which has an associated observation $y_n$. For each site $n$ we define a set $\mathrm{N}_n$ of other sites, called the neighbourhood of $n$. These neighbourhoods must be consistent, in the sense that if site $m$ is in $\mathrm{N}_n$, then site $n$ is in $\mathrm{N}_m$. The auto-models are a class of conditionally specified models introduced by Besag (1974). For an important subset of auto-models, including auto-logistic, auto-normal (uniform variance) and auto-Poisson models for binary, real-valued and count valued data respectively, the distribution for $y_n$, conditional on all other sites, takes the form (omitting normalization over $y_n$):

$$P(y_n \mid y_{-n}) \propto \mathrm{F}_n(\mathrm{y}_n) \exp(\mathrm{y}_n \sum_{m \in \mathrm{N}_n} \beta_{nm} \, \mathrm{y}_m), \qquad (1)$$

(see Besag 1974, p. 200) where the parameters $\beta_{nm}$ quantify the strength of dependence between sites. Any pair of sites that are not neighbours do not influence each-other's conditional distributions. The structure of $F_n(y_n)$ is determined by the particular auto-model and can depend on parameters at site $n$ (i.e. covariates) but does *not* depend on observations at other sites ($y_{m \neq n}$), so that (1) explicitly separates local site effects from interactions between sites. The full model is specified by the collection of all $S$ conditional distributions (1); for these to be consistent, i.e. result in a valid joint distribution $P(\mathrm{y}_1, y_2, ..., y_S \mid \{\beta_{nm}\})$, the parameters $\beta_{nm}$ must obey the symmetry relation (Besag 1974 , p. 200)

$$\beta_{nm} = \beta_{mn}. \tag{2}$$

If we define a neighbourhood system and specify a model (1) for which it happens that $\beta_{nm} = \beta_{mn} = 0$ for some pair of neighbouring sites $n$ and $m$, then we can redefine the neighbourhood of each to exclude the other without otherwise altering the form of (1). Therefore, in the context of a particular auto-model, we can *define* the neighbourhood of a site as the set of other sites on which its conditional distribution depends; this definition is adopted in Besag (1974).

Although (2) is a necessary condition, it is not always sufficient: additional conditions on the parameters $\beta_{nm}$ arise for particular auto-models. Note that $\beta_{nm} > 0$ signifies a cooperative interaction between sites $n$ and $m$, while $\beta_{nm} < 0$ indicates a competitive interaction. In Table 1 we give detailed formulae for models of particular relevance for ecology, including the Conditional Auto-Regressive (CAR) variant of the auto-normal model, along with sufficient conditions (Besag 1974) on the parameters $\beta_{nm}$ for model validity in each case.

In a further paper, Besag (1975) extended the Conditional Auto Regressive model to allow the conditional variance to differ at each site, so that the form (1) is replaced by

$$P(y_n \mid y_{-n}) \propto \mathrm{F}_n(\mathrm{y}_n)\exp(\sigma_n^{-2}\,\mathrm{y}_n \sum_{m \in \mathrm{N}_n} \beta_{nm}\,\mathrm{y}_m), \tag{3}$$

and the symmetry relation (2) must then be generalized to $\beta_{nm}\sigma_m^2 = \beta_{mn}\sigma_n^2$.

However, for the present paper we require only the common-variance CAR model used in Dormann *et al.* (2007). Thus for all auto-models investigated here, the symmetry relation (2) must hold. For the CAR model, exact maximum-likelihood methods (as distinct from MPLE) are available (Besag 1974; Cressie 1991) and are usually numerically feasible.

## Defining autocovariates

A convenient approach to estimation of $\beta_{nm}$ is to introduce a parameter $\beta_{\mathrm{auto}}$ representing the overall strength of interactions between sites, while assuming a fixed proportionality between the elements of $\beta_{nm}$, i.e. we write

$$\beta_{nm} = \beta_{\mathrm{auto}} w_{nm} , \tag{4}$$

where $w_{nm}$ are called the neighbourhood weights. Then, introducing the *autocovariate* for site $n$, comprising a *weighted sum* of neighbouring observations,

$$\text{autocov}_n = \sum_{m \in \text{N}_n} w_{nm}\, \text{y}_m, \tag{5}$$

the conditional probability for site $n$ becomes

$$P(y_n \mid y_{-n}) \propto \text{F}(\text{y}_n)\exp(\text{y}_n\, \beta_{\text{auto}} \text{autocov}_n). \tag{6}$$

Since $\beta_{\text{auto}}$ is a constant, the neighbourhood weights inherit their own symmetry relation from (2), i.e.

$$w_{nm} = w_{mn}, \tag{7}$$

Thus if sites $r$ and $s$ are neighbours, the coefficient of $y_s$ in $\text{autocov}_r$ must equal the coefficient of $y_r$ in $\text{autocov}_s$.

## Valid neighbourhood weighting schemes

Consistent neighbourhoods ($m$'s neighbour has $m$ as a neighbour), the correct autocovariate definition (5) and symmetric weights (7) are all required for a valid *neighbourhood weighting scheme*, or *'weighting'.* For the autologistic model, a valid weighting is sufficient for the model specified via (6) to be valid. For auto-normal and auto-Poisson models, a valid weighting is necessary but $\beta_{\text{auto}}$ is subject to further conditions to ensure that the parameters $\beta_{nm} = \beta_{\text{auto}} w_{nm}$ satisfy the conditions in Table 1. In particular, for auto-Poisson models the condition $\beta_{nm} \le 0$ means that all interactions must be competitive. This implies all weights have the same sign and $\beta_{\text{auto}}$ has the opposite sign: for all examples in this paper, $w_{nm} \ge 0$, so $\beta_{\text{auto}} \le 0$ for valid auto-Poisson models and this constraint must be imposed during model estimation. For auto-normal models, the positive-definiteness condition in Table 1 can always be met for $\beta_{\text{auto}}$ values of sufficiently small magnitude, but can be violated for large magnitudes. Accordingly, when estimating the model, this condition should be checked once an estimate for $\beta_{\text{auto}}$ is obtained. If it violates the condition, the estimation method will need to be modified to constrain $\beta_{\text{auto}}$ within a region that meets the condition.

## Invalid weighting schemes

In their application of the autologistic model to sparse data via Gibbs sampling of missing data, Augustin, Mugglestone and Buckland (1996) gave a *weighted-mean* construction for the autocovariate:

$$\text{autocov}^*_n = \frac{\sum_{m \in \text{N}_n} w^*_{nm}\, \text{y}_m}{\sum_{m \in \text{N}_n} w^*_{nm}}, \quad w^*_{nm} = w^*_{mn}, \tag{8}$$

where $w^*_{nm}$ is a function of the distance between sites $n$ and $m$. This expression differs from that in (5) and is not usually correct for the *finite* lattices of interest to ecologists; that is, where inference and prediction are undertaken for a collection of sites within defined edges or boundaries, beyond which data are not considered.

To see this, consider for example a square lattice of sites, each of which has a neighbourhood comprised only of its nearest adjacent sites, so that it has at most four neighbours (this is called a *first-order* neighbourhood, see Fig. 1). Since all distances between neighbours in this example are identical, $w^*_{nm}$ is a constant and cancels out of (8). Now consider a border site labelled $r$ that is not a corner site: it has three neighbours, one of which, labelled $s$, is an interior site that has four neighbours. Then we have

$$\text{autocov}^*_r = \frac{1}{3}\sum_{m \in \text{N}_r} \text{y}_m, \qquad \text{autocov}^*_s = \frac{1}{4}\sum_{m \in \text{N}_s} \text{y}_m,$$

which breaks the required symmetry (2) for the conditional probabilities for sites $r$ and $s$. That is, equating (8) with the required form (5)

$$\text{autocov}^*_n = \sum_{m \in \text{N}_n} w_{nm}\, \text{y}_m, \tag{9}$$

yields $w_{rs} = \frac{1}{3} \neq \frac{1}{4} = w_{sr}$, contradicting (7) and thus (2).

Ord (1975) had earlier proposed the weighting (8) for use in joint (i.e. simultaneously) autoregressive models (Whittle 1954), noting that " … the suggested scaling will usually produce asymmetric weights". This is permissible for models with simultaneous autoregressive (SAR) structure as distinct from the conditionally autoregressive structure of the auto-models; this distinction is discussed in Besag (1974), Ord (1975), Haining (1990) and Cressie (1991).

## Valid autologistic models on a rectangular lattice

We consider the common situation where presence-absence data occurs on a collection of $S$ sites labelled $n = 1, 2, \ldots S$, forming some subset of a finite rectangular lattice. Since sparse observations are the usual situation in ecological applications, we assume that covariate data is available at each of the $S$ sites, but any of the observations $y_n$ can be missing.

Points on the lattice occur at spacings $a_1$ and $a_2$ along the two axes respectively and we use integer lattice co-ordinates $(i, j)$ to refer to the observations: $y_n = y_{i_n, j_n} = y_{i,j}$. The symmetry requirement $w_{nm} = w_{mn}$ for a valid autologistic model is then satisfied if we define the weights $w_{nm}$ in terms of lattice separations $(i_m - i_n, j_m - j_n)$ of sites $n$ and $m$ using a neighbourhood 'template' $b_{k,l}$:

$$\begin{aligned} w_{nm} &= b_{i_m - i_n, j_m - j_n}, \\ b_{k,l} &= b_{-k,-l}, \\ b_{0,0} &= 0, \end{aligned} \tag{10}$$

where $b_{k,l}$ provides an image of the weights as a function of lattice separations $(k, l)$ from a site *with a full complement of neighbours*. However, many sites do not have a full complement, both due to the boundaries of the finite lattice and because we have not assumed that sites exist at all interior lattice points; we have only assumed that all $S$ sites lie on points of the lattice. The autocovariate definition

(5) handles this automatically by restricting the sum to actual neighbours. Implementation in software can proceed by including additional sites with observations set to zero, expanding the lattice where necessary, so that each of the $S$ actual sites appears to have a full complement of neighbours and the autocovariate becomes a simple sum over the $(k,l)$ for which $b_{k,l} \neq 0$.

The neighbourhood template $b_{k,l}$ must have central symmetry $b_{k,l} = b_{-k,-l}$ but this *does not rule out anisotropic weights*. For example Besag examined the simplest anisotropic scheme where $b_{k,l} = 0$ for all $(k,l)$ except for the four nearest neighbours: $b_{-1,0} = b_{1,0} \neq b_{0,-1} = b_{0,1}$. In applications the neighbourhood template can be defined simply in terms of the Euclidean distance $d = \sqrt{k^2 + \rho^2 l^2}$ between sites:

$$\begin{aligned} b_{k,l} &= f(d) I_{k,l}, \\ I_{k,l} &= 1, \quad 0 < d \leq r, \\ I_{0,0} &= 0, \end{aligned} \tag{11}$$

where $\rho = a_2 / a_1 > 0$ and both $d$ and the neighbourhood radius $r$ are measured in units of $a_1$. The indicator function $I_{k,l}$ defines a "radius $r$ neighbourhood" and the function $f$ can be tailored to the range-dependence of spatial autocorrelation for relevant species. We show the indicator function for commonly used small neighbourhoods in Fig. 1. In the case of a square lattice, $\rho = 1$ and the neighbourhood template (11) is then isotropic.

## Maximum pseudo-likelihood estimation (MPLE)

The joint distributions $P(y_1, y_2, ..., y_S \mid \{\beta_{nm}\})$ defined by auto-logistic and auto-Poisson variants of (1) contain intractable normalization factors that depend on the parameters $\beta_{nm}$. Maximum likelihood estimation is therefore very difficult, leading Besag (1975) to introduce an easily maximized approximation of the likelihood, known as the pseudo-likelihood (PL):

$$\mathrm{PL} = \prod_{i=1,2,...S} P(y_n \mid y_{-n}) \approx P(y_1, y_2, ..., y_S \mid \{\beta_{nm}\}).$$

Maximum Pseudo-Likelihood Estimation (MPLE) is very widely used, often under the guise of *autocovariate regression*. The autocovariate is not actually a covariate, since it contains the observations, but MPLE is equivalent (Possolo 1986) to treating $\mathrm{autocov}_n$ as though it *were* just an explanatory variable. MPLE can therefore be implemented by including $\mathrm{autocov}_n$ as an explanatory variable in the GLM corresponding to the particular auto-model (e.g. a logistic model for autologistic estimation, etc.). Evaluating $\mathrm{autocov}_n$ is impossible if neighbouring observations are missing, which led Augustin, Mugglestone and Buckland (1996) to combine Gibbs sampling of missing data with autocovariate regression.

## Estimation on simulated 'snouter' data

Dormann et al. test various models against simulated 'snouter' data (Fig. 2) in which strong SAC is driven by an environmental covariate, 'rain', and strong intrinsic SAC is also injected during the simulation process. Among those tested were autocovariate models (i.e. autocovariate-regression of auto-logistic, auto-Poisson and uncentered auto-normal models) and the Condition Autoregressive (CAR) model. A spurious covariate 'djungle' was included in each analysis along with 'rain', so for example the autologistic analysis consisted of the following logistic regression:

$$\log\left(\frac{p_n}{1-p_n}\right) = \alpha + \beta_1 \mathrm{rain}_n + \beta_2 \mathrm{djungle}_n + \beta_{\mathrm{auto}} \mathrm{autocov}_n ,$$

where $p_n$ is the probability of presence at site $n$. The principal numerical evidence presented by Dormann et al. against the validity of auto-models is that the regression parameter $\beta_1$ for 'rain', estimated in the above regression (and in the analogous auto-Poisson and uncentered auto-normal regressions) is much smaller than the 'rain' value used to generate the simulated 'snouter' data. In contrast, they found no such anomaly for the CAR model (which, we reiterate, is also an auto-model).

However, Dormann (2007) quotes the invalid weighting scheme (8) for autologistic estimation; Dormann *et al.* (2007) quote both the valid (5) and invalid (8) schemes, but from the R code in the supplementary information we find that it was the incorrect form that was used for auto-logistic, auto-Poisson and uncentered auto-normal models. Autocovariates were generated using the 'autocov_dist' function from the 'spdep' package

```
ac <- autocov_dist(snouter1.1,  coords,  nbs = 1,  type = "inverse")
```

which is equivalent to:

```
ac <- autocov_dist(snouter1.1,  coords,  nbs = 1, style="W", type = "inverse")
```

since the default option for 'style' is "W". The same applies for the related 'spdep' function 'nb2listw' that generates neighbourhood weights. In the supplement we supply R code that tests these functions and we find that for the 'snouter' data, the "W" option, and thus the default option, generates asymmetric weights that agree with the invalid scheme (8) and are therefore incorrect for auto-model estimation. Note that other types of spatial models such as joint autoregressive models (Ord 1975) do not require symmetric weights and therefore the default options may be useful for such models. Testing the option 'style= "B" ' in functions 'nb2listw' and 'autocov_dist' using the 'snouter' data, we find (see R code in supplement) that symmetric weights are generated and the autocovariate corresponds correctly to (5). Thus it appears that this option generates valid weighting schemes.

We find that we can reproduce the anomalous estimates obtained by Dormann et al. for the 'rain' parameter by using the invalid weightings generated by default in package 'spdep'. If we instead use valid weightings (option style = "B" in 'spdep'), we obtain estimates for 'rain' in the auto-logistic and uncentered auto-normal cases that are close to the values driving GLMs used in the initial generation of 'snouter' probability maps prior to the application of correlated distortions (Dormann *et al.* 2007). For the uncentered auto-normal case, the validity of each estimated model was tested (see supplement) and all were found to meet the positive-definiteness criteria (see Table 1).

Auto-models using invalid weightings severely underestimate the magnitude of the "rain" parameters for all 20 datasets comprising real-valued and binary 'snouter' simulations; our results are summarized in Table 2. See supplement for R code generating these results, along with simpler code for valid estimation of a single fully observed dataset.

The auto-Poisson case has additional complications (see below) but in the auto-logistic and auto-normal cases, the results are meaningful. We interpret them as follows: when the environmental covariate is highly spatially autocorrelated and intrinsic SAC is absent, the species distribution will tend to cluster according to the covariate. When intrinsic SAC is injected, clustering will increase and the autocovariate will be large in the clusters. Thus regions with strong covariate and strong autocovariate will tend to coincide, *except at the boundary,* where the autocovariate must necessarily be reduced. So we can envisage a situation where most of the contrast between intrinsic and extrinsic SAC occurs in the relatively small number of observations at or near (i.e. within the neighbourhood of) boundary sites. The incorrect neighbourhood weighting (8) compensates for the smaller number of neighbours at the boundary, incorrectly increasing the value of the autocovariate there and thus washing out the contrast between covariate and autocovariate at the boundary. However, since clusters occur at locations that are partly random due to intrinsic SAC, these can be mimicked more successfully by the autocovariate than the covariate. So with its main explanatory advantage (contrast at the boundary) removed, the covariate parameter is severely underestimated.

## Auto-Poisson estimation

In the auto-Poisson case, the model is valid only for competitive interactions (Besag 1974), implying $\beta_{\text{auto}} \leq 0$, so that valid weightings alone are not sufficient to ensure the model is valid (Table 1). In maximum-likelihood estimation, a constraint such as $\beta_{\text{auto}} \leq 0$ can be ignored during the model fitting procedure provided the maximum (and thus the final estimate) lies within the constraint. However, for the 'snouter' datasets, MPLE using valid weightings yields estimates with $\beta_{\text{auto}} > 0$, breaching the constraint. Thus a valid estimation of the model requires constrained optimization of the pseudolikelihood. Attempting (unconstrained) MPLE with the invalid weightings (8) used in Dormann *et al.* (2007) also results in $\beta_{\text{auto}} > 0$ and the covariate parameter estimates differ from the (unconstrained) valid-weighting estimates (see Table 2), but there is little point in pursuing the comparison since both sets of results are invalid.

The log-pseudo-likelihood for the auto-Poisson model (Cressie 1991), adapted for the 'snouter' covariates, is:

$$\begin{aligned} \log(\text{PL}) &= \sum_n y_n \log(\lambda_n) - \lambda_n - \log\left(y_n!\right), \\ \log\left(\lambda_n\right) &= \alpha + \beta_1 \text{rain}_n + \beta_2 \text{djungle}_n + \beta_{\text{auto}} \text{autocov}_n. \end{aligned} \qquad (12)$$

When (12) is optimized (see supplement) subject to the $\beta_{\text{auto}} \leq 0$ constraint, we obtain $\beta_{\text{auto}}$ estimates that are essentially zero and the model recovers the value of the 'rain' parameter used in the original GLM simulations. As the auto-Poisson model is valid only for competitive interactions and the artificial data contains strong cooperative interactions, estimates near zero for $\beta_{\text{auto}}$ are reasonable.

## CAR estimation

Conditional autoregressive (CAR) models are a form of auto-normal model and consequently, in the widely used common-variance case (see Table 1) the spatial weights matrix employed must also be symmetric. Dormann *et al.* (2007) state that “we used a row-standardised binary weights matrix …”; the code given in their supplementary information for performing CAR estimations generates a spatial weights matrix using a command

```
w<-nb2listw(nb, style="W", zero.policy=FALSE)
```

that is consistent with the text inasmuch as the R documentation for the command states that “B is the basic binary coding, W is row standardised”. However, as discussed above and shown by our R code (see supplement), the “W” option for this command generates weights with broken symmetry. When the code from the Dormann *et al.* (2007) supplement is run it generates “Non-symmetric spatial weights in CAR model” warnings and gives estimates for the `rain’ parameter that are opposite in sign to estimates quoted in the paper. Modifying the code to generate a symmetric weights matrix (substituting “B” for “W” in the above command) results in estimates very close to those in the paper. Thus it appears that CAR estimation was performed correctly in Dormann *et al.* (2007), but described incorrectly. Consequently the reported CAR estimates for ‘rain’ were not anomalous, in contrast to the other auto-model estimates, but the key difference between the codes that generated these disparate results went unreported.

## Effect of observation-grid density and covariate SAC

The underestimation (see Table 2) of the ‘rain’ parameter raises two points under our earlier interpretation. Firstly, errors due to the invalid weightings (8) are clearly a boundary effect because for ‘interior sites’, with a full complement of neighbours, these weightings are proportional to those given by the valid scheme (5). Thus, if there were no boundary sites, as for some infinite lattice models, then no errors would occur ($\beta_{\text{auto}}$ estimates for each scheme would then simply be in reciprocal proportion to the weights used for each scheme). It could be expected, therefore, that errors due to invalid weightings would decrease if the number of sites increases and the proportion of boundary sites decreases. Secondly, large errors require strong spatial autocorrelation of the ‘rain’ covariate.

To investigate these points for autologistic estimation, we generate new simulated presence-absence datsets driven by the ‘rain’ covariate on the original ‘snouter’ grid as well as on two grids of higher density (smaller lattice spacing). We create the latter by subdividing each of the original cells into either 4 or 16 cells, assigning to each of them the ‘rain’ value from the original ‘snouter’ cell they lie in. We also generate simulations using an uncorrelated version of the ‘rain’ covariate: a simple normal variate with mean and standard deviation matching that of ‘rain’ but with no inter-site dependence. For each grid density and version of ‘rain’, we generate 10 simulated datasets.

We avoid the ad-hoc approach to generating SAC in simulated data used in Dormann et al. Instead we apply Gibbs sampling to an autologistic model including as a covariate either ‘rain’ or ‘uncorrelated rain’, using a 1$^{st}$ order neighbourhood template with uniform weights. We initialize each site independently $y_n \sim \text{Bernoulli}(0.5)$ and then apply $T = 400$ Gibbs sampling sweeps according to the algorithm (see supplement for R code):

**For** $\left(t=1,2,\ldots \mathrm{T}\right)$:

**For** $\left(n=1,2,\ldots S\right)$:

**1.** Evaluate $\mathrm{autocov}_n$.

**2.** Generate $y_n$ by sampling from the conditional distribution given all other sites:

$$\Pr(y_n=1)=\frac{e^{\alpha+\beta_1 \mathrm{rain}_n+\beta_{\mathrm{auto}}\ \mathrm{autocov}_n}}{1+e^{\alpha+\beta_1 \mathrm{rain}_n+\beta_{\mathrm{auto}}\ \mathrm{autocov}_n}}\ .$$

We analyse the resulting 60 simulated datasets (see Fig. 3 for examples) using three valid autologistic models, each with a different neighbourhood template (Fig. 1) and uniform weights $w_{nm}=1$. In all models, we include the spurious covariate 'djungle' along with 'rain' (Dormann *et al.* 2007), i.e. we specify models as:

$$\Pr(y_n=1)=\frac{e^{\alpha+\beta_1 \mathrm{rain}_n+\beta_2 \mathrm{djungle}_n+\beta_{\mathrm{auto}}\ \mathrm{autocov}_n}}{1+e^{\alpha+\beta_1 \mathrm{rain}_n+\beta_2 \mathrm{djungle}_n+\beta_{\mathrm{auto}}\ \mathrm{autocov}_n}}\ .$$

Estimation was then repeated using invalid weightings (8) with $w^*_{nm}=1$. The results (Table 3) broadly support both elements of our interpretation: increasing grid density reduces errors due to invalid weighting and the 'uncorrelated rain' parameter incurs very small errors.

# Autologistic estimation of red deer data

The 'snouter' analysis showed, using simulated data generated with high intrinsic SAC and high covariate SAC, that invalid neighbourhood weightings (8) can lead to drastic underestimation of covariate parameters for each of the three auto-models considered. We now re-examine the presence-absence dataset for red deer in the Grampians region of Scotland, previously studied using logistic (Buckland & Elston 1993) and auto-logistic (Augustin, Mugglestone & Buckland 1996) models. The data consists of a full set of presence-absence observations for 1277 squares of side 1km, each aligned with the UK grid co-ordinates (*eastings* and *northings*), along with environmental covariates including *altitude, pine* and *mires* (Fig. 4). We consider an autologistic model

$$\log\frac{p_n}{1-p_n}=\alpha+\beta_1 easting_n+\beta_2 northing_n+\beta_3 altitude_n^2+\beta_4 pine_n+\beta_5 mires_n+\beta_{\mathrm{auto}}\mathrm{autocov}_n \quad (13)$$

that includes the full set of covariates used by Augustin, Mugglestone and Buckland (1996). We also examine all submodels of (13) derived by deleting covariates (always retaining the autocovariate and at least one of the covariates). Each of the resulting 31 models is estimated on the full dataset and also on four geographic subsets arising when the study area is bisected by eastings or northings. Estimation is conducted using a second order neighbourhood (Fig. 1) with valid uniform weighting $w_{nm}=1$ in (5). Estimation is then repeated with the invalid weighting derived by putting $w^*_{nm}=1$ in (8). We then examine the ratio between estimates obtained with invalid and valid weightings for each covariate parameter in each model.

Our results (Table 4 and supplementary material) show that the full model used in Augustin, Mugglestone et al. (1996) incurs only minor errors in covariate estimates when the entire dataset is estimated using invalid weightings with a second-order neighbourhood. However substantial errors can arise using subsets of the covariates or when the full model is applied to geographic subsets such as the eastern half of the study area. Larger neighbourhoods could be expected to be more sensitive to invalid weightings, since more sites will have incomplete neighbourhoods. Repeating the above analysis for a radius-2 neighbourhood (12 sites) generates more extreme variation in the effect of invalid weightings (see results in supplement) than for the second-order (eight-site) case.

From the example provided by this analysis of the red deer dataset, it appears that the extent of covariate estimation errors due to invalid weightings varies widely and is sensitive to the number and type of covariates, neighbourhood size and details of the available data, including covariate behaviour near borders.

# Autologistic estimation with missing data

### Iteration of Gibbs sampling and MPLE

The approach described by Augustin, Mugglestone and Buckland (1996) uses a series of iterations, each comprising a Gibbs sampling step for the missing data and an MPLE step for the autologistic model parameters. As we have seen already, it is essential to ensure that valid neighbourhood weightings are used in estimation. However, valid weightings are also necessary for the Gibbs sampling steps, which construct samples from the joint missing-data distribution (given the model parameter values for the current iteration) via a sequence of samples from the full conditional distributions of individual sites. If these site-wise conditional distributions use incorrect neighbourhood weightings, the Gibbs sampling procedure will not correspond to a valid joint distribution for the missing data.

Here we provide R code that implements the combined iteration of MPLE and Gibbs sampling using valid distance-based neighbourhood weighting schemes (11) for any given radius $r$. Weights can be uniform $(f=1)$ or can vary with inter-site distance $d$ as a power law $f=d^{-a}$, an exponential $f=e^{-\frac{d}{R}}$ or a product of these $f=d^{-a}e^{-\frac{d}{R}}$. Our code allows covariates and observations to be given for any subset of a rectangular grid and any number $S_{\text{miss}}$ of the presence-absence observations can be missing. Missing observations are labelled in the sequence $1,2,\ldots,\text{S}_{\text{miss}}$, according to their occurrence in the data file, so each label $l$ in the sequence indexes one site-number $n(l)$ with missing observation $y_{n(l)}$. The code implements the following procedure:

- Using the distance-based neighbourhood template (11), calculate weights $w_{nm}$ to be used in all $\text{autocov}_n$ evaluations (5), resulting in a valid weighting scheme.
- Fit a logistic model to the observed presence-absence data, obtaining initial values, $\alpha^1$ and $\boldsymbol{\beta}^1_{\text{cov}}$, for the intercept and covariate parameter vector respectively. Set $\beta^1_{\text{auto}}=0$.
- Set $y_{n(l)}=0$ for $l=1,2,\ldots S_{\text{miss}}$.
- **For** $\left(t=2,3,\ldots\text{T}\right)$:
  1. Carry out a single Gibbs sampling sweep of all missing observations:

**For** $\left(l=1,2,\ldots S_{\text{miss}}\right)$:

**1.1** Evaluate $\text{autocov}_{n(l)}$.

**1.2** Assign a value to $y_{n(l)}$ by sampling from the conditional distribution given all other sites:

$$\Pr(y_{n(l)}=1)=\frac{e^{\alpha^{t-1}+\boldsymbol{\beta}_{\text{cov}}^{t-1}\cdot\mathbf{X}_{n(l)}+\beta_{\text{auto}}^{t-1}\ \text{autocov}_{n(l)}}}{1+e^{\alpha^{t-1}+\boldsymbol{\beta}_{\text{cov}}^{t-1}\cdot\mathbf{X}_{n(l)}+\beta_{\text{auto}}^{t-1}\ \text{autocov}_{n(l)}}},$$

where $\mathbf{X}_n$ is the vector of covariates at site $n$.

**2.** Evaluate $\text{autocov}_n$ for all observed sites and include the results as a covariate in a logistic model, fitted to all observed sites, yielding a new set of model parameter estimates $\left(\alpha^t,\ \boldsymbol{\beta}_{\text{cov}}^t,\ \beta_{\text{auto}}^t\right)$.

**3.** If $\left(t>t_{\min}\right)$: Update running averages for parameter values and occupancy probabilities $\Pr(y_n=1)$ for all sites, where $t_{\min}$ is large enough to ensure adequate convergence.

**4.** **If** $\left(t\in\{t_R\}\right)$: Retain full set of $y_n$ to characterize species distribution in the estimated model at a set of selected iterations $\{t_R\}$.

As an illustration, we apply the above algorithm to a 20% subsample of the red deer data, sampled from the full study area, using a second order neighbourhood, uniform weights and 100 Gibbs-MPLE iterations. Estimated occupancy probabilities and the final stochastic occupancy realization are shown in Fig. 5.

Since this code calculates weightings and autocovariates without using 'spdep', it can be used to numerically test the `style="B"` option in 'spdep' by estimating fully observed datasets (for which 'spdep' can generate an autocovariate). For the fully observed deer dataset we find that a logistic model, containing an autocovariate generated in 'spdep' using `style="B"`, yields identical estimates to the above code.

# Discussion

### Extent and consequences of invalid weightings

The erroneous conclusions of Dormann et al., in which invalid weightings played the principal role, have led ecologists to avoid the autologistic model, notwithstanding the partial critique of Betts, Ganio et al. (2009). For example, Ficetola, Manenti et al. (2012) decide explicitly against the autologistic model and Sol, Vila et al. (2008) exclude it from their analysis. Furthermore, reviewing the 20 papers on autologistic models listed by Miller, Franklin et al. (2007), we find that at least nine used invalid weightings (three did not specify their weighting schemes). Beale et al. (2010) used invalid weightings in uncentered auto-normal models (described as 'simple auto-regressive' models).

We know on theoretical grounds that invalid neighbourhood weightings of the form (8) will lead to estimation errors; since invalid weightings break symmetry only near the boundary, small errors might be anticipated. However, re-analysis of simulated 'snouter' data and red deer census data demonstrated otherwise, yielding dramatic errors in the former case, while a variety of errors occur for the red deer example, depending on which subset of covariates and observations are used.

Invalid weightings generated incorrect estimates, similar to estimates reported in Dormann et al., when analysing new 'snouter' presence-absence data generated by Gibbs sampling of an autologistic model. Therefore the large covariate estimation errors were not artefacts of Dormann et al.'s ad-hoc approach to injecting strong intrinsic SAC into simulated 'snouter' data. Estimation errors were reduced when analysing data simulated on a denser grid using the same covariates (i.e. reducing the grid spacing over the simulated study area), as expected due to reduced boundary effects in larger lattices. Substituting a new "rain" covariate with minimal SAC yielded very small errors. Finally, using valid weighting schemes resolved autologistic estimation problems under all of the above conditions.

For the 'snouter' datasets, covariate parameters estimated by correctly implemented autocovariate regression turn out to be very close to corresponding parameters driving the generating model for the data and are also similar to GLM estimates.

However, due to the much greater complexity of auto-models compared to linear models, this simple outcome need not always occur. Further analysis can therefore be required to resolve apparent estimation anomalies. We pursue such analyses elsewhere (Bardos, Guillera-Arroita & Wintle 2015) for more complex cases (Carl & Kühn 2007; Beale *et al.* 2010) where anomalies in auto-model covariate estimation have been reported.

**Current outlook**

We have shown here that the estimation anomalies reported by Dormann et al., affecting covariate parameters in auto-models, arise from purely technical implementation errors and are easily resolved. Our results demonstrate that invalid neighbourhood weightings can generate large errors when autologistic models with spatially autocorrelated environmental covariates are estimated on occupancy and abundance data. Environmental covariates usually do exhibit SAC, however covariates can take arbitrary form and thus, while severe problems are clearly possible, we have no basis for making general assessments of the extent of poor estimation in other studies in the literature. The only practical way to make an assessment in a particular case is to reanalyse the data using correct weightings, which is straightforward where the data can still be obtained.

The situation for new analyses is much clearer: the family of auto-models are valid probability distributions that can be estimated correctly provided the conditions originally noted by Besag (1974) are enforced. This requires restricting auto-Poisson models to competitive interactions, limiting their utility in ecology. However, for the autologistic model, the conditions are met simply by using valid neighbourhood weightings.

## Acknowledgements

We thank Nicole Augustin, David Elston and Stephen Buckland for their concerted and successful efforts to recover the original red deer dataset. This work was supported by the Australian Research Council (ARC) Centre of Excellence for Environmental Decisions and an ARC Future Fellowship to BW.

# Appendix

The positive-definiteness condition (see Table 1) for uncentered auto-normal models, i.e. the requirement $\mathbf{y}^{\mathrm{T}}\mathbf{B}\mathbf{y} > 0$ for any non-zero vector $\mathbf{y}$, can be derived as follows. The relationship between joint and conditional distributions for auto models is given by eqns (4.1) and (4.6) of Besag (1974). Applying this to the uncentered auto-normal model from Table 1, the joint probability distribution for observations $y_i,\ i = 1, 2,\ \ldots N,$ must be of the form

$$\mathrm{P}(y_1, y_2,\ \ldots\ y_N) \propto \mathrm{e}^{-\left(\frac{1}{\sigma^2}\boldsymbol{\alpha}^T\mathbf{y} + \frac{1}{2\sigma^2}\mathbf{y}^T\mathbf{B}\,\mathbf{y}\right)}, \tag{A.1}$$

where $\boldsymbol{\alpha}^T = (\alpha_1, \alpha_2,\ \ldots\ \alpha_N)$ and $\mathbf{B}$, defined in Table 1, is a real symmetric matrix. For this to be a valid distribution, the $N$ - dimensional integral over observations must be finite. i.e.

$$\int \mathrm{e}^{-\left(\frac{1}{\sigma^2}\boldsymbol{\alpha}^T\mathbf{y} + \frac{1}{2\sigma^2}\mathbf{y}^T\mathbf{B}\,\mathbf{y}\right)} \mathrm{d}\mathbf{y} < \infty. \tag{A.2}$$

We now use two properties related to multivariate normal distributions (see Arnold, Castillo & Sarabia 1994, final paragraph, but note an implicit assumption of matrix symmetry). These properties are that (a) an integral of the form (A.2) can be finite only if $\mathbf{B}$ is positive-definite and (b) the integrand then defines a multivariate normal distribution. Thus when the conditions in Table 1 are met, the uncentered auto-normal model is valid and has multivariate-normal joint distribution.

On p. 232 of Besag (1974), in the discussion after the main paper, Besag gives a sufficient condition for validity of the uncentered auto-normal model on a rectangular lattice of sites where the observation $x_{i,j}$ for the site with lattice co-ordinates $(i, j)$ is normally distributed (conditional on all other sites) with mean

$$\mu_{i,j} + \sum_k \sum_l b_{k,l} x_{i-k,j-l}. \tag{A.3}$$

The notation in the above formula is from Section 5.5 of Besag (1974); accordingly we assume that $b_{0,0} = 0$, $b_{-k,-1} = b_{k,1}$, only a finite number of the coefficients $b_{k,1}$ are non-zero and the distributions have a common variance $\sigma^2$. Additionally, for (A.3) to make sense, the summations must be assumed to be restricted to exclude non-existent sites beyond the lattice boundary. A sufficient condition for validity of the above auto-normal model is given on p. 232 as

$$\sum_k \sum_l b_{k,l} z_1^k z_2^l < 1 \ \text{ wherever } |z_1| = |z_2| = 1, \tag{A.4}$$

which will clearly be satisfied if

$$\sum_{k,l} |b_{k,l}| < 1. \tag{A.5}$$

Besag states that (A.4) is derived from the validity condition for the infinite lattice case given in Section 5.5. We are unaware of any detailed proof of this derivation in the literature, however we can interpret it heuristically on the basis that the consistency conditions for a larger lattice should be more severe than for a smaller one, since more constraints must hold simultaneously. Thus, according to this reasoning, conditions sufficient for the infinite lattice would suffice for the finite one.

| Model | Conditional specification | Parameter constraints | Constraint type |
|---|---|---|---|
| auto-logistic | $y_n \sim \text{Bernoulli}\left(\dfrac{e^{\alpha_n + \sum_{m=1\ldots S} \beta_{nm} y_m}}{1+e^{\alpha_n + \sum_{m=1\ldots S} \beta_{nm} y_m}}\right)$ | $\beta_{nn} = 0$<br>$\beta_{nm} = \beta_{mn}$ | necessary and sufficient |
| auto-Poisson | $y_n \sim \text{Poisson}\left(e^{\alpha_n + \sum_{m=1\ldots S} \beta_{nm} y_m}\right)$ | $\beta_{nn} = 0$<br>$\beta_{nm} = \beta_{mn}$<br>$\beta_{mn} \le 0$ | necessary and sufficient |
| auto-normal:<br><br>uncentered with common variance<br><br>or<br><br>Conditional Autoregressive (CAR) with common variance | $y_n \sim \text{Normal}\left(\alpha_n + \sum_{m=1\ldots S} \beta_{nm} y_m,\ \sigma^2\right)$<br><br>$y_n \sim \text{Normal}\left(\alpha_n + \sum_{m=1\ldots S} \beta_{nm}(y_m - \alpha_m),\ \sigma^2\right)$<br><br>Let $\mathbf{y}$ denote the column-vector of observations, i.e. $\mathbf{y}^T = (y_1, y_2, \ldots y_S)$. Define a matrix $\mathbf{B}$ with diagonal elements $B_{nn} = 1$ and off-diagonal elements $B_{nm} = -\beta_{nm}$. | $\beta_{nn} = 0$<br>$\beta_{nm} = \beta_{mn}$<br><br>$\mathbf{y}^{\mathrm{T}}\mathbf{B}\mathbf{y} > 0$ for any non-zero vector $\mathbf{y}$.<br><br>($\mathbf{B}$ is thus 'positive definite'). | necessary and sufficient |
| auto-normal:<br><br>uncentered with common variance, on a rectangular lattice with co-ordinates $(i, j)$ | $y_{i,j} \sim \text{Normal}\left(\alpha_{i,j} + \sum_k \sum_l b_{k,l} y_{i-k,j-l},\ \sigma^2\right)$ | (i)<br>$b_{00} = 0$<br>$b_{-k,-1} = b_{k,1}$<br><br>(ii)<br>$\sum_{k,1} \lvert \mathrm{b}_{k,l} \rvert < 1$ | necessary<br><br><br>sufficient given (i) |

**Table 1.** Validity conditions for various auto-models: the symmetry condition in each case derives from eqn (4.6) of Besag (1974, p. 200); the auto-Poisson condition is given on p. 202 of Besag (1974). The positive-definiteness condition for auto-normal models is given for the CAR model on p. 201 of Besag (1974) and can be derived for the uncentered model from eqns (4.1) and (4.6) of Besag (1974) in conjunction with properties (Arnold, Castillo & Sarabia 1994) related to multivariate normal

distributions. The rectangular-lattice condition is derived from results stated on p. 232 of Besag (1974) in the Discussion following the main paper. Further details on auto-normal conditions are given in the Appendix.

| auto-model | Neighbourhood template in estimated model | ‘rain’ parameter used to generate simulated ‘snouter’ data | Mean of ‘rain’ estimates for 10 ‘snouter’ simulated datasets | |
|---|---|---|---|---|
| | | | valid weighting | invalid weighting |
| Logistic | | -0.002 | | |
| | 1st order | | -0.0017 | -0.00070 |
| | 2nd order | | -0.0019 | -0.00051 |
| | radius-2 | | -0.0020 | -0.00045 |
| | | | | |
| Poisson | | -0.001 | | |
| | 1st order | Note: All Poisson estimates are invalid (even for valid weighting) due to breaching $\beta_{\text{auto}} \leq 0$ condition | -0.00055 | -0.00011 |
| | 2nd order | | -0.00060 | -0.00010 |
| | radius-2 | | -0.00066 | -0.00012 |
| | | | | |
| Normal | | -0.015 | | |
| | 1st order | | -0.014 | -0.00038 |
| | 2nd order | | -0.015 | -0.00020 |
| | radius-2 | | -0.017 | -0.00031 |

**Table 2.** Effect of invalid neighbourhood weighting on auto-model estimates for the 'rain' covariate parameter when analysing ‘snouter’ simulated datasets (Dormann *et al.* 2007). Auto-logistic, auto-Poisson and (uncentered) auto-normal models were fitted to 10 binary, count and real-valued datasets respectively. The ‘rain’ value used in binary simulations was $-0.002$ (Betts *et al.* 2009), not the reported $-0.003$. Neighbourhood templates used are those illustrated in Fig. 1, with $w_{nm} = 1$ in eqn (5) for valid weighting and $w_{nm}^{*} = 1$ in eqn (8) for invalid weighting.

| Driving covariate for autologistic simulations | Grid density used for simulations: number of sites per original 'snouter' site | Neighbourhood template used for autologistic estimation | Mean of 'rain' estimates for 10 simulated presence-absence datasets | |
|---|---|---|---|---|
| | | | valid weighting | invalid weighting |
| 'rain' | | | | |
| | 1 | 1st order | -0.0019 | -0.00063 |
| | 4 | | -0.0021 | -0.0011 |
| | 16 | | -0.0020 | -0.0014 |
| | 1 | 2nd order | -0.0020 | -0.00032 |
| | 4 | | -0.0023 | -0.00062 |
| | 16 | | -0.0022 | -0.00094 |
| | 1 | radius-2 | -0.0020 | 0.00010 |
| | 4 | | -0.0021 | -0.000089 |
| | 16 | | -0.0020 | -0.00057 |
| | | | | |
| 'uncorrelated rain' | | | | |
| | 1 | 1st order | -0.0021 | -0.0020 |
| | 4 | | -0.0020 | -0.0019 |
| | 16 | | -0.0020 | -0.0019 |
| | 1 | 2nd order | -0.0021 | -0.0020 |
| | 4 | | -0.0023 | -0.0021 |
| | 16 | | -0.0022 | -0.0021 |
| | 1 | radius-2 | -0.0021 | -0.0020 |
| | 4 | | -0.0025 | -0.0024 |
| | 16 | | -0.0022 | -0.0021 |

**Table 3.** Effect of observation grid density and covariate spatial autocorrelation on the errors induced by invalid neighbourhood weighting. The 'rain' parameter used in the simulations is $\beta_1 = -0.002$.

| | environmental covariates | | | | |
|---|---|---|---|---|---|
| | *eastings* | *northings* | *altitude$^2$* | *pine* | *mires* |
| **whole dataset** | | | | | |
| submodel 6 | | | 89 | 98 | |
| submodel 22 | 104 | | 98 | 84 | |
| submodel 23 | 105 | | 98 | 71 | 101 |
| full model | 104 | 103 | 99 | 87 | 101 |
| | | | | | |
| **eastern sector** | | | | | |
| submodel 2 | | | | 210 | |
| submodel 3 | | | | 176 | 93 |
| submodel 6 | | | 79 | 204 | |
| submodel 14 | | -40 | 60 | 175 | |
| submodel 15 | | 779 | 53 | 157 | 96 |
| full model | 81 | 113 | 78 | 149 | 102 |
| | | | | | |
| **western sector** | | | | | |
| submodel 5 | | | 100 | | -8 |
| submodel 6 | | | 101 | 94 | |
| submodel 7 | | | 101 | 94 | 55 |
| full model | -187 | 96 | 100 | 95 | 122 |
| | | | | | |
| **northern sector** | | | | | |
| submodel 6 | | | 79 | 43 | |
| submodel 10 | | 69 | | -23 | |
| submodel 26 | -76 | 71 | | -5 | |
| full model | 110 | 73 | 98 | 97 | 98 |
| | | | | | |
| **southern sector** | | | | | |
| submodel 6 | | | 90 | 70 | |
| full model | 103 | 149 | 100 | 73 | 103 |

**Table 4**. Impact of invalid neighbourhood weightings for the analysis of red deer data. Covariate parameter estimates obtained using invalid neighbourhood weightings are expressed as a percentage of corresponding estimates using valid weightings. The analysis was carried out using the full dataset and also for various geographic divisions. Results are shown for the full model given in eqn (13) and for the *reduced model* of Augustin, Mugglestone and Buckland (1996) - submodel 6 in our analysis. Selected submodels are also included that were particularly sensitive to errors induced by invalid weightings. Full results are available as supplementary material.

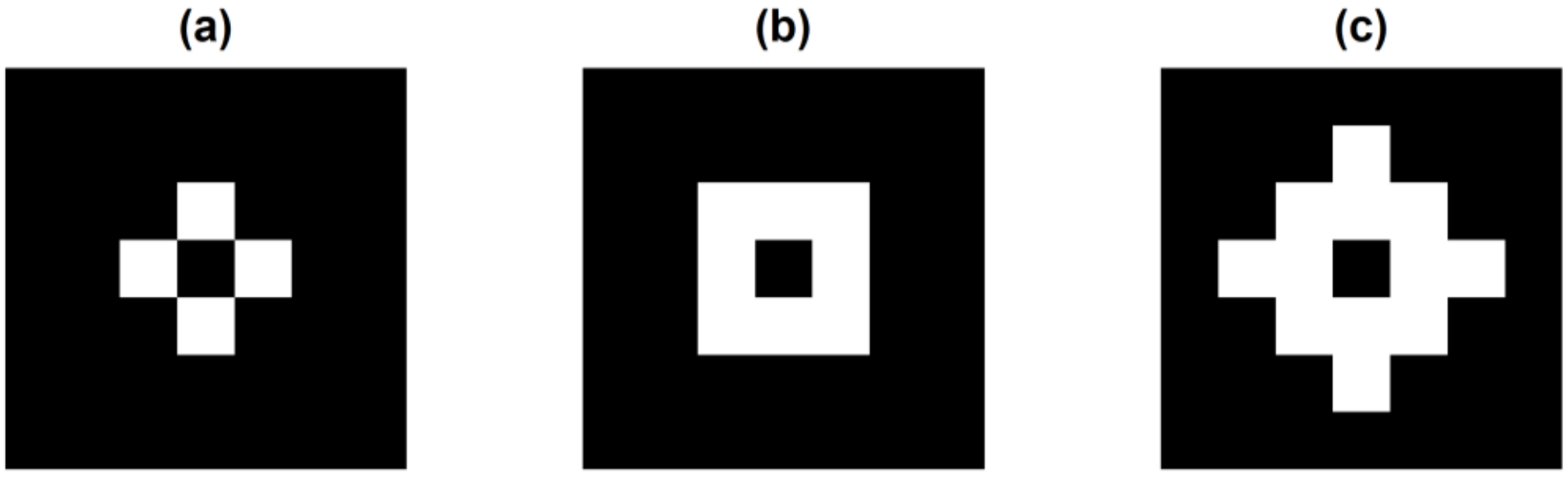

**Figure 1.** Images of neighbourhood templates on a square lattice: (a) 1st order, (b) 2nd order, and (c) radius-2.

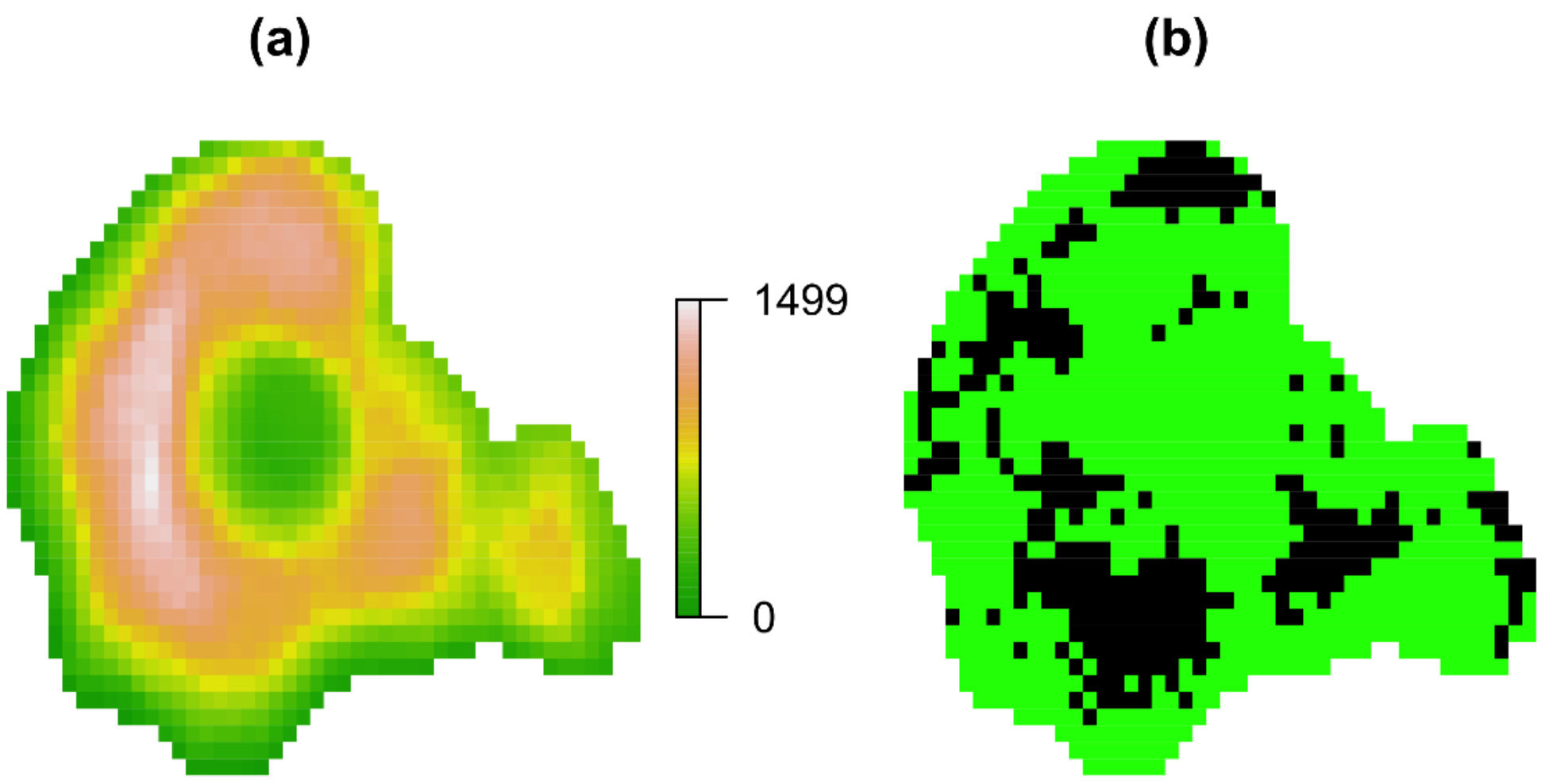

**Figure 2.** (a) Environmental covariate 'rain' used to drive simulated data generation in Dormann (2007) and Dormann et al. (2007). (b) One of the resulting presence-absence simulations for 'snouter' (green=presence; black=absence).

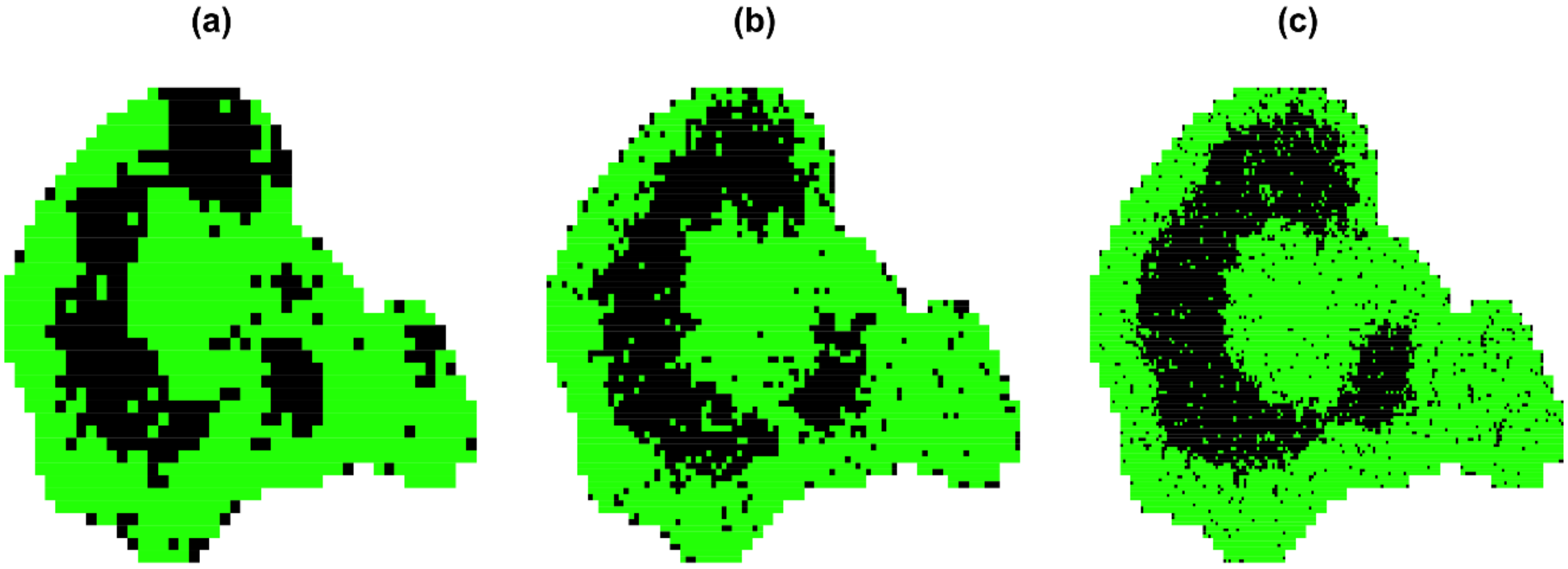

**Figure 3.** Presence-absence data simulated using the autologistic model on (a) the original 'snouter' grid (b) a grid with lattice spacing of 1/2 the original (c) a grid with lattice spacing 1/4 the original (green=presence; black=absence). The simulations used a 1st order uniform neighbourhood and parameters: intercept $\alpha = -1$, 'rain' parameter $\beta_1 = -0.002$ and autocovariate parameter $\beta_{\text{auto}} = 1.5$.

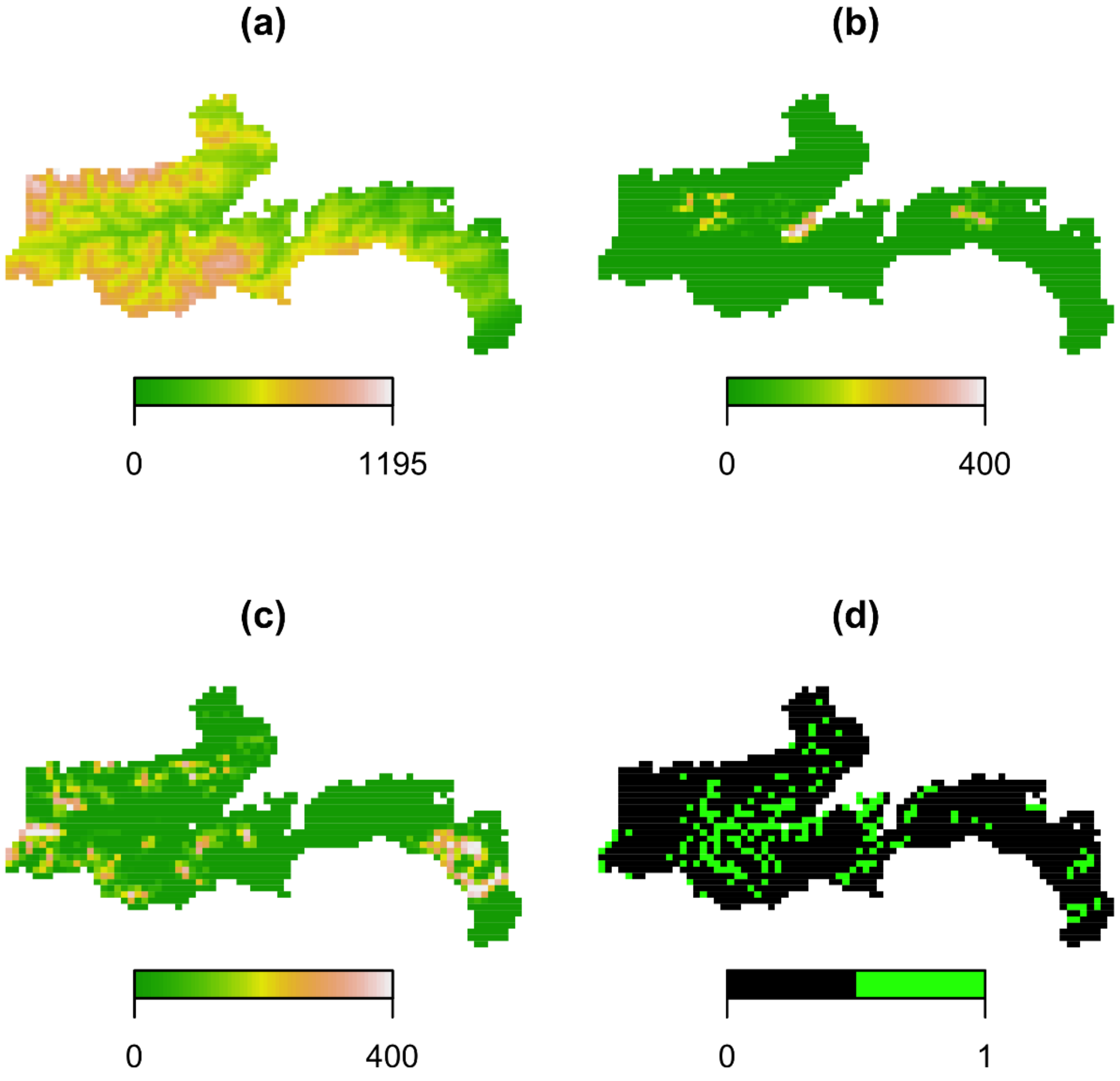

**Figure 4.** Environmental variables and presence-absence observations used in the red deer analysis: (a) altitude, (b) pine, (c) mires, (d) observations.

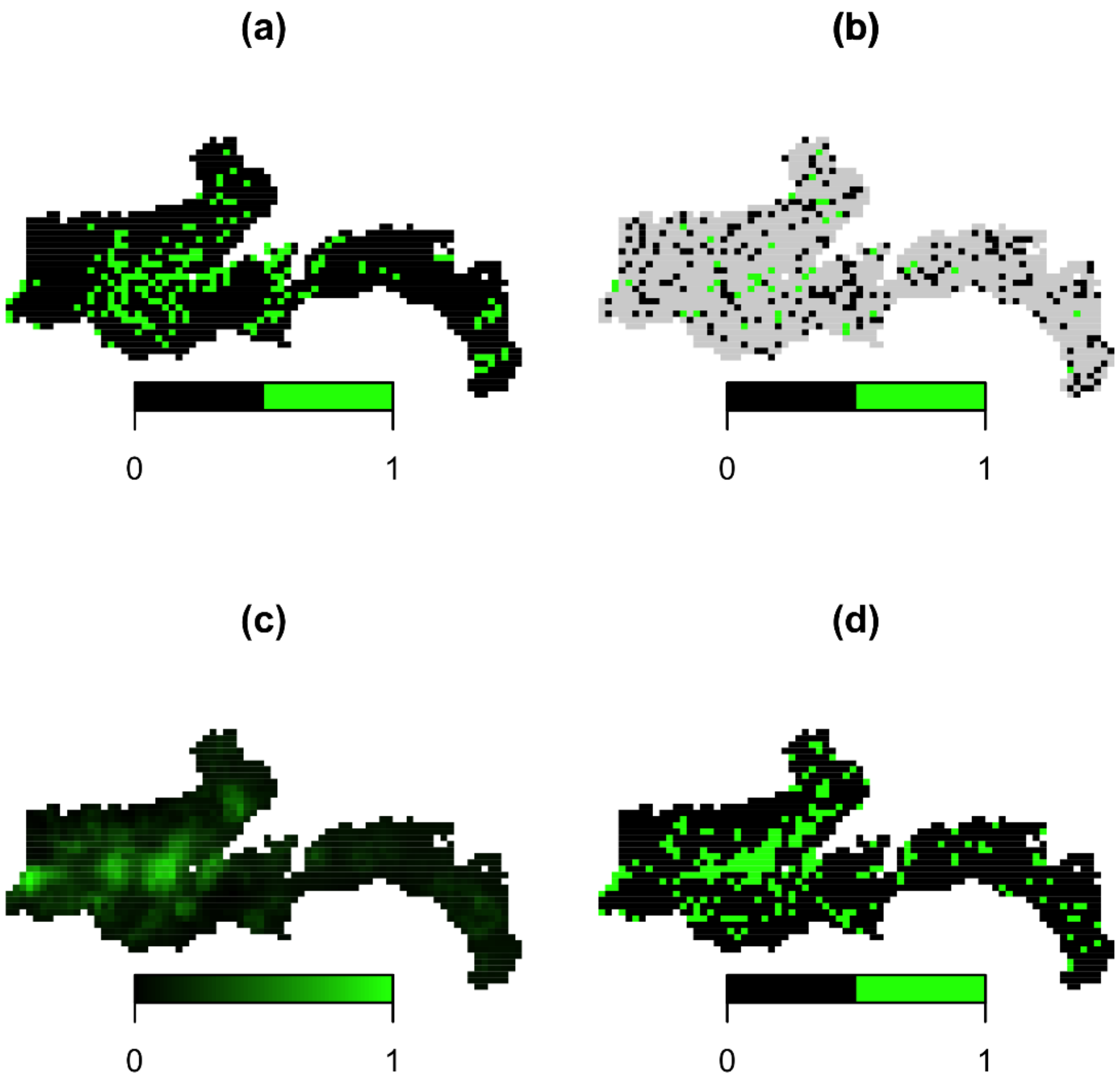

**Figure 5.** Analysis of a partial red deer dataset: (a) fully observed presence-absence dataset; (b) 20% random subsample of data; (c) estimated probabilities of presence; and (d) stochastic presence-absence map after 100 Gibbs-MPLE iterations.

# Supplementary information

## R programs and some associated data and output files.

Most of the R programs detailed below use the data file 'snouterdata.txt'

http://www.ecography.org/sites/ecography.org/files/appendix/snouterdata.txt

from the supplementary material of Dormann *et al.* (2007). However, 'Red deer estimation.R' and 'GibbsMPLE.R' use data files 'deer table.txt' and 'deer20%sample.txt' respectively; these data files are reproduced below along with the output file 'red deer estimate ratios.txt' generated in the former case.

## CheckWeights.R

```
# Checks neighbourhood weights generated for 'snouter' data by 'nb2listw' in package 'spdep'
# Author: David C Bardos
library(lattice);
setwd('C:/auto'); #change as required
snouter.df <- read.table("snouterdata.txt", header=T, sep="\t")
attach(snouter.df)
sorted <- snouter.df[order(rain),]
rain <- snouter.df[,"rain"]

require(spdep)

#Make a matrix of coordinates
coords<-as.matrix(cbind(snouter.df$X,snouter.df$Y))
n=length(coords[,1]);wmat=matrix(0,ncol=n,nrow=n); # weights matrix

nb.list <- dnearneigh(coords, 0, 1)


nb.weights <- nb2listw(nb.list) # default weights
for (i in 1:n){
  nblist=nb.list[[i]];nb.n=length(nblist);
  for (m in 1:nb.n) {j=nblist[m];wmat[i,j]=nb.weights[[3]][[i]][m];}
}
dfmat=wmat; #store default weights
DFmat=wmat-t(wmat);DFplot=levelplot(DFmat[1:100,1:100]);
max(abs(DFmat))  # =0.25  => broken symmetry

nb.weights <- nb2listw(nb.list,style="B")
for (i in 1:n){
  nblist=nb.list[[i]];nb.n=length(nblist);
  for (m in 1:nb.n) {j=nblist[m];wmat[i,j]=nb.weights[[3]][[i]][m];}
}
Bmat=wmat-t(wmat);Bplot=levelplot(Bmat[1:100,1:100]);
max(abs(Bmat))  # =0  => weights symmetric


nb.weights <- nb2listw(nb.list,style="W")
for (i in 1:n){
  nblist=nb.list[[i]];nb.n=length(nblist);
  for (m in 1:nb.n) {j=nblist[m];wmat[i,j]=nb.weights[[3]][[i]][m];}
}
wwmat=wmat; # store 'W' weights
Wmat=wmat-t(wmat);Wplot=levelplot(Wmat[1:100,1:100])
max(abs(Wmat)) # =0.25  => broken symmetry
max(abs(dfmat-wwmat))  # =0  => default weights equal 'W' weights


nb.weights <- nb2listw(nb.list,style="C")
for (i in 1:n){
  nblist=nb.list[[i]];nb.n=length(nblist);
  for (m in 1:nb.n) {j=nblist[m];wmat[i,j]=nb.weights[[3]][[i]][m];}
}
Cmat=wmat-t(wmat);Cplot=levelplot(Cmat[1:100,1:100])
```

```
max(abs(Cmat))  # =0  => weights symmetric


nb.weights <- nb2listw(nb.list,style="U")
for (i in 1:n){
  nblist=nb.list[[i]];nb.n=length(nblist);
  for (m in 1:nb.n) {j=nblist[m];wmat[i,j]=nb.weights[[3]][[i]][m];}
}
Umat=wmat-t(wmat);Uplot=levelplot(Umat[1:100,1:100])
max(abs(Umat)) # =0  => weights symmetric


nb.weights <- nb2listw(nb.list,style="S")
for (i in 1:n){
  nblist=nb.list[[i]];nb.n=length(nblist);
  for (m in 1:nb.n) {j=nblist[m];wmat[i,j]=nb.weights[[3]][[i]][m];}
}
Smat=wmat-t(wmat);Splot=levelplot(Smat[1:100,1:100])
max(abs(Smat)) # =0.105908  => broken symmetry
```

## CheckAutocov.R

```
# Checks autocovariate generated for 'snouter' data by 'autocov_dist' in package 'spdep'
# Author: David C Bardos
library(spdep);library(lattice);
setwd('C:/auto'); #change as required

obs.df=read.table("snouterdata.txt", header=T, sep="\t");
obs.df=obs.df[c(1,2,3,4,23)]

#if radius=0 then a 2nd order neighourhood is used. For radius 1,2,3 etc a neighbourhood of that radius is constructed.
radius=2;if(radius>0) {nbimage=matrix(0L,ncol=2*radius+3,nrow=2*radius+3);nbsize=radius;
nbmat=matrix(,ncol=2,nrow=0,byrow=T); # start with empty list
for(i in -radius:radius) for(j in -radius:radius) if((i^2+j^2 <= radius^2) &&(!identical(c(i,j),c(0L,0L)))) {
   tmpmat=matrix(c(i,j),ncol=2,nrow=1,byrow=T);nbmat=rbind(nbmat,tmpmat);nbimage[i+radius+2,j+radius+2]=1L;}
Nnb=length(nbmat[,1]);} else {nbsize=1.5;
#2nd order (lies between radius 1 & radius 2)
Nnb=8;nbmat=matrix(c(1L,0L,-1L,0L,0L,1L,0L,-1L,-1L,-1L,1L,1L,1L,-1L,-1L,1L),ncol=2,nrow=Nnb,byrow=T);
nbimage=matrix(0L,ncol=5,nrow=5);for(i in -1:1) for(j in -1:1) nbimage[i+3,j+3]=1L; nbimage[3,3]=0L;};levelplot(nbimage);


wvec=rep(1,Nnb) # weights w
dx=max(abs(nbmat[,1]));dy=max(abs(nbmat[,2]));border=max(dx,dy); # minimum border requirements
parameters.n=2;covflag=F; # start with autocovariate and constant term

obs.len=length(obs.df[,1]);obs.width=length(obs.df[1,]);
obs.table=matrix(0L,nrow=obs.len,ncol=obs.width);
for (k in 1:obs.len) for (j in 1:obs.width) obs.table[k,j]=obs.df[k,j];
nsites=obs.len;xmin=min(obs.table[,1]);ymin=min(obs.table[,2]);
xmax=max(obs.table[,1]);ymax=max(obs.table[,2]);Length=ymax-ymin+1;width=xmax-xmin+1;
observed=1:obs.len;observed.xy=matrix(0L,nrow=obs.len,ncol=2);
datamat=matrix(0L,nrow=width+2*border,ncol=Length+2*border); # use 1st array index for x (width)
for (k in 1:obs.len) {obs.table[k,1]=obs.table[k,1]-(xmin-1)+border;obs.table[k,2]=obs.table[k,2]-(ymin-1)+border;
                  datamat[obs.table[k,1],obs.table[k,2]]=as.integer(obs.table[k,obs.width]);
                  observed.xy[k,1]=as.integer(obs.table[k,1]);observed.xy[k,2]=as.integer(obs.table[k,2]);
                  }# k loop

sitemat=datamat;
autocov=rep(0,obs.len)
for (k in 1:obs.len){ for (m in 1:Nnb){
  i=nbmat[m,1];j=nbmat[m,2];autocov[k]=autocov[k]+sitemat[observed.xy[k,1]+i,observed.xy[k,2]+j];}}

sitemat=datamat;sitemat[]=0L;for (k in 1:obs.len){sitemat[observed.xy[k,1],observed.xy[k,2]]=1L;};
weightsum=rep(0,obs.len)
for (k in 1:obs.len){ for (m in 1:Nnb){
  i=nbmat[m,1];j=nbmat[m,2];weightsum[k]=weightsum[k]+sitemat[observed.xy[k,1]+i,observed.xy[k,2]+j];}}

autocovWeightedAvge=autocov/weightsum;   # note this is not a valid autocovariate
levelplot(datamat)

####################
coords<-as.matrix(cbind(obs.df[,1],obs.df[,2])); #Make a matrix of coordinates
GLMtype=binomial(link = "logit");
wtype="one";


wstyle="B";
acB <- autocov_dist(obs.df[,5],  coords,  nbs = nbsize, style=wstyle, type = wtype);
max(abs(acB-autocov));
# =0    => autocov_dist with  style='B' and type='one' agrees with a valid uniform weighting.

wstyle="W";
acW <- autocov_dist(obs.df[,5],  coords,  nbs = nbsize, style=wstyle, type = wtype);
max(abs(acW-autocovWeightedAvge));
# =0   => autocov_dist with  style='W' and type='one' agrees with an invalid uniform weighting

acDF <- autocov_dist(obs.df[,5],  coords,  nbs = nbsize, type = wtype);
max(abs(acDF-autocovWeightedAvge));
# =0   => autocov_dist with  default style and type='one' agrees with an invalid uniform weighting
```

## SnouterGLMwithAutocovariate.R

```
# Maximum pseudolikelihood estimation for (fully observed) 'snouter' data.
# Using GLM with an an autocovariate  => equivalent to MPLE.
# For the gaussian case, valid only if positive-definiteness criterion is met
# For the poisson case:  valid only if the estimate for beta_auto is negative or zero (=> invalid for snouter data)
# Author: David C Bardos
require(spdep);require(lattice);
setwd('C:/auto'); #change as required
snouter.df <- read.table("snouterdata.txt", header=T, sep="\t")
attach(snouter.df);
gaussflag=0L;

coords<-as.matrix(cbind(snouter.df$X,snouter.df$Y)); #Make a matrix of coordinates
len=length(coords[,1]);wmat=matrix(0,ncol=len,nrow=len);

nsims=10; #number of simulated datasets
cmat1=cmat2=matrix(0,ncol=4,nrow=nsims);
eminmax=matrix(0,ncol=2,nrow=nsims);#min and max eigenvalues from positive-definiteness test
fm1=list();fm2=list();

GLMtype=gaussian(link="identity");first=5; gaussflag=1L; # column of first gaussian dataset
#GLMtype=binomial(link = "logit"); first=15; # column of first binary dataset
#GLMtype=poisson(link = "log");    first=25; # column of first poisson dataset


wtype="one";
nbsize=1

nb <- dnearneigh(coords, 0, nbsize)
w <- nb2listw(nb,style="B",zero.policy=FALSE)
for (i in 1:len){
  nblist=nb[[i]];nb.n=length(nblist);
  for (m in 1:nb.n) {j=nblist[m];wmat[i,j]=w[[3]][[i]][m];}
}
#levelplot(wmat)


wstyle="B";  # valid weightings

for(k in 1:nsims){
ac <- autocov_dist(snouter.df[,first+k-1],  coords,  nbs = nbsize, style=wstyle, type = wtype);
fm1[[k]] <- glm(snouter.df[,first+k-1] ~ rain + djungle + ac,family=GLMtype);
cmat1[k,]=coefficients(fm1[[k]]);
if(gaussflag==1L)  {eval=eigen(diag(len)-cmat1[k,4]*wmat)$values;eminmax[k,]=c(min(eval),max(eval));}
}
#summary(fm1[[1]])
cmat1
colMeans(cmat1)

wstyle="W";  # invalid weightings

for(k in 1:nsims){
ac <- autocov_dist(snouter.df[,first+k-1],  coords,  nbs = nbsize, style=wstyle, type = wtype);
fm2[[k]] <- glm(snouter.df[,first+k-1] ~ rain + djungle + ac,family=GLMtype);
cmat2[k,]=coefficients(fm2[[k]]);
}
#summary(fm2[[1]])
cmat2
colMeans(cmat2)


if(gaussflag==1L){ eminmax}  # show upper & lower eigenvalues. Must be >0 for positive-definiteness.
if(gaussflag==1L){print("smallest eigenvalue");print(min(eminmax));}
```

## SnouterAutologistic.R

```
# Maximum pseudolikelihood estimation for (fully observed) 'snouter' data.
# Using GLM with an an autocovariate  => equivalent to MPLE.
# Author: David C Bardos
require(spdep);
setwd('C:/auto'); #change as required
snouter.df <- read.table("snouterdata.txt", header=T, sep="\t")
attach(snouter.df);

coords<-as.matrix(cbind(X,Y)); # coordinates list

wtype="one";      # weights independent of distance
nbsize=1.5        # second-order neighbourhood
wstyle="B";       # valid weighting scheme
sim=snouter2.10;  # choose simulation from data frame: snouter2.1 to snouter2.10 are presence-absence simulations
ac <- autocov_dist(sim,  coords,  nbs = nbsize, style=wstyle, type = wtype);
fm1 <- glm(sim ~ rain + djungle + ac,family=binomial);
summary(fm1)
```

## ConstrainedPoisson.R

```
# Auto-Poisson pseudolikelihood is maximized subject to beta_auto<=0. Uses 'constrOptim'
# Author: David C Bardos
require(spdep);
setwd('C:/auto'); #change as required
snouter.df <- read.table("snouterdata.txt", header=T, sep="\t")
attach(snouter.df);

coords<-as.matrix(cbind(snouter.df$X,snouter.df$Y)); #Make a matrix of coordinates

nsims=10;
first=25;# first poisson datset in file
cmat1=cmat2=matrix(0,ncol=4,nrow=nsims);
fm1=list();fm2=list();


wtype="one";
wstyle="B";
nbsize=1.5

# meaning of coeffs:
#coeffs=c(ccoeff,rcoeff,djcoeff,acoeff);

# construct log(pseudo-likelihood) function
llike=function(coeffs){llambda=coeffs[1]+coeffs[2]*rain+coeffs[3]*djungle+coeffs[4]*ac;
                                    lambda=exp(llambda);
                                     sum(-lambda+obs*llambda-lfac)}


for(k in 1:nsims){

obs=snouter.df[,first+k-1];lfac=lfactorial(obs);
ac <- autocov_dist(obs,  coords,  nbs = nbsize, style=wstyle, type = wtype);
fm1[[k]]=constrOptim(c(2,-0.0001,0.001,-0.1),llike,NULL,ui=rbind(c(0,0,0,-1)),ci=c(0),control=list(fnscale=-1))
cmat1[k,]=fm1[[k]]$par;
}
#summary(fm1[[1]])
cmat1
colMeans(cmat1)

wstyle="W";

for(k in 1:nsims){
obs=snouter.df[,first+k-1];lfac=lfactorial(obs);
ac <- autocov_dist(obs,  coords,  nbs = nbsize, style=wstyle, type = wtype);
fm2[[k]]=constrOptim(c(2,-0.0001,0.001,-0.1),llike,NULL,ui=rbind(c(0,0,0,-1)),ci=c(0),control=list(fnscale=-1))
cmat2[k,]=fm2[[k]]$par;
}
#summary(fm2[[1]])
cmat2
colMeans(cmat2)
```

## CAR.R

```
# CAR model with fully observed data
# Author: David C Bardos
require(lattice);require(spdep);
setwd('C:/auto'); #change as required
snouter.df <- read.table("snouterdata.txt", header=T, sep="\t");

attach(snouter.df);

coords<-as.matrix(cbind(snouter.df$X,snouter.df$Y)); #Make a matrix of coordinates
len=length(coords[,1]);wmat=matrix(0,ncol=len,nrow=len);

nsims=10;
first=5;#15 #column of first dataset
cmat1=cmat2=matrix(0,ncol=4,nrow=nsims);
eminmax=matrix(0,ncol=2,nrow=nsims);#min and max eigenvalues from positive-definiteness test
fm1=list();fm2=list();


nbsize=2;
nb <- dnearneigh(coords, 0, nbsize)

w <- nb2listw(nb,style="B",zero.policy=FALSE)

for (i in 1:len){
  nblist=nb[[i]];nb.n=length(nblist);
  for (m in 1:nb.n) {j=nblist[m];wmat[i,j]=w[[3]][[i]][m];}
}
#levelplot(wmat)


for(k in 1:nsims){

fm1[[k]]<-spautolm(snouter.df[,first+k-1]~rain+djungle,listw=w,family="CAR");
cmat1[k,]=coefficients(fm1[[k]]);
eval=eigen(diag(len)-cmat1[k,4]*wmat)$values;eminmax[k,]=c(min(eval),max(eval));
}
#summary(fm1[[1]])
cmat1
colMeans(cmat1)


w <- nb2listw(nb,style="W",zero.policy=FALSE)

for(k in 1:nsims){

fm2[[k]] <- spautolm(snouter.df[,first+k-1]~rain+djungle,listw=w,family="CAR");
cmat2[k,]=coefficients(fm2[[k]]);
}
#summary(fm2[[1]])
cmat2
colMeans(cmat2)
```

## SceneGen.R

```
# scene generation: simulating autologistic data
# Author: David C Bardos
library(lattice);
setwd('C:/auto'); #change as required
snouter.df <- read.table("snouterdata.txt", header=T, sep="\t")
attach(snouter.df);
snouter.occupation=mean(as.matrix(snouter.df[,15:24]));
set.seed(999);
nsims=10;
covariates.n=2;len=length(rain);rainmu=mean(rain);rainsd=sd(rain);

missingfrac=0.0;
alpha=-1.4;beta1=-0.002;beta.auto=1.4;
alpha=-1.5368776;beta1=-0.001598372;  beta.auto=1.532629;  # fit to sim1
alpha=-1;beta1=-0.002;  beta.auto=1.5;

alpha=-1;beta1=-0.002;beta.auto=1.4; # about 60% occupation with uncorrelated rain

Ymax=38;Xmax=46;border=2;
cov1=cov2=matrix(0,nrow=Xmax+2*border,ncol=Ymax+2*border);
for(k in 1:len){i=X[k]+border; j=Y[k]+border;
cov1[i,j]= rnorm(1,rainmu,rainsd);  #uncorrelated covariate
#cov1[i,j]=rain[k];
cov2[i,j]=djungle[k];
 }  #end k loop
trutharray=array(0L,dim=c(Xmax+2*border,Ymax+2*border,nsims));

for(n in 1:nsims){
randmat=matrix(0L,nrow=Xmax+2*border,ncol=Ymax+2*border);
for(k in 1:len){i=X[k]+border; j=Y[k]+border;
randmat[i,j]=as.integer(rbinom(1,1,0.5)==1); #logical coerced into integer
 }  #end k loop


# randmat=initial scene (just independent random ~~50% filling)
truthmat=randmat; # starting scene for Gibbs sampling draw of true "data"

for (iter in 1:400) {
  for(k in 1:len) {i=X[k]+border;j=Y[k]+border;
    #1st order
    prob=exp(alpha+beta1*cov1[i,j]+beta.auto*(truthmat[i+1,j] + truthmat[i-1,j] + truthmat[i,j+1] + truthmat[i,j-1] ));

    # 2nd order
 #   prob=exp(alpha+beta1*cov1[i,j]+beta.auto*(truthmat[i+1,j] + truthmat[i-1,j] + truthmat[i,j+1] + truthmat[i,j-1]  +
  #                         truthmat[i+1,j-1]  +  truthmat[i-1,j+1]   +  truthmat[i-1,j-1]  +  truthmat[i+1,j+1] ));
    # n=2
   # prob=exp(alpha+beta1*cov1[i,j]+beta.auto*(truthmat[i+1,j] + truthmat[i-1,j] + truthmat[i,j+1] + truthmat[i,j-1]  +
    #                        truthmat[i+1,j-1]  +  truthmat[i-1,j+1]   +  truthmat[i-1,j-1]  +  truthmat[i+1,j+1] +
     #                       truthmat[i+2,j] + truthmat[i-2,j] + truthmat[i,j+2] + truthmat[i,j-2]                  ));


    prob=prob/(1+prob);
    truthmat[i,j]<-as.integer(rbinom(1,1,prob)==1);    #logical coerced into integer
  } # k loop
} #  iter loop
trutharray[,,n]=truthmat;
} # n loop
#levelplot(cov1);
levelplot(truthmat);
# this is now the true scene resulting from gibbs sampling at true parameter values, starting with initial scene "randmat"

tableoftruth=matrix(0L,ncol=2+covariates.n+nsims,nrow=len);#ncol=4=> (x,y,cov,obs)
for(k in 1:len) {i=X[k]+border;j=Y[k]+border;
  tableoftruth[k,]=c(X[k],Y[k],cov1[i,j],cov2[i,j],trutharray[i,j,]);
  if(rbinom(1,1,missingfrac)==1) tableoftruth[k,4]=NA; }


write.table(tableoftruth,"1stOrderAuto.C-1B0.002B_auto1.4.uncorrRain.txt",col.names=c("X","Y","cov1","cov2",rep("obs",nsims)))
#write.table(tableoftruth,"alpha=-1beta1=-0.002beta.auto=1.5.txt",col.names=c("X","Y","cov1","cov2",rep("obs",nsims)))

sum(tableoftruth[,5])/len  #occupied fraction for first simulation
sum(tableoftruth[,(2+covariates.n)+1:nsims])/(nsims*len) # occupied fraction across all simulations
```

## SceneGen.g.R

```
# scene generation: simulating autologistic data on grids differing from the supplied covariate grid
# Author: David C Bardos
library(lattice);
setwd('C:/auto'); #change as required
snouter.df <- read.table("snouterdata.txt", header=T, sep="\t")
attach(snouter.df);
snouter.occupation=mean(as.matrix(snouter.df[,15:24]));
set.seed(999);
nsims=10;
covariates.n=2;len=length(rain);rainmu=mean(rain);rainsd=sd(rain);

missingfrac=0.0;
alpha=-1.4;beta1=-0.002;beta.auto=1.4;
alpha=-1.5368776;beta1=-0.001598372;  beta.auto=1.532629;  # fit to sim1
alpha=-1;beta1=-0.002;  beta.auto=1.5;

alpha=-1;beta1=-0.002;beta.auto=1.4; # about 60% occupation with uncorrelated rain

Ymax=38;Xmax=46;border=2;
cov1=cov2=matrix(0,nrow=Xmax+2*border,ncol=Ymax+2*border);
keymat=matrix(0L,nrow=Xmax+2*border,ncol=Ymax+2*border);
for(k in 1:len){i=X[k]+border; j=Y[k]+border;
cov1[i,j]= rnorm(1,rainmu,rainsd); #uncorrelated covariate
#cov1[i,j]=rain[k];
cov2[i,j]=djungle[k];
keymat[i,j]=1L;
 }  #end k loop
levelplot(cov1);


#now create new grid with associated objects
border.g=2;
scalefac=4; # increase linear grid density by this factor (number of cells increases by its square)
Ymax.g=scalefac*Ymax;Xmax.g=scalefac*Xmax;
#Ymax.g=20;Xmax.g=17; # or can just give numbers here.
Yfac=Ymax.g/Ymax;Xfac=Xmax.g/Xmax;
keymat.g=matrix(0L,nrow=Xmax.g+2*border.g,ncol=Ymax.g+2*border.g);
cov1.g=cov2.g=matrix(0,nrow=Xmax.g+2*border.g,ncol=Ymax.g+2*border.g);
X.g=Y.g=1L[-1]; #empty integer vectors
for(i.g in 1:Xmax.g) for(j.g in 1:Ymax.g)
{i=ceiling((i.g-0.5)/Xfac);j=ceiling((j.g-0.5)/Yfac);
 if(keymat[border+i,border+j]==1) {keymat.g[border.g+i.g,border.g+j.g]=1;
                                 cov1.g[border.g+i.g,border.g+j.g]=cov1[border+i,border+j];
                                 cov2.g[border.g+i.g,border.g+j.g]=cov2[border+i,border+j];
                                X.g=append(X.g,i.g);Y.g=append(Y.g,j.g)
                                }
 }  #end x,y loops
len.g=length(X.g);
trutharray=array(0L,dim=c(Xmax.g+2*border.g,Ymax.g+2*border.g,nsims));

for(n in 1:nsims){
randmat.g=matrix(0L,nrow=Xmax.g+2*border.g,ncol=Ymax.g+2*border.g);
for(k in 1:len.g) {i=X.g[k]+border.g;j=Y.g[k]+border.g;
randmat.g[i,j]=as.integer(rbinom(1,1,0.5)==1);#logical coerced into integer
 }  #end k loop

# randmat.g=initial scene (just independent random ~~50% filling)
truthmat.g=randmat.g; # starting scene for Gibbs sampling draw of true "data"

for (iter in 1:400) {
  for(k in 1:len.g) {i=X.g[k]+border.g;j=Y.g[k]+border.g;
    #1st order
    prob=exp(alpha+beta1*cov1.g[i,j]+beta.auto*(truthmat.g[i+1,j] + truthmat.g[i-1,j] + truthmat.g[i,j+1] + truthmat.g[i,j-1] ));

    # 2nd order
 #   prob=exp(alpha+beta1*cov1.g[i,j]+beta.auto*(truthmat.g[i+1,j] + truthmat.g[i-1,j] + truthmat.g[i,j+1] + truthmat.g[i,j-1]  +
 #                         truthmat.g[i+1,j-1]  +  truthmat.g[i-1,j+1]   +  truthmat.g[i-1,j-1]  +  truthmat.g[i+1,j+1] ));
    # n=2
   # prob=exp(alpha+beta1*cov1.g[i,j]+beta.auto*(truthmat.g[i+1,j] + truthmat.g[i-1,j] + truthmat.g[i,j+1] + truthmat.g[i,j-1]  +
    #                       truthmat.g[i+1,j-1]  +  truthmat.g[i-1,j+1]   +  truthmat.g[i-1,j-1]  +  truthmat.g[i+1,j+1] +
     #                      truthmat.g[i+2,j] + truthmat.g[i-2,j] + truthmat.g[i,j+2] + truthmat.g[i,j-2]                 ));
```

```
    prob=prob/(1+prob);
    truthmat.g[i,j]<-as.integer(rbinom(1,1,prob)==1);    #logical coerced into integer
  } # k loop
} #  iter loop
trutharray[,,n]=truthmat.g;
} # n loop

levelplot(truthmat.g);
# this is now the true scene resulting from gibbs sampling at true parameter values, starting with initial scene "randmat"


# new grid
tableoftruth=matrix(0L,ncol=2+covariates.n+nsims,nrow=len.g);#ncol=4=> (x,y,cov,obs)
for(k in 1:len.g) {i=X.g[k]+border.g;j=Y.g[k]+border.g;
  tableoftruth[k,]=c(X.g[k],Y.g[k],cov1.g[i,j],cov2.g[i,j],trutharray[i,j,]);
  if(rbinom(1,1,missingfrac)==1) tableoftruth[k,4]=NA; }

write.table(tableoftruth,"1stOrderAuto.C-1B0.002B_auto1.4.uncorrRain.g4.txt",col.names=c("X","Y","cov1","cov2",rep("obs",nsims)))
#write.table(tableoftruth,"alpha=-1beta1=-0.002beta.auto=1.5.g2.txt",col.names=c("X","Y","cov1","cov2",rep("obs",nsims)))
#write.table(tableoftruth,"alpha=-1beta1=-0.002beta.auto=1.5.g4.txt",col.names=c("X","Y","cov1","cov2",rep("obs",nsims)))


sum(tableoftruth[,5])/len.g  #occupied fraction for first simulation
sum(tableoftruth[,(2+covariates.n)+1:nsims])/(nsims*len.g) # occupied fraction across all simulations
```

## SimGLMauto.R

```
# Maximum Pseudo-Likelihood Estimation for autologistic model with fully observed data
# Author: David C Bardos
require(spdep);
setwd('C:/auto'); #change as required
#snouter.df <- read.table("snouterdata.txt", header=T, sep="\t");
#snouter.df <- read.table("alpha=-1beta1=-0.002beta.auto=1.5.txt", header=T, sep=" ");
#snouter.df <- read.table("alpha=-1beta1=-0.002beta.auto=1.5.g2.txt", header=T, sep=" ");
snouter.df <- read.table("alpha=-1beta1=-0.002beta.auto=1.5.g4.txt", header=T, sep=" ");
attach(snouter.df);
rain <- cov1; djungle <- cov2;
coords<-as.matrix(cbind(snouter.df$X,snouter.df$Y)); #Make a matrix of coordinates

nsims=10;
first=5;#15 #column of first dataset
cmat1=cmat2=matrix(0,ncol=4,nrow=nsims);
fm1=list();fm2=list();

wtype="one";
wstyle="B";
nbsize=1

for(k in 1:nsims){
ac <- autocov_dist(snouter.df[,first+k-1],  coords,  nbs = nbsize, style=wstyle, type = wtype);
fm1[[k]] <- glm(snouter.df[,first+k-1] ~ rain + djungle + ac,family=binomial, data=snouter.df);
cmat1[k,]=coefficients(fm1[[k]]);
}
#summary(fm1[[1]])
cmat1
colMeans(cmat1)

wstyle="W";

for(k in 1:nsims){
ac <- autocov_dist(snouter.df[,first+k-1],  coords,  nbs = nbsize, style=wstyle, type = wtype);
fm2[[k]] <- glm(snouter.df[,first+k-1] ~ rain + djungle + ac,family=binomial, data=snouter.df);
cmat2[k,]=coefficients(fm2[[k]]);
}
#summary(fm2[[1]])
cmat2
colMeans(cmat2)
```

## Red deer estimation.R

## (Requires ‘deer table.txt’ as input, generates ‘red deer estimate ratios.txt’ as output)

```
# Calculate relative percentage errors in covariate estimates due to invalid neighbourhoods.
# Loops over neighbourhood size and geographic subsets of red deer data.
# Author: David C. Bardos
require(lattice);require(spdep);
setwd('C:/deer'); #change as required
dat.df <- read.table("deer table.txt", header=T, sep="");
data.mat=unname(as.matrix(dat.df));
X=data.mat[,1];Y=data.mat[,2];
maxX=max(X);minX=min(X);midX= minX+(maxX-minX)/2;
maxY=max(Y);minY=min(Y);midY= minY+(maxY-minY)/2;
len=length(X); #numberof data points
subsets=list();subsets[['F']]=1L:len; # full dataset
subsets[['E']]=subsets[['W']]=subsets[['N']]=subsets[['S']]=0L[-1]; #empty integer vectors
for(k in 1L:len){
  if(X[k]<midX) subsets[['W']]=append(subsets[['W']],k) else subsets[['E']]=append(subsets[['E']],k);
   if(Y[k]<midY) subsets[['S']]=append(subsets[['S']],k) else subsets[['N']]=append(subsets[['N']],k);
} #k loop

ncovs=5;covlist=c("east + ","north + ","alt2 + ","pine + ","mires + ");
nmodels=2^ncovs;abin=2^((ncovs-1):0);bbin=2*abin;
combins=matrix(0L,ncol=ncovs,nrow=2^ncovs);
for(i in 1:2^ncovs) combins[i,]=as.integer((i %% bbin)>=abin);

models=rep("ac",2^ncovs);
for(k in 1:nmodels) {
for (j in ncovs:1) if(combins[k,j]==1L) models[k]=paste(covlist[j],models[k]);
models[k]=paste("pa ~ ",models[k]);
} #k loop, model list
elist1=elist2=dlist=plist1=plist2=plistm=ratiolist=sgnlist=list();

results=list()
write("","red deer estimate ratios.txt",append=F);

for (nbsize in c(1.5,2)){
for (sset in c('F','E','W','N','S')) { dat.mat=data.mat[subsets[[sset]],];

X=east=dat.mat[,1];Y=north=dat.mat[,2];alt=dat.mat[,3];pine=dat.mat[,4];mires=dat.mat[,5];obs=dat.mat[,6];
maxX=max(X);minX=min(X);maxY=max(Y);minY=min(Y);
len=length(X); #numberof data points
coords<-as.matrix(cbind(X,Y))
pa=obs; alt2=alt^2;
mat=matPA=list();

for(m in 3:6){
mat[[m]]=matPA[[m]]=matrix(0,nrow=maxX-minX+1,ncol=maxY-minY+1);
for(k in 1:len){
  mat[[m]][X[k]-minX+1,Y[k]-minY+1]=dat.mat[k,m]; if(dat.mat[k,m]>0) matPA[[m]][X[k]-minX+1,Y[k]-minY+1]=1;
} #k loop (map entries)
}# m loop

#levelplot(mat[[3]]); #alt
#levelplot(matPA[[3]]);
#levelplot(mat[[4]]); #pine
#levelplot(matPA[[4]]);
#levelplot(mat[[5]]); #mires
#levelplot(matPA[[5]]);
#levelplot(mat[[6]]); #pa

wtype="one";
#nbsize=1.5;

for(k in 1:nmodels) {
wstyle="B";
ac <- autocov_dist(pa,  coords,  nbs = nbsize, style=wstyle, type = wtype);
fmA <- glm(models[k],family=binomial);
elist1[[k]]=coefficients(fmA);

wstyle="W";
ac <- autocov_dist(pa,  coords,  nbs = nbsize, style=wstyle, type = wtype);
fmB <- glm(models[k],family=binomial);
```

```
elist2[[k]]=coefficients(fmB);

dlist[[k]]=(elist2[[k]]-elist1[[k]] ); # difference between incorrect and correct estimates
ratiolist[[k]]=elist2[[k]]/elist1[[k]];
sgnlist[[k]]=sign(ratiolist[[k]]);
plist1[[k]]=sgnlist[[k]]*abs((elist2[[k]]-elist1[[k]] )/elist1[[k]]); # proportional difference
plist2[[k]]=sgnlist[[k]]*abs((elist2[[k]]-elist1[[k]] )/elist2[[k]]); # proportional difference
plistm[[k]]=sgnlist[[k]]*abs(elist2[[k]]-elist1[[k]] )/pmin( abs(elist1[[k]]), abs(elist2[[k]])  ); # max proportional difference
} #k (models)

df0=data.frame(east=rep(0,nmodels),north=rep(0,nmodels),alt.squared=rep(0,nmodels),pine=rep(0,nmodels),mires=rep(0,nmodels))

ratio=em1=em2=dm=pc1=pc2=pcm=pm1=pm2=sgm=df0;sigfigs=3;
for(k in 1:nmodels) {count=1;
for (j in 1:ncovs) if(combins[k,j]==1L) {count=count+1;
        ratio[k,j]=round(unname(100*ratiolist[[k]][count]));#invalid estimates as percentages of valid estimates
        em1[k,j]=signif(unname(elist1[[k]][count]),sigfigs);#correct estimates
        em2[k,j]=signif(unname(elist2[[k]][count]),sigfigs);#invalid estimates
        dm[k,j]=signif(unname(dlist[[k]][count]),sigfigs);# error (difference)
        pm1[k,j]=signif(unname(100*plist1[[k]][count]),sigfigs);# % error relative to correct estimate
        pm2[k,j]=signif(unname(100*plist2[[k]][count]),sigfigs);# % error relative to invalid estimate
        sgm[k,j]=unname(sgnlist[[k]][count]);            # relative sign of the two estimates
        pc1[k,j]=round(unname(100*plist1[[k]][count]));# rounded % error relative to correct estimate
        pc2[k,j]=round(unname(100*plist2[[k]][count]));# rounded % error relative to invalid estimate
        pcm[k,j]=round(unname(100*plistm[[k]][count]));};# rounded % error relative to smallest of the two estimates
} #k loop, model list
for(k in 1:nmodels) for (j in 1:ncovs) {if(combins[k,j]==0L) ratio[k,j]=em1[k,j]=em2[k,j]=dm[k,j]=pc1[k,j]=pc2[k,j]=pcm[k,j]=sgm[k,j]="";
} #k & j loops

results[[paste(sset,nbsize)]]=list(ratio,em1,em2,dm,pm1,pm2,pc1,pc2,pcm);
if(sset=='F') {
write("#################################################################\n","red deer estimate ratios.txt",append=T);
write(paste("Invalid estimates as percentages of valid estimates: radius=",nbsize),"red deer estimate ratios.txt",append=T);
write("\n\n Full red deer dataset\n","red deer estimate ratios.txt",append=T);
}
else {write(paste(sset,"half of red deer dataset: \n\n"),"red deer estimate ratios.txt",append=T);};
write("model# east   north   alt.sq   pine   mires","red deer estimate ratios.txt",append=T);
write.table(cbind(1:nmodels,ratio)[1:(nmodels-1),],"red deer estimate ratios.txt",append=T,row.names=F,col.names=F,quote=F,sep='\t');
write("\n\n","red deer estimate ratios.txt",append=T);
} #sset  (subsets)
} #nbsize

results[['F 1.5']][[8]]
```

## deer table.txt

| "east" | "north" | "alt" | "pine" | "mires" | "obs" |
|---|---|---|---|---|---|
| 346 | 798 | 220.44 | 0 | 0 | 0 |
| 347 | 798 | 217.08 | 0 | 0 | 0 |
| 350 | 798 | 162.6 | 0 | 0 | 0 |
| 340 | 797 | 316.8 | 34 | 0 | 0 |
| 341 | 797 | 213.36 | 0 | 0 | 0 |
| 344 | 797 | 287.52 | 0 | 0 | 0 |
| 345 | 797 | 332.7 | 78 | 0 | 0 |
| 346 | 797 | 314.22 | 61 | 0 | 0 |
| 347 | 797 | 215.58 | 0 | 0 | 0 |
| 348 | 797 | 199.38 | 0 | 0 | 0 |
| 349 | 797 | 215.58 | 0 | 0 | 0 |
| 351 | 797 | 298.2 | 0 | 0 | 0 |
| 352 | 797 | 258.42 | 0 | 0 | 0 |
| 353 | 797 | 175.56 | 0 | 0 | 0 |
| 354 | 797 | 158.34 | 0 | 0 | 0 |
| 355 | 797 | 173.1 | 0 | 0 | 0 |
| 338 | 796 | 421.56 | 0 | 0 | 0 |
| 339 | 796 | 457.32 | 0 | 0 | 1 |
| 340 | 796 | 357.12 | 0 | 0 | 1 |
| 341 | 796 | 273.6 | 0 | 0 | 0 |
| 342 | 796 | 285.54 | 0 | 0 | 0 |
| 343 | 796 | 335.7 | 0 | 0 | 0 |
| 344 | 796 | 414 | 0 | 0 | 0 |
| 345 | 796 | 356.88 | 11 | 0 | 0 |
| 346 | 796 | 319.02 | 91 | 0 | 0 |
| 347 | 796 | 220.68 | 31 | 0 | 0 |
| 348 | 796 | 213.18 | 0 | 0 | 0 |
| 349 | 796 | 301.74 | 0 | 0 | 0 |
| 350 | 796 | 299.22 | 0 | 0 | 0 |
| 351 | 796 | 328.68 | 0 | 0 | 0 |
| 353 | 796 | 212.28 | 0 | 0 | 0 |
| 354 | 796 | 230.34 | 0 | 0 | 0 |
| 329 | 795 | 402.12 | 78 | 0 | 1 |
| 332 | 795 | 360.36 | 56 | 0 | 1 |
| 333 | 795 | 339.48 | 111 | 0 | 0 |
| 337 | 795 | 344.16 | 0 | 0 | 0 |
| 338 | 795 | 500.04 | 0 | 0 | 1 |
| 339 | 795 | 510.12 | 0 | 0 | 0 |
| 340 | 795 | 409.68 | 0 | 0 | 0 |
| 341 | 795 | 392.1 | 0 | 0 | 0 |
| 342 | 795 | 445.8 | 0 | 0 | 0 |
| 343 | 795 | 461.76 | 0 | 0 | 0 |
| 344 | 795 | 364.62 | 2 | 0 | 0 |
| 345 | 795 | 284.52 | 74 | 0 | 0 |
| 346 | 795 | 234.42 | 24 | 0 | 0 |
| 347 | 795 | 213.78 | 9 | 0 | 0 |
| 348 | 795 | 300.54 | 1 | 0 | 0 |
| 349 | 795 | 385.56 | 89 | 0 | 0 |
| 350 | 795 | 397.26 | 0 | 0 | 0 |
| 351 | 795 | 370.2 | 0 | 0 | 0 |
| 352 | 795 | 361.74 | 0 | 0 | 0 |
| 353 | 795 | 295.56 | 0 | 0 | 0 |
| 329 | 794 | 418.74 | 17 | 0 | 1 |
| 330 | 794 | 458.16 | 0 | 0 | 0 |
| 331 | 794 | 352.56 | 0 | 0 | 1 |
| 332 | 794 | 367.68 | 0 | 0 | 1 |
| 333 | 794 | 356.76 | 0 | 0 | 0 |
| 336 | 794 | 350.7 | 0 | 0 | 0 |
| 337 | 794 | 460.92 | 0 | 0 | 0 |
| 338 | 794 | 540.54 | 0 | 0 | 0 |
| 339 | 794 | 517.26 | 0 | 0 | 0 |
| 340 | 794 | 492.48 | 0 | 0 | 0 |
| 341 | 794 | 495.48 | 0 | 0 | 0 |
| 342 | 794 | 446.46 | 0 | 0 | 0 |
| 343 | 794 | 350.94 | 147 | 0 | 0 |
| 344 | 794 | 308.04 | 267 | 0 | 0 |
| 345 | 794 | 351.9 | 95 | 0 | 0 |
| 346 | 794 | 254.4 | 11 | 0 | 0 |
| 347 | 794 | 288.3 | 0 | 0 | 0 |
| 348 | 794 | 348.96 | 34 | 0 | 0 |
| 349 | 794 | 436.32 | 53 | 0 | 0 |
| 350 | 794 | 436.32 | 0 | 0 | 0 |
| 351 | 794 | 424.2 | 28 | 0 | 0 |
| 352 | 794 | 421.32 | 0 | 0 | 0 |
| 353 | 794 | 287.34 | 0 | 0 | 0 |
| 356 | 794 | 226.8 | 0 | 0 | 0 |
| 326 | 793 | 450.24 | 0 | 0 | 0 |
| 327 | 793 | 411.18 | 0 | 0 | 1 |
| 330 | 793 | 413.52 | 0 | 0 | 1 |
| 331 | 793 | 337.2 | 0 | 0 | 1 |
| 332 | 793 | 452.34 | 0 | 0 | 0 |
| 335 | 793 | 306.36 | 0 | 0 | 0 |
| 336 | 793 | 398.94 | 0 | 0 | 0 |
| 337 | 793 | 536.34 | 0 | 0 | 1 |
| 338 | 793 | 550.92 | 0 | 0 | 0 |
| 339 | 793 | 552.12 | 0 | 0 | 0 |
| 340 | 793 | 520.14 | 0 | 0 | 0 |
| 341 | 793 | 475.98 | 0 | 0 | 0 |
| 342 | 793 | 380.1 | 3 | 0 | 0 |
| 343 | 793 | 335.94 | 293 | 0 | 0 |
| 344 | 793 | 421.98 | 175 | 0 | 0 |
| 345 | 793 | 398.16 | 167 | 0 | 0 |
| 346 | 793 | 298.32 | 298 | 0 | 0 |
| 347 | 793 | 294.9 | 168 | 0 | 0 |
| 348 | 793 | 315.78 | 16 | 0 | 0 |
| 349 | 793 | 372.3 | 5 | 0 | 0 |
| 350 | 793 | 435.24 | 0 | 0 | 0 |
| 351 | 793 | 416.94 | 0 | 0 | 0 |
| 352 | 793 | 402.66 | 0 | 0 | 0 |
| 353 | 793 | 317.64 | 0 | 0 | 0 |
| 354 | 793 | 325.8 | 0 | 0 | 1 |
| 355 | 793 | 349.14 | 0 | 0 | 1 |
| 356 | 793 | 311.58 | 0 | 0 | 1 |
| 315 | 792 | 382.44 | 0 | 0 | 0 |
| 322 | 792 | 358.5 | 286 | 0 | 0 |
| 324 | 792 | 397.86 | 0 | 0 | 1 |
| 325 | 792 | 419.4 | 0 | 0 | 0 |
| 326 | 792 | 533.64 | 0 | 0 | 0 |
| 327 | 792 | 482.88 | 0 | 0 | 0 |
| 328 | 792 | 424.08 | 0 | 0 | 0 |
| 329 | 792 | 367.62 | 0 | 0 | 1 |
| 330 | 792 | 347.16 | 0 | 0 | 1 |
| 331 | 792 | 375.9 | 0 | 0 | 0 |
| 332 | 792 | 491.7 | 0 | 0 | 0 |
| 335 | 792 | 351.48 | 0 | 0 | 0 |
| 336 | 792 | 444.42 | 0 | 0 | 1 |
| 337 | 792 | 575.76 | 0 | 0 | 1 |
| 338 | 792 | 597.06 | 0 | 0 | 0 |
| 339 | 792 | 582.96 | 0 | 0 | 0 |
| 340 | 792 | 467.22 | 0 | 0 | 0 |
| 341 | 792 | 368.58 | 0 | 0 | 0 |
| 342 | 792 | 401.94 | 0 | 0 | 0 |
| 343 | 792 | 455.34 | 6 | 0 | 1 |
| 344 | 792 | 521.82 | 0 | 0 | 0 |
| 345 | 792 | 388.02 | 57 | 0 | 0 |
| 346 | 792 | 381.3 | 96 | 0 | 0 |
| 347 | 792 | 385.56 | 322 | 0 | 0 |
| 348 | 792 | 373.5 | 173 | 0 | 0 |
| 349 | 792 | 476.22 | 70 | 0 | 0 |
| 350 | 792 | 479.82 | 0 | 0 | 0 |
| 351 | 792 | 388.86 | 0 | 0 | 0 |
| 352 | 792 | 349.08 | 0 | 0 | 0 |
| 353 | 792 | 332.46 | 0 | 0 | 0 |
| 354 | 792 | 310.92 | 0 | 0 | 0 |
| 355 | 792 | 279.54 | 0 | 0 | 0 |
| 356 | 792 | 286.8 | 0 | 0 | 0 |
| 360 | 792 | 124.38 | 0 | 0 | 0 |
| 361 | 792 | 114.66 | 0 | 0 | 0 |
| 362 | 792 | 102.54 | 0 | 0 | 0 |
| 363 | 792 | 100.2 | 0 | 0 | 0 |
| 315 | 791 | 389.64 | 0 | 0 | 0 |
| 316 | 791 | 562.74 | 0 | 0 | 1 |
| 317 | 791 | 524.88 | 42 | 0 | 0 |
| 321 | 791 | 447.54 | 378 | 0 | 1 |
| 322 | 791 | 478.68 | 188 | 0 | 1 |
| 323 | 791 | 444.72 | 70 | 0 | 1 |
| 324 | 791 | 440.4 | 0 | 0 | 0 |
| 325 | 791 | 434.1 | 0 | 19 | 0 |
| 326 | 791 | 502.14 | 0 | 0 | 0 |
| 327 | 791 | 479.7 | 0 | 0 | 0 |
| 328 | 791 | 399.72 | 0 | 0 | 0 |
| 329 | 791 | 382.86 | 0 | 0 | 1 |
| 330 | 791 | 374.22 | 0 | 0 | 0 |
| 331 | 791 | 417.84 | 0 | 0 | 0 |
| 332 | 791 | 424.38 | 0 | 0 | 1 |
| 335 | 791 | 415.62 | 0 | 0 | 0 |
| 336 | 791 | 557.7 | 0 | 0 | 1 |
| 337 | 791 | 651.54 | 0 | 0 | 0 |
| 338 | 791 | 633.06 | 0 | 0 | 0 |
| 339 | 791 | 606.96 | 0 | 0 | 0 |
| 340 | 791 | 478.5 | 0 | 0 | 0 |
| 341 | 791 | 437.46 | 0 | 0 | 0 |
| 342 | 791 | 556.32 | 0 | 0 | 0 |
| 343 | 791 | 555.84 | 0 | 0 | 1 |
| 344 | 791 | 489.6 | 0 | 0 | 0 |
| 345 | 791 | 419.82 | 108 | 0 | 0 |
| 346 | 791 | 467.88 | 6 | 0 | 1 |
| 347 | 791 | 451.14 | 98 | 0 | 0 |
| 348 | 791 | 464.52 | 105 | 0 | 0 |
| 349 | 791 | 510.9 | 0 | 0 | 0 |
| 350 | 791 | 408.24 | 0 | 0 | 0 |
| 351 | 791 | 299.46 | 0 | 0 | 0 |
| 352 | 791 | 254.34 | 0 | 0 | 0 |
| 353 | 791 | 253.32 | 0 | 0 | 0 |
| 354 | 791 | 261.66 | 0 | 0 | 0 |
| 355 | 791 | 246.9 | 0 | 0 | 0 |
| 356 | 791 | 226.8 | 0 | 0 | 0 |
| 359 | 791 | 218.4 | 0 | 0 | 0 |
| 360 | 791 | 177.72 | 0 | 0 | 0 |
| 362 | 791 | 144.42 | 0 | 0 | 0 |
| 363 | 791 | 138.9 | 0 | 0 | 0 |
| 315 | 790 | 401.16 | 0 | 0 | 1 |
| 316 | 790 | 596.34 | 0 | 0 | 1 |
| 317 | 790 | 638.46 | 0 | 0 | 1 |
| 319 | 790 | 392.22 | 362 | 0 | 1 |
| 320 | 790 | 433.32 | 400 | 0 | 1 |
| 321 | 790 | 541.32 | 266 | 0 | 1 |
| 322 | 790 | 585.72 | 2 | 68 | 0 |
| 323 | 790 | 558.12 | 0 | 21 | 1 |
| 324 | 790 | 511.86 | 0 | 0 | 0 |
| 325 | 790 | 469.26 | 0 | 95 | 0 |
| 326 | 790 | 506.22 | 0 | 124 | 0 |
| 327 | 790 | 494.28 | 0 | 0 | 0 |
| 328 | 790 | 424.44 | 0 | 0 | 0 |
| 329 | 790 | 444.24 | 0 | 0 | 0 |
| 330 | 790 | 412.26 | 0 | 0 | 1 |
| 331 | 790 | 464.58 | 0 | 0 | 1 |
| 332 | 790 | 387.6 | 0 | 0 | 0 |
| 335 | 790 | 475.92 | 0 | 0 | 0 |
| 336 | 790 | 617.7 | 0 | 0 | 0 |
| 337 | 790 | 616.2 | 0 | 0 | 0 |
| 338 | 790 | 541.86 | 0 | 0 | 0 |
| 339 | 790 | 510.96 | 0 | 0 | 0 |
| 340 | 790 | 428.28 | 0 | 0 | 0 |
| 341 | 790 | 476.76 | 0 | 0 | 0 |
| 342 | 790 | 565.8 | 0 | 0 | 0 |
| 343 | 790 | 515.58 | 0 | 0 | 0 |
| 344 | 790 | 479.04 | 0 | 0 | 0 |
| 345 | 790 | 546.9 | 0 | 0 | 0 |
| 346 | 790 | 580.86 | 0 | 0 | 0 |
| 347 | 790 | 451.98 | 0 | 0 | 0 |
| 348 | 790 | 514.68 | 0 | 0 | 1 |
| 349 | 790 | 526.5 | 0 | 0 | 0 |
| 350 | 790 | 432.66 | 0 | 0 | 0 |
| 351 | 790 | 317.7 | 0 | 0 | 0 |
| 352 | 790 | 310.44 | 0 | 0 | 0 |
| 353 | 790 | 377.64 | 0 | 0 | 0 |
| 354 | 790 | 379.32 | 0 | 14 | 0 |
| 355 | 790 | 429.9 | 0 | 20 | 0 |
| 356 | 790 | 365.04 | 0 | 60 | 0 |
| 357 | 790 | 432.06 | 0 | 0 | 0 |
| 358 | 790 | 432.36 | 0 | 0 | 0 |
| 359 | 790 | 356.4 | 0 | 76 | 0 |
| 360 | 790 | 225 | 0 | 0 | 0 |
| 315 | 789 | 399.6 | 0 | 0 | 0 |
| 316 | 789 | 550.74 | 0 | 0 | 0 |
| 317 | 789 | 701.88 | 0 | 0 | 0 |
| 318 | 789 | 579.06 | 141 | 0 | 1 |
| 319 | 789 | 522 | 193 | 44 | 0 |
| 320 | 789 | 542.7 | 182 | 106 | 0 |
| 321 | 789 | 633.6 | 47 | 120 | 1 |
| 322 | 789 | 754.26 | 0 | 119 | 0 |
| 323 | 789 | 724.68 | 0 | 95 | 1 |
| 324 | 789 | 694.32 | 0 | 3 | 0 |
| 325 | 789 | 555.48 | 0 | 197 | 0 |
| 326 | 789 | 537.9 | 0 | 372 | 0 |
| 327 | 789 | 570.06 | 0 | 40 | 0 |
| 328 | 789 | 528.6 | 0 | 0 | 0 |
| 329 | 789 | 464.04 | 0 | 0 | 0 |
| 330 | 789 | 455.46 | 0 | 0 | 0 |
| 331 | 789 | 413.4 | 0 | 0 | 1 |
| 333 | 789 | 454.32 | 0 | 0 | 0 |
| 334 | 789 | 489.66 | 0 | 0 | 0 |
| 335 | 789 | 552.9 | 0 | 0 | 0 |
| 336 | 789 | 664.38 | 0 | 0 | 0 |
| 337 | 789 | 599.34 | 0 | 0 | 0 |
| 338 | 789 | 560.04 | 0 | 0 | 0 |
| 339 | 789 | 565.26 | 0 | 0 | 0 |
| 340 | 789 | 531.42 | 0 | 0 | 0 |
| 341 | 789 | 621.9 | 0 | 0 | 0 |
| 342 | 789 | 716.58 | 0 | 0 | 0 |
| 343 | 789 | 573 | 0 | 0 | 0 |
| 344 | 789 | 614.16 | 0 | 0 | 0 |
| 345 | 789 | 670.56 | 0 | 0 | 0 |
| 346 | 789 | 649.68 | 0 | 0 | 0 |
| 347 | 789 | 541.26 | 0 | 0 | 0 |
| 348 | 789 | 575.46 | 0 | 0 | 0 |
| 349 | 789 | 648.48 | 0 | 0 | 0 |
| 350 | 789 | 520.98 | 0 | 0 | 0 |
| 351 | 789 | 355.32 | 0 | 0 | 0 |
| 352 | 789 | 424.02 | 0 | 0 | 0 |
| 353 | 789 | 535.08 | 0 | 40 | 0 |
| 354 | 789 | 511.38 | 0 | 259 | 0 |
| 355 | 789 | 526.86 | 0 | 222 | 0 |
| 356 | 789 | 499.38 | 0 | 295 | 0 |
| 357 | 789 | 534.54 | 0 | 13 | 0 |
| 358 | 789 | 402.42 | 0 | 2 | 0 |
| 359 | 789 | 300.72 | 0 | 46 | 0 |
| 360 | 789 | 322.68 | 0 | 0 | 0 |
| 361 | 789 | 311.28 | 0 | 0 | 0 |
| 362 | 789 | 244.44 | 0 | 0 | 0 |
| 363 | 789 | 290.34 | 0 | 0 | 0 |
| 364 | 789 | 188.04 | 0 | 0 | 0 |
| 315 | 788 | 454.56 | 0 | 0 | 0 |
| 316 | 788 | 504 | 0 | 0 | 1 |
| 317 | 788 | 673.14 | 0 | 0 | 0 |
| 318 | 788 | 589.74 | 20 | 0 | 1 |
| 319 | 788 | 630.6 | 0 | 172 | 0 |
| 320 | 788 | 605.88 | 0 | 254 | 1 |
| 321 | 788 | 666.06 | 0 | 147 | 1 |
| 322 | 788 | 804.36 | 0 | 25 | 0 |
| 323 | 788 | 818.94 | 0 | 36 | 0 |
| 324 | 788 | 823.56 | 0 | 4 | 0 |
| 325 | 788 | 670.32 | 0 | 24 | 0 |
| 326 | 788 | 591.12 | 0 | 356 | 0 |
| 327 | 788 | 703.2 | 0 | 142 | 0 |
| 328 | 788 | 735.6 | 0 | 0 | 0 |
| 329 | 788 | 507.66 | 0 | 0 | 0 |
| 330 | 788 | 436.68 | 0 | 0 | 1 |
| 331 | 788 | 408.12 | 0 | 0 | 1 |
| 332 | 788 | 472.44 | 0 | 0 | 0 |
| 333 | 788 | 566.76 | 0 | 0 | 0 |
| 334 | 788 | 623.82 | 0 | 0 | 0 |
| 335 | 788 | 621.18 | 0 | 0 | 0 |
| 336 | 788 | 650.7 | 0 | 0 | 1 |
| 337 | 788 | 642.12 | 0 | 0 | 0 |
| 338 | 788 | 650.1 | 0 | 0 | 0 |
| 339 | 788 | 696.06 | 0 | 0 | 0 |
| 340 | 788 | 722.76 | 0 | 0 | 0 |
| 341 | 788 | 723.72 | 0 | 0 | 0 |
| 342 | 788 | 810.24 | 0 | 0 | 0 |
| 343 | 788 | 696.78 | 0 | 0 | 0 |
| 344 | 788 | 711.96 | 0 | 0 | 0 |
| 347 | 788 | 629.7 | 0 | 0 | 0 |
| 348 | 788 | 669.24 | 0 | 0 | 0 |
| 349 | 788 | 692.58 | 0 | 1 | 0 |
| 350 | 788 | 534.84 | 0 | 111 | 0 |
| 351 | 788 | 390.12 | 0 | 1 | 0 |
| 352 | 788 | 516.96 | 0 | 0 | 0 |
| 353 | 788 | 563.1 | 0 | 78 | 0 |
| 354 | 788 | 505.14 | 0 | 384 | 0 |
| 355 | 788 | 477.36 | 0 | 269 | 0 |
| 356 | 788 | 427.5 | 0 | 219 | 0 |
| 357 | 788 | 398.04 | 0 | 143 | 0 |
| 358 | 788 | 392.82 | 0 | 291 | 0 |
| 359 | 788 | 438.3 | 0 | 160 | 0 |
| 360 | 788 | 448.32 | 0 | 132 | 0 |
| 361 | 788 | 380.7 | 0 | 0 | 0 |
| 362 | 788 | 395.46 | 0 | 0 | 0 |
| 363 | 788 | 351.84 | 0 | 0 | 0 |
| 315 | 787 | 539.58 | 0 | 0 | 1 |
| 316 | 787 | 499.02 | 0 | 0 | 0 |
| 317 | 787 | 689.1 | 0 | 91 | 0 |
| 318 | 787 | 710.4 | 0 | 0 | 0 |
| 319 | 787 | 702.48 | 0 | 305 | 0 |
| 320 | 787 | 683.04 | 0 | 275 | 0 |
| 321 | 787 | 741.96 | 0 | 117 | 0 |
| 322 | 787 | 799.98 | 0 | 112 | 0 |
| 323 | 787 | 905.58 | 0 | 8 | 0 |
| 324 | 787 | 891.72 | 0 | 0 | 0 |
| 325 | 787 | 781.2 | 0 | 0 | 0 |
| 326 | 787 | 686.58 | 0 | 160 | 0 |
| 327 | 787 | 742.74 | 0 | 137 | 0 |
| 328 | 787 | 707.28 | 0 | 0 | 0 |
| 329 | 787 | 501.42 | 0 | 0 | 0 |
| 330 | 787 | 410.22 | 0 | 0 | 0 |
| 331 | 787 | 424.38 | 0 | 0 | 0 |
| 332 | 787 | 501.6 | 0 | 0 | 1 |
| 333 | 787 | 624.6 | 0 | 0 | 0 |
| 334 | 787 | 642.36 | 0 | 0 | 1 |
| 335 | 787 | 636.36 | 0 | 0 | 1 |
| 336 | 787 | 645 | 0 | 0 | 0 |
| 337 | 787 | 636.72 | 0 | 0 | 0 |
| 338 | 787 | 711.18 | 0 | 0 | 0 |
| 339 | 787 | 728.28 | 0 | 0 | 0 |
| 340 | 787 | 803.94 | 0 | 0 | 0 |
| 341 | 787 | 781.92 | 0 | 0 | 0 |
| 342 | 787 | 798.42 | 0 | 0 | 0 |
| 349 | 787 | 664.08 | 0 | 0 | 0 |
| 350 | 787 | 567.06 | 0 | 43 | 0 |
| 351 | 787 | 543.84 | 0 | 58 | 0 |
| 352 | 787 | 525.36 | 0 | 214 | 0 |
| 353 | 787 | 521.04 | 0 | 274 | 0 |
| 354 | 787 | 486.66 | 0 | 224 | 0 |
| 355 | 787 | 495.96 | 0 | 281 | 0 |
| 356 | 787 | 430.32 | 0 | 391 | 0 |
| 357 | 787 | 491.22 | 0 | 296 | 0 |
| 358 | 787 | 530.22 | 0 | 378 | 0 |
| 359 | 787 | 518.64 | 0 | 394 | 0 |
| 360 | 787 | 514.92 | 0 | 400 | 0 |
| 361 | 787 | 480.9 | 0 | 62 | 0 |
| 362 | 787 | 345.72 | 0 | 0 | 0 |
| 363 | 787 | 229.74 | 0 | 0 | 0 |
| 364 | 787 | 193.8 | 0 | 0 | 0 |
| 315 | 786 | 629.4 | 0 | 0 | 1 |
| 316 | 786 | 510 | 0 | 0 | 0 |
| 317 | 786 | 627 | 0 | 56 | 0 |
| 318 | 786 | 742.74 | 0 | 105 | 0 |
| 319 | 786 | 723.96 | 0 | 336 | 0 |
| 320 | 786 | 802.02 | 0 | 18 | 0 |
| 321 | 786 | 917.82 | 0 | 0 | 0 |
| 322 | 786 | 915.6 | 0 | 0 | 0 |
| 323 | 786 | 955.74 | 0 | 0 | 0 |
| 324 | 786 | 1022.28 | 0 | 0 | 0 |
| 325 | 786 | 941.28 | 0 | 0 | 0 |
| 326 | 786 | 803.22 | 0 | 0 | 0 |
| 327 | 786 | 673.44 | 0 | 0 | 0 |
| 328 | 786 | 576.12 | 0 | 0 | 0 |
| 329 | 786 | 456.42 | 0 | 0 | 1 |
| 330 | 786 | 400.98 | 0 | 0 | 0 |
| 331 | 786 | 480.96 | 0 | 0 | 1 |
| 332 | 786 | 541.14 | 0 | 0 | 0 |
| 333 | 786 | 655.74 | 0 | 0 | 0 |
| 334 | 786 | 689.64 | 0 | 0 | 0 |
| 350 | 786 | 594.18 | 0 | 9 | 0 |
| 351 | 786 | 608.04 | 0 | 20 | 0 |
| 352 | 786 | 566.4 | 0 | 84 | 0 |
| 353 | 786 | 515.1 | 0 | 160 | 0 |
| 354 | 786 | 537.24 | 0 | 289 | 0 |
| 355 | 786 | 571.98 | 0 | 134 | 0 |
| 356 | 786 | 520.38 | 0 | 320 | 0 |
| 357 | 786 | 534.42 | 0 | 178 | 0 |
| 358 | 786 | 541.08 | 0 | 359 | 0 |
| 359 | 786 | 542.28 | 0 | 400 | 0 |
| 360 | 786 | 508.86 | 0 | 396 | 0 |
| 361 | 786 | 414.36 | 0 | 183 | 0 |
| 362 | 786 | 332.22 | 0 | 0 | 1 |
| 363 | 786 | 239.1 | 0 | 0 | 0 |
| 364 | 786 | 205.26 | 0 | 0 | 0 |
| 314 | 785 | 474.42 | 0 | 0 | 1 |
| 315 | 785 | 713.16 | 0 | 0 | 0 |
| 316 | 785 | 601.68 | 0 | 0 | 1 |
| 317 | 785 | 552 | 0 | 8 | 0 |
| 318 | 785 | 644.7 | 0 | 63 | 0 |
| 319 | 785 | 754.92 | 0 | 112 | 0 |
| 320 | 785 | 953.82 | 0 | 0 | 0 |
| 321 | 785 | 971.58 | 0 | 0 | 0 |
| 322 | 785 | 1044.3 | 0 | 0 | 0 |
| 323 | 785 | 1022.22 | 0 | 0 | 0 |
| 324 | 785 | 994.38 | 0 | 0 | 0 |
| 325 | 785 | 943.02 | 0 | 0 | 0 |
| 326 | 785 | 839.64 | 0 | 0 | 0 |
| 327 | 785 | 694.92 | 0 | 0 | 0 |
| 328 | 785 | 647.4 | 0 | 0 | 0 |
| 329 | 785 | 441.78 | 0 | 0 | 0 |
| 330 | 785 | 444.24 | 0 | 0 | 0 |
| 331 | 785 | 566.52 | 0 | 0 | 0 |
| 332 | 785 | 614.82 | 0 | 0 | 0 |
| 333 | 785 | 669.54 | 0 | 0 | 0 |
| 354 | 785 | 678 | 0 | 37 | 0 |
| 355 | 785 | 654.12 | 0 | 197 | 0 |
| 356 | 785 | 557.88 | 0 | 353 | 0 |
| 357 | 785 | 453.84 | 0 | 185 | 0 |
| 358 | 785 | 411.9 | 0 | 28 | 0 |
| 359 | 785 | 419.7 | 0 | 108 | 0 |
| 360 | 785 | 418.68 | 0 | 336 | 0 |
| 361 | 785 | 394.8 | 0 | 208 | 0 |
| 362 | 785 | 311.52 | 0 | 0 | 0 |
| 363 | 785 | 271.62 | 0 | 0 | 1 |
| 364 | 785 | 227.1 | 0 | 0 | 0 |
| 314 | 784 | 561.12 | 0 | 0 | 0 |
| 315 | 784 | 662.04 | 0 | 18 | 1 |
| 316 | 784 | 662.64 | 0 | 171 | 0 |
| 317 | 784 | 604.2 | 0 | 124 | 1 |
| 318 | 784 | 556.38 | 0 | 39 | 0 |
| 319 | 784 | 700.26 | 0 | 23 | 0 |
| 320 | 784 | 917.52 | 0 | 12 | 0 |
| 321 | 784 | 919.98 | 0 | 0 | 0 |
| 322 | 784 | 884.16 | 0 | 0 | 0 |
| 323 | 784 | 906.06 | 0 | 0 | 0 |
| 324 | 784 | 901.32 | 0 | 0 | 0 |
| 325 | 784 | 845.46 | 0 | 0 | 0 |
| 326 | 784 | 753.36 | 0 | 0 | 0 |
| 327 | 784 | 672.3 | 0 | 0 | 0 |
| 328 | 784 | 526.26 | 0 | 0 | 0 |
| 329 | 784 | 406.92 | 0 | 0 | 1 |
| 330 | 784 | 543.3 | 0 | 0 | 1 |
| 331 | 784 | 618 | 0 | 0 | 1 |
| 332 | 784 | 624.96 | 0 | 0 | 0 |
| 355 | 784 | 590.64 | 0 | 243 | 0 |
| 356 | 784 | 475.14 | 0 | 324 | 0 |
| 357 | 784 | 386.1 | 0 | 20 | 0 |
| 358 | 784 | 348.78 | 0 | 12 | 0 |
| 359 | 784 | 340.02 | 0 | 22 | 0 |
| 360 | 784 | 374.4 | 0 | 194 | 0 |
| 361 | 784 | 306.12 | 0 | 73 | 1 |
| 362 | 784 | 274.86 | 0 | 0 | 1 |
| 363 | 784 | 264.48 | 0 | 78 | 1 |
| 364 | 784 | 267.18 | 0 | 4 | 0 |
| 314 | 783 | 613.44 | 0 | 0 | 0 |
| 315 | 783 | 714.84 | 0 | 19 | 1 |
| 316 | 783 | 705.48 | 0 | 258 | 0 |
| 317 | 783 | 733.92 | 0 | 350 | 1 |
| 318 | 783 | 695.22 | 0 | 77 | 0 |
| 319 | 783 | 658.92 | 0 | 18 | 0 |
| 320 | 783 | 875.52 | 0 | 79 | 0 |
| 321 | 783 | 955.26 | 0 | 0 | 0 |
| 322 | 783 | 910.26 | 0 | 0 | 0 |
| 323 | 783 | 729.06 | 0 | 0 | 0 |
| 324 | 783 | 695.1 | 0 | 0 | 0 |
| 325 | 783 | 631.08 | 0 | 0 | 0 |
| 326 | 783 | 562.2 | 0 | 0 | 0 |
| 327 | 783 | 446.7 | 0 | 0 | 0 |
| 328 | 783 | 419.94 | 0 | 0 | 1 |
| 329 | 783 | 558.42 | 0 | 0 | 0 |
| 330 | 783 | 664.56 | 0 | 0 | 1 |
| 331 | 783 | 647.94 | 0 | 0 | 0 |
| 332 | 783 | 648.36 | 0 | 0 | 0 |
| 356 | 783 | 408.48 | 0 | 303 | 0 |
| 357 | 783 | 368.46 | 0 | 59 | 0 |
| 358 | 783 | 393.48 | 0 | 96 | 0 |
| 359 | 783 | 361.68 | 0 | 203 | 0 |
| 360 | 783 | 299.52 | 0 | 30 | 1 |
| 361 | 783 | 302.52 | 0 | 0 | 0 |
| 362 | 783 | 400.14 | 0 | 5 | 1 |
| 363 | 783 | 388.98 | 0 | 395 | 0 |
| 364 | 783 | 357.84 | 0 | 92 | 0 |
| 314 | 782 | 605.58 | 0 | 0 | 0 |
| 315 | 782 | 769.02 | 0 | 0 | 0 |
| 316 | 782 | 770.58 | 0 | 44 | 0 |
| 317 | 782 | 854.1 | 0 | 273 | 0 |
| 318 | 782 | 825.9 | 0 | 65 | 0 |
| 319 | 782 | 664.68 | 0 | 33 | 0 |
| 320 | 782 | 744.9 | 0 | 1 | 0 |
| 321 | 782 | 920.22 | 0 | 0 | 0 |
| 322 | 782 | 921.96 | 0 | 0 | 0 |
| 323 | 782 | 907.08 | 0 | 0 | 0 |
| 324 | 782 | 806.76 | 0 | 0 | 0 |
| 325 | 782 | 671.4 | 0 | 0 | 0 |
| 326 | 782 | 605.58 | 0 | 0 | 1 |
| 327 | 782 | 576.9 | 0 | 0 | 1 |
| 328 | 782 | 605.7 | 0 | 0 | 0 |
| 329 | 782 | 633.42 | 0 | 0 | 0 |
| 330 | 782 | 678.3 | 0 | 0 | 0 |
| 331 | 782 | 667.14 | 0 | 0 | 0 |
| 332 | 782 | 719.16 | 0 | 0 | 0 |
| 356 | 782 | 439.02 | 0 | 148 | 0 |
| 357 | 782 | 435.06 | 0 | 313 | 0 |
| 358 | 782 | 429.36 | 0 | 383 | 0 |
| 359 | 782 | 353.52 | 0 | 161 | 0 |
| 360 | 782 | 346.98 | 0 | 0 | 1 |
| 361 | 782 | 332.4 | 0 | 3 | 0 |
| 362 | 782 | 420.84 | 0 | 148 | 0 |
| 363 | 782 | 430.5 | 0 | 400 | 0 |
| 364 | 782 | 396.24 | 0 | 338 | 0 |
| 314 | 781 | 602.34 | 0 | 4 | 0 |
| 315 | 781 | 633.96 | 0 | 0 | 1 |
| 316 | 781 | 732.66 | 0 | 0 | 0 |
| 317 | 781 | 955.26 | 0 | 0 | 0 |
| 318 | 781 | 908.46 | 0 | 0 | 0 |
| 319 | 781 | 877.86 | 0 | 234 | 0 |
| 320 | 781 | 763.98 | 0 | 83 | 0 |
| 321 | 781 | 863.04 | 0 | 0 | 0 |
| 325 | 781 | 683.94 | 0 | 0 | 0 |
| 326 | 781 | 707.7 | 0 | 0 | 0 |
| 327 | 781 | 652.2 | 0 | 0 | 1 |
| 328 | 781 | 651.24 | 0 | 0 | 0 |
| 329 | 781 | 672.96 | 0 | 0 | 0 |
| 330 | 781 | 708.06 | 0 | 0 | 0 |
| 331 | 781 | 693.78 | 0 | 0 | 0 |
| 357 | 781 | 410.82 | 0 | 10 | 0 |
| 358 | 781 | 441.18 | 0 | 279 | 0 |
| 359 | 781 | 405.6 | 0 | 360 | 0 |
| 360 | 781 | 383.04 | 0 | 150 | 0 |
| 361 | 781 | 354.96 | 0 | 280 | 0 |
| 362 | 781 | 409.56 | 0 | 400 | 0 |
| 363 | 781 | 448.68 | 0 | 348 | 0 |
| 364 | 781 | 408.54 | 0 | 212 | 0 |
| 314 | 780 | 608.7 | 0 | 0 | 0 |
| 315 | 780 | 615.42 | 0 | 0 | 0 |
| 316 | 780 | 700.92 | 0 | 0 | 0 |
| 317 | 780 | 855.78 | 0 | 0 | 0 |
| 318 | 780 | 1006.26 | 0 | 0 | 0 |
| 319 | 780 | 945.42 | 0 | 19 | 0 |
| 320 | 780 | 882.96 | 0 | 10 | 0 |
| 321 | 780 | 856.68 | 0 | 0 | 0 |
| 327 | 780 | 484.92 | 0 | 0 | 0 |
| 328 | 780 | 614.22 | 0 | 0 | 0 |
| 329 | 780 | 679.26 | 0 | 0 | 0 |
| 330 | 780 | 742.14 | 0 | 0 | 0 |
| 331 | 780 | 704.52 | 0 | 0 | 0 |
| 358 | 780 | 455.88 | 0 | 131 | 0 |
| 359 | 780 | 434.34 | 0 | 400 | 0 |
| 360 | 780 | 386.1 | 0 | 400 | 0 |
| 361 | 780 | 456.3 | 0 | 400 | 0 |
| 362 | 780 | 462.12 | 0 | 230 | 0 |
| 363 | 780 | 419.52 | 0 | 12 | 0 |
| 364 | 780 | 346.74 | 0 | 0 | 0 |
| 315 | 779 | 717.84 | 0 | 0 | 0 |
| 316 | 779 | 804 | 0 | 0 | 0 |
| 317 | 779 | 879.6 | 0 | 0 | 0 |
| 318 | 779 | 984.96 | 0 | 0 | 0 |
| 331 | 779 | 757.62 | 0 | 0 | 0 |
| 332 | 779 | 754.32 | 0 | 0 | 0 |
| 359 | 779 | 475.02 | 0 | 373 | 0 |
| 360 | 779 | 465.54 | 0 | 386 | 0 |
| 361 | 779 | 483.66 | 0 | 285 | 0 |
| 362 | 779 | 378.78 | 0 | 3 | 0 |
| 363 | 779 | 317.34 | 0 | 0 | 0 |
| 364 | 779 | 287.58 | 0 | 0 | 0 |
| 365 | 779 | 233.16 | 0 | 0 | 0 |
| 314 | 778 | 723.84 | 0 | 0 | 0 |
| 315 | 778 | 820.26 | 0 | 0 | 0 |
| 316 | 778 | 910.26 | 0 | 0 | 0 |
| 317 | 778 | 894.54 | 0 | 0 | 0 |
| 359 | 778 | 455.82 | 0 | 50 | 0 |
| 360 | 778 | 454.86 | 0 | 77 | 0 |
| 361 | 778 | 402.42 | 0 | 34 | 0 |
| 362 | 778 | 311.88 | 0 | 0 | 0 |
| 363 | 778 | 279.54 | 0 | 0 | 0 |
| 364 | 778 | 230.94 | 0 | 0 | 1 |
| 365 | 778 | 204.48 | 0 | 0 | 0 |
| 366 | 778 | 150.42 | 0 | 0 | 0 |
| 316 | 777 | 1012.38 | 0 | 0 | 0 |
| 359 | 777 | 315.12 | 0 | 0 | 0 |
| 360 | 777 | 328.98 | 0 | 0 | 1 |
| 361 | 777 | 299.52 | 0 | 0 | 1 |
| 362 | 777 | 243 | 0 | 0 | 1 |
| 363 | 777 | 223.86 | 0 | 0 | 0 |
| 364 | 777 | 168.6 | 0 | 0 | 1 |
| 365 | 777 | 151.98 | 0 | 0 | 0 |
| 366 | 777 | 130.38 | 0 | 0 | 0 |
| 360 | 776 | 291.06 | 0 | 0 | 0 |
| 361 | 776 | 240.9 | 0 | 0 | 0 |
| 362 | 776 | 182.1 | 0 | 0 | 1 |
| 363 | 776 | 144.36 | 0 | 0 | 0 |
| 364 | 776 | 122.7 | 0 | 0 | 0 |
| 365 | 776 | 115.32 | 0 | 0 | 0 |
| 360 | 775 | 223.62 | 0 | 0 | 1 |
| 361 | 775 | 172.2 | 0 | 0 | 1 |
| 362 | 775 | 133.56 | 0 | 0 | 0 |
| 363 | 775 | 108.18 | 0 | 0 | 0 |
| 364 | 775 | 93.84 | 0 | 0 | 0 |
| 365 | 775 | 86.46 | 0 | 0 | 0 |
| 360 | 774 | 143.64 | 0 | 0 | 0 |
| 361 | 774 | 115.38 | 0 | 0 | 0 |
| 362 | 774 | 108.12 | 0 | 0 | 0 |
| 363 | 774 | 103.14 | 0 | 0 | 0 |
| 364 | 774 | 84.06 | 0 | 0 | 0 |
| 365 | 774 | 73.08 | 0 | 0 | 0 |
| 359 | 773 | 100.02 | 0 | 0 | 0 |
| 360 | 773 | 100.86 | 0 | 0 | 0 |
| 361 | 773 | 103.98 | 0 | 0 | 0 |
| 362 | 773 | 85.44 | 0 | 0 | 0 |
| 359 | 772 | 91.26 | 0 | 0 | 0 |
| 360 | 772 | 85.86 | 0 | 0 | 0 |
| 361 | 772 | 80.58 | 0 | 0 | 0 |
| 362 | 772 | 66.96 | 0 | 0 | 0 |
| 364 | 772 | 60.78 | 0 | 0 | 0 |
| 360 | 771 | 64.26 | 0 | 0 | 0 |
| 361 | 771 | 67.98 | 0 | 0 | 0 |
| 350 | 797 | 242.46 | 0 | 0 | 0 |
| 356 | 797 | 175.08 | 0 | 0 | 0 |
| 352 | 796 | 298.38 | 0 | 0 | 0 |
| 355 | 796 | 244.44 | 0 | 0 | 0 |
| 356 | 796 | 188.88 | 0 | 0 | 0 |
| 355 | 795 | 259.26 | 0 | 0 | 0 |
| 361 | 791 | 177.36 | 0 | 0 | 0 |
| 362 | 790 | 175.38 | 0 | 0 | 0 |
| 363 | 790 | 160.02 | 0 | 0 | 0 |
| 366 | 776 | 96.36 | 0 | 0 | 0 |
| 363 | 773 | 68.46 | 0 | 0 | 0 |
| 364 | 773 | 64.26 | 0 | 0 | 0 |
| 363 | 772 | 58.86 | 0 | 0 | 0 |
| 321 | 812 | 587.04 | 0 | 0 | 0 |
| 323 | 812 | 699.54 | 0 | 0 | 0 |
| 324 | 812 | 554.1 | 0 | 0 | 0 |
| 320 | 811 | 548.46 | 0 | 0 | 0 |
| 321 | 811 | 523.98 | 0 | 0 | 0 |
| 322 | 811 | 546.42 | 0 | 0 | 0 |
| 323 | 811 | 639.42 | 0 | 0 | 0 |
| 324 | 811 | 514.92 | 0 | 0 | 0 |
| 319 | 810 | 468.36 | 0 | 0 | 0 |
| 320 | 810 | 489.66 | 0 | 0 | 0 |
| 321 | 810 | 498.96 | 0 | 0 | 0 |
| 322 | 810 | 466.44 | 0 | 0 | 0 |
| 323 | 810 | 457.2 | 0 | 0 | 1 |
| 324 | 810 | 451.14 | 0 | 12 | 0 |
| 318 | 809 | 485.34 | 0 | 0 | 0 |
| 319 | 809 | 495.12 | 0 | 0 | 0 |
| 320 | 809 | 485.64 | 0 | 0 | 0 |
| 321 | 809 | 458.22 | 0 | 0 | 0 |
| 322 | 809 | 466.02 | 0 | 0 | 0 |
| 323 | 809 | 512.7 | 0 | 0 | 0 |
| 324 | 809 | 490.2 | 0 | 0 | 0 |
| 325 | 809 | 450.9 | 0 | 0 | 0 |
| 319 | 808 | 582.12 | 0 | 0 | 0 |
| 320 | 808 | 628.62 | 0 | 58 | 0 |
| 321 | 808 | 526.56 | 0 | 0 | 0 |
| 322 | 808 | 524.34 | 0 | 0 | 0 |
| 323 | 808 | 629.4 | 0 | 47 | 0 |
| 324 | 808 | 588.12 | 0 | 0 | 0 |
| 325 | 808 | 472.5 | 0 | 0 | 0 |
| 319 | 807 | 650.22 | 0 | 0 | 0 |
| 320 | 807 | 622.74 | 0 | 9 | 0 |
| 321 | 807 | 572.58 | 0 | 0 | 0 |
| 322 | 807 | 618.42 | 0 | 0 | 0 |
| 323 | 807 | 647.34 | 0 | 0 | 0 |
| 324 | 807 | 600.48 | 0 | 0 | 1 |
| 325 | 807 | 506.7 | 0 | 0 | 0 |
| 326 | 807 | 563.22 | 0 | 0 | 0 |
| 327 | 807 | 530.82 | 0 | 0 | 0 |
| 328 | 807 | 449.94 | 0 | 0 | 0 |
| 319 | 806 | 672.66 | 0 | 0 | 0 |
| 320 | 806 | 705.84 | 0 | 0 | 0 |
| 321 | 806 | 670.38 | 0 | 0 | 0 |
| 322 | 806 | 678.78 | 0 | 0 | 1 |
| 323 | 806 | 730.44 | 0 | 0 | 0 |
| 324 | 806 | 657.54 | 0 | 0 | 0 |
| 325 | 806 | 593.1 | 0 | 0 | 0 |
| 326 | 806 | 639.42 | 0 | 0 | 0 |
| 327 | 806 | 628.26 | 0 | 0 | 0 |
| 328 | 806 | 566.04 | 0 | 0 | 0 |
| 329 | 806 | 501.66 | 0 | 0 | 0 |
| 320 | 805 | 710.52 | 0 | 0 | 0 |
| 321 | 805 | 793.86 | 0 | 0 | 0 |
| 322 | 805 | 798.48 | 0 | 0 | 1 |
| 323 | 805 | 750 | 0 | 0 | 0 |
| 324 | 805 | 650.7 | 0 | 0 | 0 |
| 325 | 805 | 587.58 | 0 | 0 | 0 |
| 326 | 805 | 604.56 | 0 | 0 | 1 |
| 327 | 805 | 624.36 | 0 | 0 | 0 |
| 328 | 805 | 623.16 | 0 | 22 | 0 |
| 329 | 805 | 624.24 | 0 | 40 | 0 |
| 320 | 804 | 619.08 | 0 | 0 | 0 |
| 321 | 804 | 681.9 | 0 | 0 | 0 |
| 322 | 804 | 679.86 | 0 | 0 | 0 |
| 323 | 804 | 581.82 | 0 | 0 | 0 |
| 324 | 804 | 519 | 0 | 0 | 0 |
| 325 | 804 | 506.82 | 0 | 82 | 0 |
| 326 | 804 | 497.28 | 0 | 77 | 0 |
| 327 | 804 | 499.08 | 0 | 120 | 0 |
| 328 | 804 | 578.1 | 0 | 124 | 0 |
| 329 | 804 | 632.28 | 0 | 1 | 0 |
| 330 | 804 | 516.18 | 0 | 0 | 0 |
| 319 | 803 | 490.08 | 0 | 0 | 1 |
| 320 | 803 | 504.3 | 0 | 0 | 0 |
| 321 | 803 | 509.4 | 0 | 0 | 0 |
| 322 | 803 | 501.66 | 0 | 0 | 1 |
| 323 | 803 | 502.2 | 0 | 0 | 0 |
| 324 | 803 | 487.68 | 0 | 0 | 0 |
| 325 | 803 | 454.02 | 0 | 81 | 0 |
| 326 | 803 | 426.96 | 0 | 7 | 0 |
| 327 | 803 | 455.34 | 0 | 37 | 0 |
| 328 | 803 | 579.24 | 0 | 0 | 0 |
| 329 | 803 | 488.82 | 0 | 62 | 0 |
| 330 | 803 | 478.26 | 0 | 0 | 0 |
| 317 | 802 | 618.6 | 0 | 0 | 0 |
| 318 | 802 | 528.12 | 0 | 0 | 0 |
| 319 | 802 | 545.16 | 0 | 0 | 0 |
| 320 | 802 | 485.7 | 0 | 0 | 0 |
| 321 | 802 | 492.18 | 0 | 0 | 1 |
| 322 | 802 | 492.84 | 0 | 0 | 0 |
| 323 | 802 | 460.26 | 0 | 22 | 1 |
| 324 | 802 | 443.46 | 0 | 0 | 1 |
| 325 | 802 | 411.9 | 0 | 23 | 0 |
| 326 | 802 | 403.62 | 0 | 67 | 0 |
| 327 | 802 | 402.78 | 0 | 0 | 0 |
| 328 | 802 | 455.22 | 0 | 0 | 0 |
| 329 | 802 | 414.06 | 0 | 42 | 0 |
| 330 | 802 | 392.58 | 0 | 0 | 0 |
| 313 | 801 | 999.9 | 0 | 0 | 0 |
| 314 | 801 | 901.68 | 0 | 0 | 0 |
| 315 | 801 | 762.48 | 0 | 0 | 0 |
| 316 | 801 | 633.12 | 0 | 0 | 0 |
| 317 | 801 | 547.86 | 0 | 0 | 0 |
| 318 | 801 | 612.66 | 0 | 0 | 0 |
| 319 | 801 | 653.04 | 0 | 0 | 0 |
| 320 | 801 | 578.22 | 0 | 0 | 0 |
| 321 | 801 | 578.28 | 0 | 0 | 0 |
| 322 | 801 | 584.22 | 0 | 119 | 0 |
| 323 | 801 | 473.52 | 0 | 223 | 0 |
| 324 | 801 | 452.76 | 0 | 0 | 0 |
| 325 | 801 | 453.72 | 0 | 0 | 0 |
| 326 | 801 | 429.42 | 0 | 0 | 0 |
| 327 | 801 | 398.88 | 0 | 0 | 0 |
| 328 | 801 | 377.82 | 0 | 46 | 0 |
| 329 | 801 | 405.36 | 0 | 4 | 0 |
| 297 | 800 | 845.58 | 0 | 155 | 0 |
| 298 | 800 | 1195.02 | 0 | 0 | 0 |
| 302 | 800 | 800.22 | 0 | 0 | 0 |
| 303 | 800 | 737.16 | 0 | 30 | 0 |
| 304 | 800 | 825.36 | 0 | 238 | 0 |
| 306 | 800 | 824.94 | 0 | 317 | 0 |
| 307 | 800 | 886.44 | 0 | 47 | 0 |
| 309 | 800 | 1005.72 | 0 | 0 | 0 |
| 310 | 800 | 911.4 | 0 | 0 | 0 |
| 311 | 800 | 882.3 | 0 | 0 | 0 |
| 312 | 800 | 1039.98 | 0 | 0 | 0 |
| 313 | 800 | 928.56 | 0 | 0 | 0 |
| 314 | 800 | 722.16 | 0 | 3 | 0 |
| 315 | 800 | 582.72 | 0 | 142 | 0 |
| 316 | 800 | 565.56 | 0 | 208 | 0 |
| 317 | 800 | 569.88 | 0 | 345 | 0 |
| 318 | 800 | 690.3 | 0 | 222 | 0 |
| 319 | 800 | 774.12 | 0 | 158 | 0 |
| 320 | 800 | 643.02 | 0 | 129 | 0 |
| 321 | 800 | 612.3 | 0 | 6 | 0 |
| 322 | 800 | 627.48 | 0 | 3 | 0 |
| 323 | 800 | 550.26 | 0 | 67 | 0 |
| 324 | 800 | 522 | 0 | 0 | 0 |
| 325 | 800 | 478.08 | 0 | 13 | 0 |
| 326 | 800 | 409.86 | 0 | 38 | 1 |
| 327 | 800 | 424.08 | 0 | 0 | 0 |
| 328 | 800 | 455.7 | 0 | 0 | 0 |
| 329 | 800 | 569.1 | 0 | 0 | 0 |
| 294 | 799 | 1097.64 | 0 | 0 | 0 |
| 295 | 799 | 896.7 | 0 | 0 | 0 |
| 296 | 799 | 757.62 | 0 | 31 | 0 |
| 297 | 799 | 674.7 | 0 | 231 | 0 |
| 298 | 799 | 1033.98 | 0 | 0 | 0 |
| 300 | 799 | 983.52 | 0 | 0 | 0 |
| 301 | 799 | 1063.8 | 0 | 0 | 0 |
| 302 | 799 | 808.26 | 0 | 0 | 0 |
| 303 | 799 | 606.42 | 0 | 30 | 0 |
| 304 | 799 | 848.88 | 0 | 223 | 0 |
| 305 | 799 | 836.1 | 0 | 47 | 0 |
| 306 | 799 | 768.78 | 0 | 114 | 0 |
| 307 | 799 | 894.78 | 0 | 11 | 0 |
| 308 | 799 | 1060.62 | 0 | 0 | 0 |
| 309 | 799 | 1087.26 | 0 | 0 | 0 |
| 310 | 799 | 832.14 | 0 | 69 | 0 |
| 311 | 799 | 772.32 | 0 | 142 | 0 |
| 312 | 799 | 918 | 0 | 0 | 0 |
| 313 | 799 | 858.12 | 0 | 0 | 0 |
| 314 | 799 | 638.7 | 0 | 74 | 0 |
| 315 | 799 | 748.32 | 0 | 9 | 0 |
| 316 | 799 | 710.22 | 0 | 45 | 0 |
| 317 | 799 | 658.56 | 0 | 148 | 0 |
| 318 | 799 | 698.76 | 0 | 113 | 0 |
| 319 | 799 | 782.16 | 0 | 0 | 0 |
| 320 | 799 | 679.14 | 0 | 35 | 0 |
| 321 | 799 | 619.92 | 0 | 235 | 0 |
| 322 | 799 | 584.76 | 0 | 6 | 1 |
| 323 | 799 | 544.02 | 0 | 230 | 0 |
| 324 | 799 | 481.32 | 0 | 100 | 1 |
| 325 | 799 | 443.76 | 0 | 72 | 0 |
| 326 | 799 | 473.58 | 0 | 14 | 0 |
| 327 | 799 | 505.44 | 0 | 0 | 0 |
| 294 | 798 | 1053.48 | 0 | 0 | 0 |
| 295 | 798 | 1120.2 | 0 | 0 | 0 |
| 296 | 798 | 1035.3 | 0 | 0 | 0 |
| 297 | 798 | 674.28 | 0 | 177 | 0 |
| 298 | 798 | 741.36 | 0 | 15 | 0 |
| 299 | 798 | 926.34 | 0 | 0 | 0 |
| 300 | 798 | 865.26 | 0 | 95 | 0 |
| 301 | 798 | 949.2 | 0 | 0 | 0 |
| 302 | 798 | 839.46 | 0 | 0 | 0 |
| 303 | 798 | 587.76 | 4 | 19 | 1 |
| 304 | 798 | 803.28 | 0 | 8 | 0 |
| 305 | 798 | 865.26 | 0 | 12 | 0 |
| 306 | 798 | 652.38 | 0 | 0 | 0 |
| 307 | 798 | 794.28 | 0 | 0 | 0 |
| 308 | 798 | 984 | 0 | 0 | 0 |
| 309 | 798 | 971.1 | 0 | 0 | 0 |
| 310 | 798 | 729.72 | 0 | 117 | 0 |
| 311 | 798 | 642.96 | 0 | 280 | 0 |
| 312 | 798 | 752.52 | 0 | 27 | 0 |
| 313 | 798 | 667.2 | 0 | 185 | 0 |
| 314 | 798 | 658.26 | 0 | 136 | 0 |
| 315 | 798 | 759.06 | 0 | 0 | 0 |
| 316 | 798 | 797.16 | 0 | 0 | 0 |
| 317 | 798 | 682.56 | 0 | 23 | 1 |
| 318 | 798 | 592.98 | 0 | 380 | 1 |
| 319 | 798 | 588 | 0 | 149 | 0 |
| 320 | 798 | 636 | 0 | 175 | 0 |
| 321 | 798 | 674.4 | 0 | 39 | 0 |
| 322 | 798 | 591.84 | 0 | 61 | 0 |
| 323 | 798 | 526.14 | 0 | 0 | 0 |
| 324 | 798 | 466.68 | 0 | 53 | 0 |
| 325 | 798 | 417.78 | 0 | 40 | 0 |
| 326 | 798 | 484.38 | 0 | 0 | 0 |
| 294 | 797 | 984.96 | 0 | 0 | 0 |
| 295 | 797 | 1070.16 | 0 | 0 | 0 |
| 296 | 797 | 1092.6 | 0 | 0 | 0 |
| 297 | 797 | 742.86 | 0 | 42 | 0 |
| 298 | 797 | 688.38 | 0 | 153 | 0 |
| 299 | 797 | 797.1 | 0 | 55 | 0 |
| 300 | 797 | 771.18 | 0 | 132 | 0 |
| 301 | 797 | 805.92 | 0 | 66 | 0 |
| 302 | 797 | 809.4 | 0 | 0 | 0 |
| 303 | 797 | 560.58 | 63 | 91 | 0 |
| 304 | 797 | 662.28 | 18 | 0 | 0 |
| 305 | 797 | 792.72 | 0 | 0 | 0 |
| 306 | 797 | 649.32 | 20 | 0 | 0 |
| 307 | 797 | 670.98 | 2 | 0 | 1 |
| 308 | 797 | 743.1 | 0 | 0 | 1 |
| 309 | 797 | 772.74 | 0 | 56 | 0 |
| 310 | 797 | 635.46 | 0 | 199 | 0 |
| 311 | 797 | 608.76 | 0 | 267 | 0 |
| 312 | 797 | 693.96 | 0 | 136 | 0 |
| 313 | 797 | 646.32 | 0 | 73 | 0 |
| 314 | 797 | 631.2 | 0 | 14 | 0 |
| 315 | 797 | 640.08 | 0 | 0 | 0 |
| 316 | 797 | 722.88 | 0 | 0 | 0 |
| 317 | 797 | 587.46 | 0 | 19 | 0 |
| 318 | 797 | 544.02 | 0 | 137 | 0 |
| 319 | 797 | 525.84 | 0 | 32 | 0 |
| 320 | 797 | 603.72 | 0 | 0 | 0 |
| 321 | 797 | 592.86 | 0 | 0 | 0 |
| 322 | 797 | 559.44 | 0 | 0 | 0 |
| 323 | 797 | 483.54 | 0 | 0 | 1 |
| 324 | 797 | 450.84 | 0 | 0 | 0 |
| 325 | 797 | 393.54 | 0 | 0 | 0 |
| 294 | 796 | 837.48 | 0 | 52 | 0 |
| 295 | 796 | 933.48 | 0 | 0 | 0 |
| 296 | 796 | 951.06 | 0 | 0 | 0 |
| 297 | 796 | 738.96 | 0 | 0 | 0 |
| 298 | 796 | 626.88 | 0 | 196 | 0 |
| 299 | 796 | 871.68 | 0 | 10 | 0 |
| 300 | 796 | 687.06 | 0 | 191 | 0 |
| 301 | 796 | 694.2 | 0 | 214 | 0 |
| 302 | 796 | 785.16 | 0 | 71 | 0 |
| 303 | 796 | 562.38 | 34 | 101 | 0 |
| 304 | 796 | 554.76 | 222 | 116 | 0 |
| 305 | 796 | 663.36 | 0 | 110 | 0 |
| 306 | 796 | 566.4 | 52 | 0 | 0 |
| 307 | 796 | 523.92 | 204 | 0 | 1 |
| 308 | 796 | 529.98 | 135 | 0 | 0 |
| 309 | 796 | 551.22 | 244 | 16 | 0 |
| 310 | 796 | 575.34 | 207 | 3 | 0 |
| 311 | 796 | 603.9 | 0 | 118 | 0 |
| 312 | 796 | 650.76 | 0 | 74 | 0 |
| 313 | 796 | 623.64 | 0 | 108 | 0 |
| 314 | 796 | 594 | 0 | 0 | 1 |
| 315 | 796 | 622.5 | 0 | 0 | 0 |
| 316 | 796 | 639 | 0 | 0 | 0 |
| 317 | 796 | 578.58 | 45 | 0 | 0 |
| 318 | 796 | 448.92 | 13 | 0 | 0 |
| 319 | 796 | 432.48 | 0 | 0 | 1 |
| 320 | 796 | 455.04 | 0 | 43 | 0 |
| 321 | 796 | 480.06 | 0 | 144 | 0 |
| 322 | 796 | 494.52 | 0 | 0 | 1 |
| 323 | 796 | 397.08 | 0 | 0 | 0 |
| 324 | 796 | 360.96 | 10 | 0 | 0 |
| 325 | 796 | 324.18 | 0 | 0 | 0 |
| 294 | 795 | 827.7 | 0 | 0 | 0 |
| 295 | 795 | 692.34 | 0 | 43 | 0 |
| 296 | 795 | 709.98 | 0 | 80 | 0 |
| 297 | 795 | 629.94 | 0 | 134 | 0 |
| 298 | 795 | 591.36 | 0 | 292 | 0 |
| 299 | 795 | 831.84 | 0 | 4 | 0 |
| 300 | 795 | 704.34 | 0 | 0 | 0 |
| 301 | 795 | 638.4 | 0 | 107 | 0 |
| 302 | 795 | 657.84 | 0 | 0 | 0 |
| 303 | 795 | 580.8 | 60 | 0 | 0 |
| 304 | 795 | 522.36 | 253 | 29 | 0 |
| 305 | 795 | 687 | 0 | 146 | 0 |
| 306 | 795 | 680.58 | 0 | 62 | 0 |
| 307 | 795 | 503.4 | 8 | 9 | 1 |
| 308 | 795 | 478.86 | 212 | 0 | 1 |
| 309 | 795 | 578.1 | 53 | 0 | 0 |
| 310 | 795 | 648.42 | 2 | 0 | 0 |
| 311 | 795 | 680.46 | 0 | 35 | 0 |
| 312 | 795 | 628.62 | 0 | 8 | 0 |
| 313 | 795 | 496.38 | 0 | 0 | 0 |
| 314 | 795 | 473.22 | 0 | 0 | 0 |
| 315 | 795 | 505.86 | 3 | 0 | 0 |
| 316 | 795 | 543.48 | 47 | 0 | 1 |
| 317 | 795 | 497.34 | 0 | 0 | 0 |
| 318 | 795 | 536.88 | 13 | 0 | 0 |
| 319 | 795 | 506.04 | 58 | 0 | 0 |
| 320 | 795 | 392.4 | 13 | 65 | 0 |
| 321 | 795 | 377.58 | 0 | 0 | 0 |
| 322 | 795 | 403.74 | 0 | 0 | 0 |
| 323 | 795 | 373.44 | 0 | 0 | 0 |
| 324 | 795 | 300.36 | 0 | 0 | 0 |
| 294 | 794 | 907.98 | 0 | 0 | 0 |
| 295 | 794 | 721.98 | 0 | 0 | 0 |
| 296 | 794 | 779.22 | 0 | 0 | 0 |
| 297 | 794 | 662.28 | 0 | 21 | 0 |
| 298 | 794 | 551.28 | 0 | 339 | 0 |
| 299 | 794 | 580.02 | 0 | 338 | 0 |
| 300 | 794 | 572.16 | 0 | 289 | 0 |
| 301 | 794 | 516.42 | 17 | 316 | 0 |
| 302 | 794 | 475.5 | 112 | 96 | 0 |
| 303 | 794 | 464.52 | 198 | 0 | 0 |
| 304 | 794 | 489.3 | 165 | 0 | 0 |
| 305 | 794 | 595.14 | 0 | 110 | 0 |
| 306 | 794 | 591.42 | 0 | 147 | 1 |
| 307 | 794 | 484.14 | 1 | 0 | 0 |
| 308 | 794 | 484.92 | 76 | 0 | 0 |
| 309 | 794 | 553.2 | 0 | 0 | 0 |
| 310 | 794 | 576.54 | 0 | 0 | 0 |
| 311 | 794 | 643.2 | 0 | 0 | 0 |
| 312 | 794 | 743.82 | 0 | 0 | 0 |
| 313 | 794 | 634.74 | 0 | 0 | 1 |
| 314 | 794 | 527.82 | 31 | 0 | 0 |
| 315 | 794 | 395.82 | 0 | 0 | 0 |
| 316 | 794 | 405 | 11 | 0 | 0 |
| 317 | 794 | 435.54 | 12 | 0 | 0 |
| 318 | 794 | 519.54 | 23 | 0 | 1 |
| 319 | 794 | 443.28 | 0 | 0 | 1 |
| 320 | 794 | 372.3 | 2 | 0 | 0 |
| 321 | 794 | 369 | 0 | 0 | 0 |
| 322 | 794 | 309.36 | 6 | 0 | 0 |
| 323 | 794 | 302.34 | 169 | 0 | 0 |
| 294 | 793 | 970.68 | 0 | 0 | 0 |
| 295 | 793 | 1055.34 | 0 | 0 | 0 |
| 296 | 793 | 1021.56 | 0 | 0 | 0 |
| 297 | 793 | 833.7 | 0 | 0 | 0 |
| 298 | 793 | 552.9 | 0 | 153 | 0 |
| 299 | 793 | 647.88 | 0 | 151 | 0 |
| 300 | 793 | 605.52 | 0 | 400 | 0 |
| 301 | 793 | 580.5 | 0 | 302 | 0 |
| 302 | 793 | 594.12 | 0 | 219 | 0 |
| 303 | 793 | 620.88 | 15 | 0 | 0 |
| 304 | 793 | 485.04 | 0 | 0 | 1 |
| 305 | 793 | 486 | 0 | 0 | 0 |
| 306 | 793 | 539.82 | 8 | 17 | 1 |
| 307 | 793 | 539.28 | 0 | 0 | 0 |
| 308 | 793 | 455.28 | 216 | 0 | 0 |
| 309 | 793 | 446.46 | 173 | 0 | 0 |
| 310 | 793 | 454.26 | 14 | 0 | 0 |
| 311 | 793 | 530.76 | 0 | 0 | 1 |
| 312 | 793 | 562.44 | 14 | 0 | 0 |
| 313 | 793 | 503.46 | 2 | 0 | 0 |
| 314 | 793 | 414 | 30 | 0 | 0 |
| 315 | 793 | 344.88 | 0 | 0 | 1 |
| 316 | 793 | 346.5 | 0 | 0 | 1 |
| 317 | 793 | 387.6 | 5 | 0 | 1 |
| 318 | 793 | 480.06 | 70 | 0 | 1 |
| 319 | 793 | 361.68 | 0 | 0 | 1 |
| 320 | 793 | 391.44 | 0 | 0 | 0 |
| 321 | 793 | 348.72 | 117 | 0 | 0 |
| 322 | 793 | 324.72 | 339 | 0 | 0 |
| 294 | 792 | 895.2 | 0 | 0 | 0 |
| 295 | 792 | 995.1 | 0 | 0 | 0 |
| 296 | 792 | 938.46 | 0 | 0 | 0 |
| 297 | 792 | 748.2 | 0 | 0 | 0 |
| 298 | 792 | 550.2 | 0 | 141 | 0 |
| 299 | 792 | 624.24 | 0 | 116 | 0 |
| 300 | 792 | 709.14 | 0 | 41 | 0 |
| 301 | 792 | 697.98 | 0 | 0 | 0 |
| 302 | 792 | 669.36 | 0 | 0 | 0 |
| 303 | 792 | 671.52 | 0 | 0 | 0 |
| 304 | 792 | 583.44 | 0 | 0 | 0 |
| 305 | 792 | 478.86 | 2 | 0 | 0 |
| 306 | 792 | 469.08 | 221 | 0 | 1 |
| 307 | 792 | 545.76 | 201 | 0 | 1 |
| 308 | 792 | 561 | 22 | 0 | 1 |
| 309 | 792 | 533.16 | 31 | 0 | 0 |
| 310 | 792 | 465.42 | 132 | 0 | 0 |
| 311 | 792 | 392.1 | 50 | 0 | 1 |
| 312 | 792 | 367.5 | 19 | 0 | 1 |
| 313 | 792 | 347.52 | 0 | 0 | 1 |
| 314 | 792 | 346.08 | 0 | 0 | 0 |
| 316 | 792 | 425.82 | 47 | 0 | 0 |
| 317 | 792 | 376.98 | 10 | 0 | 1 |
| 318 | 792 | 365.28 | 0 | 0 | 0 |
| 319 | 792 | 390.84 | 0 | 0 | 0 |
| 320 | 792 | 435.3 | 18 | 0 | 0 |
| 321 | 792 | 346.44 | 280 | 0 | 0 |
| 294 | 791 | 756.36 | 0 | 90 | 0 |
| 295 | 791 | 807.06 | 0 | 71 | 0 |
| 296 | 791 | 852.6 | 0 | 11 | 0 |
| 297 | 791 | 709.74 | 0 | 0 | 0 |
| 298 | 791 | 595.74 | 0 | 26 | 0 |
| 299 | 791 | 562.38 | 0 | 33 | 0 |
| 300 | 791 | 637.98 | 0 | 0 | 0 |
| 301 | 791 | 620.1 | 0 | 0 | 0 |
| 302 | 791 | 603.12 | 0 | 0 | 0 |
| 303 | 791 | 590.34 | 0 | 0 | 1 |
| 304 | 791 | 589.14 | 0 | 0 | 0 |
| 305 | 791 | 479.4 | 72 | 0 | 0 |
| 306 | 791 | 403.38 | 200 | 0 | 0 |
| 307 | 791 | 441.84 | 103 | 0 | 1 |
| 308 | 791 | 503.58 | 138 | 0 | 0 |
| 309 | 791 | 521.58 | 59 | 0 | 0 |
| 310 | 791 | 481.08 | 24 | 0 | 1 |
| 311 | 791 | 358.38 | 2 | 0 | 0 |
| 312 | 791 | 387.42 | 0 | 0 | 0 |
| 313 | 791 | 407.82 | 0 | 14 | 0 |
| 314 | 791 | 389.34 | 0 | 0 | 0 |
| 318 | 791 | 376.38 | 105 | 0 | 0 |
| 319 | 791 | 328.32 | 240 | 0 | 0 |
| 320 | 791 | 342.72 | 369 | 0 | 0 |
| 294 | 790 | 678.24 | 0 | 324 | 0 |
| 295 | 790 | 628.2 | 0 | 370 | 0 |
| 296 | 790 | 649.98 | 0 | 195 | 0 |
| 297 | 790 | 617.64 | 0 | 327 | 0 |
| 298 | 790 | 578.46 | 0 | 257 | 0 |
| 299 | 790 | 497.04 | 0 | 5 | 0 |
| 300 | 790 | 499.62 | 0 | 0 | 0 |
| 301 | 790 | 533.76 | 0 | 0 | 0 |
| 302 | 790 | 480.48 | 0 | 0 | 0 |
| 303 | 790 | 446.34 | 0 | 0 | 0 |
| 304 | 790 | 426 | 32 | 0 | 0 |
| 305 | 790 | 413.16 | 85 | 0 | 1 |
| 306 | 790 | 405.72 | 0 | 0 | 0 |
| 307 | 790 | 372.54 | 13 | 0 | 0 |
| 308 | 790 | 355.08 | 9 | 0 | 1 |
| 310 | 790 | 381.66 | 0 | 0 | 1 |
| 311 | 790 | 397.44 | 0 | 0 | 1 |
| 312 | 790 | 537.54 | 48 | 0 | 0 |
| 313 | 790 | 644.7 | 22 | 0 | 1 |
| 314 | 790 | 511.26 | 0 | 0 | 1 |
| 293 | 789 | 638.28 | 0 | 354 | 1 |
| 294 | 789 | 611.46 | 0 | 400 | 0 |
| 295 | 789 | 593.46 | 0 | 400 | 1 |
| 296 | 789 | 586.44 | 0 | 398 | 0 |
| 297 | 789 | 588.9 | 0 | 400 | 0 |
| 298 | 789 | 573.3 | 0 | 385 | 0 |
| 299 | 789 | 567.48 | 0 | 188 | 0 |
| 300 | 789 | 465.42 | 0 | 0 | 0 |
| 301 | 789 | 440.04 | 0 | 0 | 0 |
| 302 | 789 | 433.8 | 0 | 2 | 0 |
| 303 | 789 | 489.18 | 0 | 17 | 0 |
| 304 | 789 | 505.14 | 0 | 40 | 0 |
| 305 | 789 | 561.24 | 0 | 40 | 0 |
| 306 | 789 | 573.3 | 0 | 43 | 1 |
| 307 | 789 | 518.4 | 0 | 44 | 0 |
| 308 | 789 | 400.68 | 0 | 0 | 1 |
| 309 | 789 | 550.44 | 0 | 0 | 0 |
| 310 | 789 | 492.24 | 0 | 0 | 0 |
| 311 | 789 | 542.4 | 0 | 0 | 0 |
| 312 | 789 | 689.94 | 0 | 0 | 1 |
| 313 | 789 | 738.54 | 0 | 0 | 0 |
| 314 | 789 | 520.14 | 0 | 0 | 1 |
| 293 | 788 | 553.92 | 0 | 304 | 1 |
| 294 | 788 | 549.36 | 0 | 184 | 0 |
| 295 | 788 | 545.46 | 0 | 166 | 0 |
| 296 | 788 | 514.74 | 0 | 119 | 0 |
| 297 | 788 | 512.28 | 0 | 215 | 0 |
| 298 | 788 | 513.36 | 0 | 159 | 0 |
| 299 | 788 | 517.38 | 0 | 6 | 0 |
| 300 | 788 | 462.48 | 0 | 0 | 0 |
| 301 | 788 | 482.94 | 0 | 0 | 0 |
| 302 | 788 | 583.08 | 0 | 17 | 0 |
| 303 | 788 | 668.52 | 0 | 32 | 0 |
| 304 | 788 | 655.26 | 0 | 2 | 1 |
| 305 | 788 | 687.66 | 0 | 4 | 0 |
| 306 | 788 | 637.2 | 0 | 0 | 0 |
| 307 | 788 | 502.68 | 0 | 0 | 1 |
| 308 | 788 | 429.12 | 0 | 0 | 1 |
| 309 | 788 | 569.88 | 0 | 0 | 0 |
| 310 | 788 | 578.64 | 0 | 0 | 1 |
| 311 | 788 | 600.24 | 0 | 90 | 1 |
| 312 | 788 | 742.8 | 0 | 174 | 0 |
| 313 | 788 | 687.3 | 0 | 26 | 1 |
| 314 | 788 | 487.38 | 0 | 0 | 0 |
| 292 | 787 | 564.18 | 0 | 161 | 1 |
| 293 | 787 | 569.52 | 0 | 55 | 0 |
| 294 | 787 | 562.92 | 0 | 329 | 0 |
| 295 | 787 | 551.94 | 0 | 131 | 0 |
| 296 | 787 | 569.16 | 0 | 230 | 0 |
| 297 | 787 | 585 | 0 | 46 | 0 |
| 298 | 787 | 544.8 | 0 | 66 | 0 |
| 299 | 787 | 483.54 | 0 | 0 | 0 |
| 300 | 787 | 471.24 | 0 | 0 | 0 |
| 301 | 787 | 573.72 | 0 | 55 | 0 |
| 302 | 787 | 660.24 | 0 | 9 | 0 |
| 303 | 787 | 751.92 | 0 | 87 | 1 |
| 304 | 787 | 661.5 | 0 | 0 | 0 |
| 305 | 787 | 604.08 | 0 | 0 | 0 |
| 306 | 787 | 551.1 | 0 | 0 | 1 |
| 307 | 787 | 490.32 | 0 | 0 | 0 |
| 308 | 787 | 463.02 | 0 | 0 | 1 |
| 309 | 787 | 571.02 | 0 | 0 | 0 |
| 310 | 787 | 637.62 | 0 | 0 | 0 |
| 311 | 787 | 668.34 | 0 | 242 | 1 |
| 312 | 787 | 747.18 | 0 | 172 | 0 |
| 313 | 787 | 681.06 | 0 | 0 | 0 |
| 314 | 787 | 451.2 | 0 | 0 | 0 |
| 291 | 786 | 643.92 | 0 | 346 | 1 |
| 292 | 786 | 637.5 | 0 | 241 | 0 |
| 293 | 786 | 688.86 | 0 | 22 | 0 |
| 294 | 786 | 625.92 | 0 | 274 | 0 |
| 295 | 786 | 636.18 | 0 | 0 | 0 |
| 296 | 786 | 650.4 | 0 | 5 | 0 |
| 297 | 786 | 597.72 | 0 | 190 | 0 |
| 298 | 786 | 560.22 | 0 | 74 | 0 |
| 299 | 786 | 501.84 | 0 | 0 | 0 |
| 300 | 786 | 498.42 | 0 | 0 | 0 |
| 301 | 786 | 631.62 | 0 | 0 | 0 |
| 302 | 786 | 669.24 | 0 | 11 | 0 |
| 303 | 786 | 743.94 | 0 | 315 | 1 |
| 304 | 786 | 665.16 | 0 | 69 | 0 |
| 305 | 786 | 679.68 | 0 | 51 | 0 |
| 306 | 786 | 594.78 | 0 | 0 | 1 |
| 307 | 786 | 578.58 | 0 | 3 | 0 |
| 308 | 786 | 545.16 | 0 | 0 | 1 |
| 309 | 786 | 475.14 | 0 | 0 | 0 |
| 310 | 786 | 559.2 | 0 | 7 | 0 |
| 311 | 786 | 671.52 | 0 | 46 | 0 |
| 312 | 786 | 709.2 | 0 | 7 | 1 |
| 313 | 786 | 570.96 | 0 | 0 | 1 |
| 314 | 786 | 437.64 | 0 | 0 | 0 |
| 291 | 785 | 699.18 | 0 | 345 | 0 |
| 292 | 785 | 720 | 0 | 274 | 0 |
| 293 | 785 | 814.8 | 0 | 95 | 0 |
| 294 | 785 | 748.86 | 0 | 16 | 0 |
| 295 | 785 | 710.04 | 0 | 0 | 0 |
| 296 | 785 | 652.98 | 0 | 66 | 0 |
| 297 | 785 | 608.22 | 0 | 317 | 0 |
| 298 | 785 | 620.34 | 0 | 107 | 0 |
| 299 | 785 | 529.44 | 0 | 78 | 0 |
| 300 | 785 | 522.78 | 0 | 18 | 0 |
| 301 | 785 | 619.02 | 0 | 69 | 1 |
| 302 | 785 | 652.5 | 0 | 45 | 0 |
| 303 | 785 | 742.92 | 0 | 222 | 0 |
| 304 | 785 | 669.24 | 0 | 165 | 0 |
| 305 | 785 | 608.04 | 0 | 68 | 0 |
| 306 | 785 | 616.56 | 0 | 26 | 1 |
| 307 | 785 | 685.5 | 0 | 206 | 1 |
| 308 | 785 | 730.68 | 0 | 0 | 0 |
| 309 | 785 | 560.1 | 0 | 0 | 0 |
| 310 | 785 | 576.84 | 0 | 22 | 0 |
| 311 | 785 | 721.02 | 0 | 0 | 0 |
| 312 | 785 | 712.2 | 0 | 0 | 0 |
| 313 | 785 | 533.16 | 0 | 0 | 0 |
| 291 | 784 | 753.9 | 0 | 185 | 1 |
| 292 | 784 | 860.4 | 0 | 73 | 0 |
| 293 | 784 | 935.4 | 0 | 0 | 0 |
| 294 | 784 | 892.32 | 0 | 0 | 0 |
| 295 | 784 | 784.56 | 0 | 9 | 0 |
| 296 | 784 | 627.48 | 0 | 170 | 0 |
| 297 | 784 | 649.2 | 0 | 236 | 0 |
| 298 | 784 | 651.6 | 0 | 22 | 0 |
| 299 | 784 | 521.76 | 0 | 86 | 0 |
| 300 | 784 | 550.74 | 0 | 81 | 0 |
| 301 | 784 | 657.12 | 0 | 234 | 0 |
| 302 | 784 | 734.16 | 0 | 103 | 0 |
| 303 | 784 | 794.94 | 0 | 78 | 1 |
| 304 | 784 | 763.14 | 0 | 117 | 0 |
| 305 | 784 | 777.9 | 0 | 93 | 1 |
| 306 | 784 | 744.42 | 0 | 228 | 0 |
| 307 | 784 | 823.74 | 0 | 64 | 1 |
| 308 | 784 | 678.6 | 0 | 0 | 1 |
| 309 | 784 | 635.82 | 0 | 0 | 0 |
| 310 | 784 | 653.34 | 0 | 0 | 1 |
| 311 | 784 | 783.48 | 0 | 0 | 0 |
| 312 | 784 | 628.08 | 0 | 0 | 0 |
| 313 | 784 | 455.52 | 0 | 0 | 0 |
| 291 | 783 | 765.96 | 0 | 9 | 1 |
| 292 | 783 | 801.24 | 0 | 0 | 0 |
| 294 | 783 | 824.04 | 0 | 4 | 0 |
| 295 | 783 | 798.42 | 0 | 152 | 0 |
| 296 | 783 | 676.5 | 0 | 207 | 0 |
| 297 | 783 | 688.26 | 0 | 95 | 0 |
| 298 | 783 | 584.58 | 0 | 0 | 0 |
| 299 | 783 | 503.58 | 0 | 8 | 0 |
| 300 | 783 | 574.98 | 0 | 34 | 0 |
| 301 | 783 | 621.48 | 0 | 33 | 0 |
| 303 | 783 | 832.38 | 0 | 0 | 0 |
| 304 | 783 | 835.5 | 0 | 139 | 0 |
| 305 | 783 | 772.02 | 0 | 196 | 0 |
| 306 | 783 | 758.88 | 0 | 167 | 0 |
| 307 | 783 | 680.7 | 0 | 34 | 0 |
| 308 | 783 | 583.56 | 0 | 0 | 1 |
| 309 | 783 | 728.4 | 0 | 11 | 1 |
| 310 | 783 | 740.64 | 0 | 0 | 1 |
| 311 | 783 | 803.76 | 0 | 0 | 1 |
| 312 | 783 | 542.64 | 0 | 0 | 0 |
| 313 | 783 | 452.76 | 0 | 0 | 0 |
| 295 | 782 | 735.42 | 0 | 0 | 1 |
| 296 | 782 | 745.8 | 0 | 0 | 0 |
| 297 | 782 | 710.1 | 0 | 0 | 0 |
| 304 | 782 | 765.72 | 0 | 167 | 0 |
| 305 | 782 | 724.02 | 0 | 282 | 0 |
| 306 | 782 | 625.08 | 0 | 251 | 0 |
| 307 | 782 | 589.68 | 0 | 71 | 0 |
| 308 | 782 | 679.68 | 0 | 35 | 0 |
| 309 | 782 | 783.42 | 0 | 0 | 0 |
| 310 | 782 | 739.8 | 0 | 0 | 0 |
| 311 | 782 | 623.1 | 0 | 0 | 1 |
| 312 | 782 | 484.2 | 0 | 0 | 1 |
| 313 | 782 | 532.02 | 0 | 0 | 0 |
| 304 | 781 | 787.62 | 0 | 45 | 1 |
| 305 | 781 | 795.72 | 0 | 151 | 0 |
| 306 | 781 | 667.44 | 0 | 52 | 0 |
| 307 | 781 | 734.64 | 0 | 18 | 0 |
| 308 | 781 | 808.8 | 0 | 34 | 0 |
| 309 | 781 | 867.78 | 0 | 0 | 0 |
| 310 | 781 | 715.74 | 0 | 0 | 1 |
| 311 | 781 | 590.4 | 0 | 0 | 0 |
| 312 | 781 | 600.06 | 0 | 0 | 0 |
| 313 | 781 | 663.54 | 0 | 48 | 0 |
| 305 | 780 | 869.1 | 0 | 41 | 0 |
| 306 | 780 | 804.72 | 0 | 5 | 0 |
| 307 | 780 | 780.42 | 0 | 129 | 0 |
| 308 | 780 | 815.58 | 0 | 50 | 0 |
| 309 | 780 | 701.52 | 0 | 106 | 0 |
| 310 | 780 | 598.26 | 0 | 147 | 0 |
| 311 | 780 | 609.96 | 0 | 21 | 0 |
| 312 | 780 | 751.5 | 0 | 54 | 0 |
| 313 | 780 | 810.72 | 0 | 53 | 1 |
| 305 | 779 | 852.48 | 0 | 97 | 0 |
| 306 | 779 | 820.56 | 0 | 80 | 1 |
| 307 | 779 | 796.92 | 0 | 116 | 0 |
| 308 | 779 | 743.7 | 0 | 199 | 0 |
| 309 | 779 | 625.5 | 0 | 369 | 0 |
| 310 | 779 | 631.98 | 0 | 214 | 0 |
| 311 | 779 | 705.72 | 0 | 282 | 0 |
| 312 | 779 | 782.4 | 0 | 79 | 0 |
| 313 | 779 | 739.32 | 0 | 0 | 0 |
| 314 | 779 | 664.74 | 0 | 0 | 1 |
| 308 | 778 | 711 | 0 | 32 | 0 |
| 309 | 778 | 749.94 | 0 | 252 | 0 |
| 310 | 778 | 810.9 | 0 | 15 | 0 |
| 311 | 778 | 822 | 0 | 124 | 0 |
| 312 | 778 | 800.16 | 0 | 3 | 0 |
| 313 | 778 | 785.82 | 0 | 0 | 0 |
| 309 | 777 | 830.7 | 0 | 0 | 0 |
| 310 | 777 | 863.28 | 0 | 0 | 0 |
| 311 | 777 | 829.38 | 0 | 0 | 0 |
| 309 | 790 | 383.58 | 0 | 0 | 1 |

## red deer estimate ratios.txt

#################################################################

Invalid estimates as percentages of valid estimates: radius= 1.5

Full red deer dataset

| model# | east | north | alt.sq | pine | mires | |
|---|---|---|---|---|---|---|
| 1 | | | | | | 95 |
| 2 | | | | | 95 | |
| 3 | | | | | 95 | 95 |
| 4 | | | | 89 | | |
| 5 | | | | 89 | | 96 |
| 6 | | | | 89 | 98 | |
| 7 | | | | 89 | 98 | 96 |
| 8 | | | 107 | | | |
| 9 | | | 105 | | | 96 |
| 10 | | | 106 | | 96 | |

| | | | | | |
|---|---|---|---|---|---|
| 11 | | 104 | | 96 | 96 |
| 12 | | 109 | 90 | | |
| 13 | | 107 | 90 | | 97 |
| 14 | | 108 | 90 | 99 | |
| 15 | | 106 | 89 | 99 | 97 |
| 16 | 118 | | | | |
| 17 | 117 | | | | 99 |
| 18 | 119 | | | 94 | |
| 19 | 117 | | | 93 | 99 |
| 20 | 104 | | 97 | | |
| 21 | 104 | | 98 | | 101 |
| 22 | 104 | | 98 | 84 | |
| 23 | 105 | | 98 | 71 | 101 |
| 24 | 115 | 107 | | | |
| 25 | 114 | 105 | | | 99 |
| 26 | 115 | 106 | | 95 | |
| 27 | 113 | 105 | | 95 | 100 |
| 28 | 103 | 104 | 98 | | |
| 29 | 103 | 104 | 99 | | 101 |
| 30 | 103 | 104 | 99 | 91 | |
| 31 | 104 | 103 | 99 | 87 | 101 |

E half of red deer dataset:

| model# | east | north | alt.sq | pine | mires |
|---|---|---|---|---|---|
| 1 | | | | | 92 |
| 2 | | | | 210 | |
| 3 | | | | 176 | 93 |
| 4 | | | 63 | | |
| 5 | | | 60 | | 92 |
| 6 | | | 79 | 204 | |
| 7 | | | 77 | 173 | 93 |
| 8 | | -175 | | | |
| 9 | | 316 | | | 96 |
| 10 | | -59 | | 182 | |
| 11 | | 453 | | 161 | 96 |
| 12 | | -140 | 40 | | |
| 13 | | 371 | 24 | | 96 |
| 14 | | -40 | 60 | 175 | |
| 15 | | 779 | 53 | 157 | 96 |
| 16 | 68 | | | | |
| 17 | 50 | | | | 98 |
| 18 | 71 | | | 193 | |
| 19 | 56 | | | 172 | 99 |
| 20 | 71 | | 69 | | |
| 21 | 55 | | 56 | | 103 |
| 22 | 76 | | 77 | 161 | |
| 23 | 64 | | 67 | 154 | 102 |
| 24 | 86 | 149 | | | |
| 25 | 83 | 142 | | | 98 |
| 26 | 85 | 143 | | 187 | |
| 27 | 81 | 138 | | 165 | 99 |
| 28 | 86 | 117 | 81 | | |
| 29 | 80 | 117 | 74 | | 102 |
| 30 | 87 | 113 | 84 | 157 | |
| 31 | 81 | 113 | 78 | 149 | 102 |

W half of red deer dataset:

| model# | east | north | alt.sq | pine | mires |
|---|---|---|---|---|---|
| 1 | | | | | 120 |
| 2 | | | | 97 | |
| 3 | | | | 97 | 138 |
| 4 | | | 101 | | |
| 5 | | | 100 | | -8 |
| 6 | | | 101 | 94 | |
| 7 | | | 101 | 94 | 55 |
| 8 | | 90 | | | |
| 9 | | 90 | | | 108 |
| 10 | | 91 | | 97 | |
| 11 | | 92 | | 97 | 112 |
| 12 | | 96 | 100 | | |
| 13 | | 97 | 100 | | 108 |
| 14 | | 97 | 100 | 95 | |
| 15 | | 97 | 100 | 95 | 111 |
| 16 | 101 | | | | |
| 17 | 86 | | | | 125 |
| 18 | 101 | | | 97 | |
| 19 | 82 | | | 97 | 150 |
| 20 | 103 | | 101 | | |
| 21 | 106 | | 101 | | 129 |
| 22 | 103 | | 101 | 94 | |
| 23 | 106 | | 101 | 93 | 148 |
| 24 | 94 | 90 | | | |
| 25 | 89 | 91 | | | 113 |
| 26 | 94 | 91 | | 97 | |
| 27 | 89 | 92 | | 97 | 121 |
| 28 | 29 | 96 | 100 | | |
| 29 | -397 | 96 | 100 | | 116 |
| 30 | 29 | 96 | 100 | 95 | |
| 31 | -187 | 96 | 100 | 95 | 122 |

N half of red deer dataset:

| model# | east | north | alt.sq | pine | mires |
|---|---|---|---|---|---|
| 1 | | | | | 83 |
| 2 | | | | 86 | |
| 3 | | | | 90 | 83 |
| 4 | | | 79 | | |
| 5 | | | 80 | | 90 |
| 6 | | | 79 | 43 | |
| 7 | | | 80 | 65 | 90 |
| 8 | | 69 | | | |
| 9 | | 71 | | | 87 |
| 10 | | 69 | | -23 | |
| 11 | | 71 | | 47 | 87 |
| 12 | | 70 | 83 | | |
| 13 | | 72 | 84 | | 91 |
| 14 | | 70 | 83 | 65 | |
| 15 | | 72 | 84 | 71 | 91 |
| 16 | -31 | | | | |
| 17 | -443 | | | | 94 |
| 18 | -28 | | | 80 | |
| 19 | -391 | | | 70 | 95 |
| 20 | 114 | | 93 | | |
| 21 | 110 | | 94 | | 98 |
| 22 | 113 | | 94 | 107 | |
| 23 | 109 | | 95 | 105 | 98 |
| 24 | -75 | 72 | | | |
| 25 | 618 | 74 | | | 97 |
| 26 | -76 | 71 | | -5 | |
| 27 | 539 | 74 | | 71 | 97 |
| 28 | 117 | 64 | 98 | | |
| 29 | 112 | 69 | 98 | | 98 |
| 30 | 115 | 67 | 98 | 96 | |
| 31 | 110 | 73 | 98 | 97 | 98 |

S half of red deer dataset:

| model# | east | north | alt.sq | pine | mires |
|---|---|---|---|---|---|
| 1 | | | | | 94 |
| 2 | | | | 70 | |
| 3 | | | | 70 | 98 |
| 4 | | | 85 | | |
| 5 | | | 85 | | 95 |
| 6 | | | 90 | 70 | |
| 7 | | | 90 | 70 | 98 |
| 8 | | 67 | | | |
| 9 | | 68 | | | 95 |
| 10 | | 69 | | 71 | |
| 11 | | 70 | | 71 | 98 |
| 12 | | 66 | 85 | | |
| 13 | | 67 | 85 | | 96 |
| 14 | | 68 | 89 | 70 | |
| 15 | | 70 | 90 | 70 | 99 |
| 16 | 100 | | | | |
| 17 | 101 | | | | 97 |
| 18 | 101 | | | 72 | |
| 19 | 102 | | | 72 | 100 |
| 20 | 98 | | 94 | | |
| 21 | 99 | | 95 | | 100 |
| 22 | 100 | | 97 | 69 | |
| 23 | 101 | | 98 | 69 | 103 |
| 24 | 104 | 50 | | | |
| 25 | 104 | 49 | | | 98 |
| 26 | 104 | 31 | | 74 | |
| 27 | 104 | 28 | | 74 | 100 |
| 28 | 101 | 405 | 97 | | |
| 29 | 102 | 325 | 97 | | 101 |
| 30 | 102 | 153 | 99 | 73 | |
| 31 | 103 | 149 | 100 | 73 | 103 |

#################################################################

Invalid estimates as percentages of valid estimates: radius= 2

Full red deer dataset

| model# | east | north | alt.sq | pine | mires |
|---|---|---|---|---|---|
| 1 | | | | | 94 |
| 2 | | | | 98 | |
| 3 | | | | 98 | 94 |
| 4 | | | 84 | | |
| 5 | | | 83 | | 97 |
| 6 | | | 83 | 106 | |
| 7 | | | 83 | 107 | 97 |
| 8 | | 113 | | | |
| 9 | | 110 | | | 96 |
| 10 | | 112 | | 99 | |
| 11 | | 109 | | 100 | 95 |
| 12 | | 114 | 85 | | |
| 13 | | 111 | 84 | | 98 |
| 14 | | 114 | 84 | 107 | |
| 15 | | 111 | 83 | 108 | 98 |
| 16 | 140 | | | | |
| 17 | 137 | | | | 100 |
| 18 | 141 | | | 96 | |
| 19 | 137 | | | 95 | 100 |
| 20 | 108 | | 96 | | |
| 21 | 108 | | 96 | | 103 |
| 22 | 108 | | 96 | 96 | |
| 23 | 108 | | 97 | 86 | 103 |
| 24 | 133 | 116 | | | |
| 25 | 129 | 112 | | | 101 |
| 26 | 133 | 114 | | 98 | |
| 27 | 129 | 111 | | 98 | 101 |
| 28 | 106 | 107 | 98 | | |
| 29 | 106 | 106 | 98 | | 103 |
| 30 | 106 | 107 | 98 | 100 | |
| 31 | 107 | 106 | 98 | 97 | 103 |

E half of red deer dataset:

| model# | east | north | alt.sq | pine | mires |
|---|---|---|---|---|---|
| 1 | | | | | 91 |
| 2 | | | | 234 | |
| 3 | | | | 194 | 92 |
| 4 | | | 80 | | |
| 5 | | | 74 | | 91 |
| 6 | | | 93 | 227 | |
| 7 | | | 88 | 190 | 92 |
| 8 | | -152 | | | |
| 9 | | 455 | | | 95 |
| 10 | | -60 | | 198 | |
| 11 | | 975 | | 175 | 95 |
| 12 | | -119 | 51 | | |
| 13 | | 617 | 38 | | 95 |
| 14 | | -39 | 68 | 190 | |
| 15 | | -6164 | 62 | 170 | 95 |
| 16 | 58 | | | | |
| 17 | 24 | | | | 97 |
| 18 | 62 | | | 213 | |
| 19 | 33 | | | 188 | 98 |
| 20 | 65 | | 67 | | |
| 21 | 40 | | 48 | | 102 |
| 22 | 72 | | 75 | 176 | |
| 23 | 53 | | 62 | 167 | 102 |
| 24 | 84 | 176 | | | |
| 25 | 79 | 167 | | | 98 |
| 26 | 82 | 172 | | 202 | |
| 27 | 75 | 165 | | 177 | 98 |
| 28 | 86 | 126 | 82 | | |
| 29 | 78 | 128 | 72 | | 101 |
| 30 | 87 | 120 | 84 | 167 | |
| 31 | 78 | 122 | 76 | 160 | 101 |

W half of red deer dataset:

| model# | east | north | alt.sq | pine | mires |
|---|---|---|---|---|---|
| 1 | | | | | -504 |
| 2 | | | | 96 | |
| 3 | | | | 96 | 25 |
| 4 | | | 100 | | |
| 5 | | | 99 | | 78 |
| 6 | | | 100 | 94 | |
| 7 | | | 100 | 93 | 80 |
| 8 | | 86 | | | |
| 9 | | 87 | | | 143 |
| 10 | | 87 | | 96 | |
| 11 | | 88 | | 96 | 216 |
| 12 | | 94 | 99 | | |
| 13 | | 95 | 99 | | 465 |
| 14 | | 95 | 99 | 95 | |
| 15 | | 95 | 99 | 95 | -41 |
| 16 | 83 | | | | |
| 17 | 69 | | | | -49 |
| 18 | 73 | | | 96 | |
| 19 | 63 | | | 96 | 36 |
| 20 | 107 | | 100 | | |
| 21 | 113 | | 100 | | 27 |
| 22 | 107 | | 101 | 93 | |
| 23 | 113 | | 101 | 93 | 46 |
| 24 | 85 | 86 | | | |
| 25 | 80 | 87 | | | 339 |
| 26 | 84 | 88 | | 96 | |
| 27 | 80 | 88 | | 96 | -58 |
| 28 | -25 | 93 | 99 | | |
| 29 | -48 | 94 | 99 | | -618 |
| 30 | -33 | 94 | 99 | 95 | |
| 31 | -34 | 94 | 99 | 95 | -39 |

N half of red deer dataset:

| model# | east | north | alt.sq | pine | mires |
|---|---|---|---|---|---|
| 1 | | | | | 83 |
| 2 | | | | 93 | |
| 3 | | | | 102 | 83 |
| 4 | | | 80 | | |
| 5 | | | 80 | | 90 |
| 6 | | | 80 | 49 | |
| 7 | | | 80 | 65 | 90 |
| 8 | | 73 | | | |
| 9 | | 75 | | | 87 |
| 10 | | 72 | | -2 | |
| 11 | | 74 | | 50 | 87 |
| 12 | | 75 | 83 | | |
| 13 | | 77 | 84 | | 92 |
| 14 | | 75 | 83 | 68 | |
| 15 | | 77 | 84 | 74 | 91 |
| 16 | -15 | | | | |
| 17 | -183 | | | | 95 |
| 18 | -13 | | | 83 | |
| 19 | -169 | | | 76 | 96 |
| 20 | 120 | | 96 | | |
| 21 | 115 | | 97 | | 99 |
| 22 | 119 | | 97 | 109 | |
| 23 | 113 | | 97 | 106 | 99 |
| 24 | -41 | 76 | | | |
| 25 | -2951 | 78 | | | 98 |
| 26 | -42 | 75 | | 14 | |
| 27 | 43121 | 78 | | 74 | 98 |
| 28 | 123 | 68 | 100 | | |
| 29 | 116 | 74 | 100 | | 100 |
| 30 | 120 | 71 | 100 | 98 | |
| 31 | 113 | 77 | 100 | 99 | 99 |

S half of red deer dataset:

| model# | east | north | alt.sq | pine | mires |
|---|---|---|---|---|---|
| 1 | | | | | 92 |
| 2 | | | | 53 | |
| 3 | | | | 53 | 96 |
| 4 | | | 71 | | |
| 5 | | | 69 | | 94 |
| 6 | | | 76 | 52 | |
| 7 | | | 74 | 52 | 98 |
| 8 | | 47 | | | |
| 9 | | 48 | | | 93 |
| 10 | | 47 | | 54 | |
| 11 | | 49 | | 53 | 97 |
| 12 | | 47 | 71 | | |
| 13 | | 48 | 70 | | 95 |
| 14 | | 48 | 76 | 53 | |
| 15 | | 50 | 74 | 52 | 98 |
| 16 | 102 | | | | |
| 17 | 103 | | | | 96 |
| 18 | 103 | | | 55 | |
| 19 | 104 | | | 55 | 101 |
| 20 | 95 | | 88 | | |
| 21 | 96 | | 88 | | 101 |
| 22 | 98 | | 92 | 50 | |
| 23 | 99 | | 93 | 50 | 105 |
| 24 | 109 | 8 | | | |
| 25 | 109 | 5 | | | 97 |
| 26 | 107 | -88 | | 58 | |
| 27 | 108 | -109 | | 58 | 101 |
| 28 | 100 | 316 | 91 | | |
| 29 | 101 | 291 | 92 | | 102 |
| 30 | 101 | 157 | 95 | 56 | |
| 31 | 102 | 155 | 95 | 56 | 105 |

## GibbsMPLE.R

## (Uses ‘deer20%sample.txt’ as input)

```
# Maximum pseudo-likelihood estimation of autologistic model with Gibbs sampling of missing data.
# Author: David C Bardos
library(lattice);set.seed(999);
setwd('C:/auto'); #change as required

nML=100L;# number  >2 of MPLE phases (= number of iterations `T' in the pseudocode)
Tmin=50;  # start running averages (must have Tmin > 1)
Tmin=min(Tmin,nML-1); #     ensures Tmin < nML
t.R=95:100; # iteration numbers to retain full set of y_n maps
pa.maps=list(); # store presence-absence maps here
nMiss=1L; #number of Gibbs cycles after each MPLE phase

covCentFlag=F; # center covariates or not

obs.df=read.table("deer20%sample.txt",header=T,sep="");
# data file must have columns in the format: X, Y, covariate1, covariate2, ... , observations
# observations can be 0, 1 or 'NA'


radius=1.5;   # neighbourhood radius r
lattice.ratio=1;  #ratio of lattice spacings,i.e. lattice.ratio=a.y/a.x where a.x & a.y are lattice spacing in the x & y directions
# for square lattice set lattice.ratio=1. Putting radius=1.5 with lattice.ratio=1 gives 2nd-order (8-cell) neighbourhood
nbimage=matrix(0L,ncol=2*radius+3,nrow=2*radius+3);
l.x=as.integer(ceiling(radius));l.y=as.integer(ceiling(radius/lattice.ratio));lr2=lattice.ratio^2;
nbimage=matrix(0L,ncol=2*l.y+3,nrow=2*l.x+3);
nbmat=matrix(,ncol=2,nrow=0,byrow=T); # start with empty list
for (i in (-l.x:l.x)) for (j in (-l.y:l.y))
  if ((i^2 + lr2*j^2<=radius^2)&&  (!identical(c(i,j),c(0L,0L))) ){ nbmat=rbind(nbmat,t(c(i,j)));nbimage[i+l.x+2,j+l.y+2]=1L; }
Nnb=length(nbmat[,1]);levelplot(nbimage)

power=2.0;range=2.5; # power law and exponential decay parameters
r.nb= sqrt(nbmat[,1]^2+lr2*nbmat[,2]^2);
# 4 otions for specifying weight vs distance:
# wvec=r.nb^(-power); # weights w,  power-law decline vs. central distance
# wvec=exp(-r.nb/range);# weights w,  exponential decline vs. central distance
#wvec=r.nb^(-power)*exp(-r.nb/range);# weights w,  exponential and powerlaw decline vs. central distance
wvec=rep(1,Nnb) # or just use uniform weights


dx=max(abs(nbmat[,1]));dy=max(abs(nbmat[,2]));border=max(dx,dy); # minimum border requirements
parameters.n=2;covflag=F; # start with autocovariate and constant term

obs.len=length(obs.df[,1]);obs.width=length(obs.df[1,]);covariate.data.n=obs.width-3;# number of columns aside from x,y and observation.

# user specifies covariates included in estimation:

#covariates.estimated=c("rain","djungle");
covariates.estimated=c("east","north","alt.squared","pine","mires");
#covariates.estimated=c("alt.squared","pine","mires");
#covariates.estimated=c("alt.squared","pine","mires");
#covariates.estimated=c("alt.squared")

if(covariate.data.n > 0) {
  covariates.list=colnames(obs.df)[3:(obs.width-1)];
  if(length(which(covariates.estimated%in%covariates.list))==length(covariates.estimated)) {
  covindex=match(covariates.estimated,colnames(obs.df));covindex=unique(covindex);
  covflag=T;covariate.param.n=length(covindex);parameters.n=2+covariate.param.n} else warning("Covariate mismatch");
}

obs.table=matrix(0L,nrow=obs.len,ncol=obs.width);
for (j in 1:obs.width) obs.table[,j]=obs.df[,j];
nsites=obs.len;xmin=min(obs.table[,1]);ymin=min(obs.table[,2]);
xmax=max(obs.table[,1]);ymax=max(obs.table[,2]);Length=ymax-ymin+1;width=xmax-xmin+1;
observed=missing=1L[-1]; #initially empty index vectors
datamat=matrix(0L,nrow=width+2*border,ncol=Length+2*border); # use 1st array index for x (width)
p.hat=p.hat.sum=p.hat.mean=0.0*datamat;p.hat=p.hat-1; #  occupation probabilities
for (k in t.R) pa.maps[[k]]=datamat-1L;
if(covCentFlag) for (l in covindex) obs.table[,l]=obs.table[,l]-mean(obs.table[,l]); # center covariates
```

```
obs.table[,1]=obs.table[,1]-(xmin-1)+border;  obs.table[,2]=obs.table[,2]-(ymin-1)+border;
for (k in 1:obs.len) {if(is.na(obs.table[k,obs.width])) missing=append(missing,k) else
                    {observed=append(observed,k);datamat[obs.table[k,1],obs.table[k,2]]=as.integer(obs.table[k,obs.width]);}
                    #x is 1st coord in matrix
}# k loop

observed.n=length(observed);missing.n=length(missing);
observed.xy=matrix(0L,nrow=observed.n,ncol=2);missing.xy=matrix(0L,nrow=missing.n,ncol=2);
observed.pa=as.integer(obs.table[observed,obs.width]);
missing.pa=as.integer(obs.table[missing,obs.width]);#init to NA so we can check all values are set


if(observed.n>0) { for (k in 1:observed.n) {# generate observed.xy using "observed" to give sequence of entries in data file
  observed.xy[k,1]=as.integer(obs.table[observed[k],1]);observed.xy[k,2]=as.integer(obs.table[observed[k],2]);}# end k loop
} #endif observed.n>0

if(missing.n>0) {for (k in 1:missing.n) {# generate missing.xy using "missing" to give sequences of entries in data file
  missing.xy[k,1]=as.integer(obs.table[missing[k],1]);missing.xy[k,2]=as.integer(obs.table[missing[k],2]);}# end k loop
} #endif missing.n>0

observed.covariates=as.matrix(obs.table[observed,covindex]);
# as.matrix so %*% product with parameter vector is OK when only 1 covariate.
missing.covariates=as.matrix(obs.table[missing,covindex]);#nb: it is observations that are missing, these are the associated covariates
h.observed=h.observed.proposed=rep(0,observed.n);h.missing=h.missing.proposed=rep(0,missing.n);

theta=matrix(0,nrow=nML,ncol=parameters.n);

f1=glm(observed.pa~observed.covariates,family=binomial);c1=as.vector(coefficients(f1));
theta[1,]=c(c1,0);

if(covflag){ # initialize objects depending on parameters
covariate.params=theta[1,2:(1+covariate.param.n)];auto=theta[1,parameters.n];
h.observed[]=observed.covariates%*%covariate.params+theta[1,1]; # X_i.beta + alpha
h.missing[]=missing.covariates%*%covariate.params+theta[1,1]} else{
auto=theta[1,parameters.n]; h.observed[]=h.missing[]=theta[1,1];}


full.n=observed.n+missing.n;
full.xy=rbind(observed.xy,missing.xy);
observed.ac=rep(0,observed.n);


for(n in 2L:nML){  # n='t' in the pseudocode

# update missing data via gibbs sampling
if(missing.n>0){ pvec=runif(nMiss*missing.n,0,1);pcount=0L; # Bernoulli draws for nMiss Gibbs cycles

for(k in 1:nMiss) {
for(m in 1:missing.n)  {

pcount=pcount+1;
i=missing.xy[m,1]; j=missing.xy[m,2];
autocov=0;
for(l in 1:Nnb) {autocov=autocov+datamat[i+nbmat[l,1],j+nbmat[l,2]]*wvec[l] ;}

prob=exp(h.missing[m]+auto*autocov);
prob=prob/(1+prob);

if(pvec[pcount]<prob) {datamat[i,j]=1L;missing.pa[m]=1L;} else {datamat[i,j]=0L;missing.pa[m]=0L;}


} # m loop
} # k loop


  } # if(missing.n>0)

#ML step

observed.ac[]=0;
for(m in 1:observed.n)  {

i=observed.xy[m,1]; j=observed.xy[m,2];
for(l in 1:Nnb) {observed.ac[m]=observed.ac[m]+datamat[i+nbmat[l,1],j+nbmat[l,2]]*wvec[l] ;}
```

```
                    }

f1=glm(observed.pa~observed.covariates+observed.ac,family=binomial);c1=as.vector(coefficients(f1));
theta[n,]=c1;

print("n=");print(n);
print("theta[n,]=");print(theta[n,]);
if(n %in% t.R) for(m in 1:full.n)  {i=full.xy[m,1]; j=full.xy[m,2]; pa.maps[[n]][i,j]=datamat[i,j];}

if(covflag){ # calculate objects depending on parameters
covariate.params=theta[n,2:(1+covariate.param.n)];auto=theta[n,parameters.n];
h.observed[]=observed.covariates%*%covariate.params+theta[n,1]; # X_i.beta + alpha
h.missing[]=missing.covariates%*%covariate.params+theta[n,1]} else{
auto=theta[n,parameters.n]; h.observed[]=h.missing[]=theta[n,1];}


if(n>Tmin)    {h.full=c(h.observed,h.missing);
for(m in 1:full.n)  {
i=full.xy[m,1]; j=full.xy[m,2];
autocov=0;
for(l in 1:Nnb) {autocov=autocov+datamat[i+nbmat[l,1],j+nbmat[l,2]]*wvec[l] ;}
prob=exp(h.full[m]+auto*autocov);
prob=prob/(1+prob); p.hat[i,j]=prob;
} # m loop
p.hat.sum=p.hat.sum+p.hat;
            }# if(n>Tmin)


} #end main loop (n)

p.hat.mean=p.hat.sum/(nML-Tmin)

( parameter.means=colMeans(theta[Tmin:nML,]) );
levelplot(pa.maps[[max(t.R)]]) #show last retained presence-absence map
sum(datamat) # number of occupied sites at final iteration
levelplot(p.hat.mean)   # show map of estimated occupation probabilities
```

## deer20%sample.txt

| "x" | "y" | "east" | "north" | "alt.squared" | "pine" | "mires" | "obs" |
|---|---|---|---|---|---|---|---|
| 346 | 798 | 346 | 798 | 48593.7936 | 0 | 0 | NA |
| 347 | 798 | 347 | 798 | 47123.7264 | 0 | 0 | NA |
| 350 | 798 | 350 | 798 | 26438.76   0 | 0 | NA | |
| 340 | 797 | 340 | 797 | 100362.24 34 | 0 | NA | |
| 341 | 797 | 341 | 797 | 45522.4896 | 0 | 0 | NA |
| 344 | 797 | 344 | 797 | 82667.7504 | 0 | 0 | NA |
| 345 | 797 | 345 | 797 | 110689.29 78 | 0 | NA | |
| 346 | 797 | 346 | 797 | 98734.2084 | 61 | 0 | 0 |
| 347 | 797 | 347 | 797 | 46474.7364 | 0 | 0 | NA |
| 348 | 797 | 348 | 797 | 39752.3844 | 0 | 0 | 0 |
| 349 | 797 | 349 | 797 | 46474.7364 | 0 | 0 | NA |
| 351 | 797 | 351 | 797 | 88923.24   0 | 0 | NA | |
| 352 | 797 | 352 | 797 | 66780.8964 | 0 | 0 | NA |
| 353 | 797 | 353 | 797 | 30821.3136 | 0 | 0 | NA |
| 354 | 797 | 354 | 797 | 25071.5556 | 0 | 0 | NA |
| 355 | 797 | 355 | 797 | 29963.61   0 | 0 | NA | |
| 338 | 796 | 338 | 796 | 177712.8336 | 0 | 0 | NA |
| 339 | 796 | 339 | 796 | 209141.5824 | 0 | 0 | NA |
| 340 | 796 | 340 | 796 | 127534.6944 | 0 | 0 | NA |
| 341 | 796 | 341 | 796 | 74856.96   0 | 0 | NA | |
| 342 | 796 | 342 | 796 | 81533.0916 | 0 | 0 | NA |
| 343 | 796 | 343 | 796 | 112694.49 0 | 0 | NA | |
| 344 | 796 | 344 | 796 | 171396      0 | 0 | NA | |
| 345 | 796 | 345 | 796 | 127363.3344 | 11 | 0 | NA |
| 346 | 796 | 346 | 796 | 101773.7604 | 91 | 0 | NA |
| 347 | 796 | 347 | 796 | 48699.6624 | 31 | 0 | NA |
| 348 | 796 | 348 | 796 | 45445.7124 | 0 | 0 | 0 |
| 349 | 796 | 349 | 796 | 91047.0276 | 0 | 0 | 0 |
| 350 | 796 | 350 | 796 | 89532.6084 | 0 | 0 | NA |
| 351 | 796 | 351 | 796 | 108030.5424 | 0 | 0 | NA |
| 353 | 796 | 353 | 796 | 45062.7984 | 0 | 0 | 0 |
| 354 | 796 | 354 | 796 | 53056.5156 | 0 | 0 | 0 |
| 329 | 795 | 329 | 795 | 161700.4944 | 78 | 0 | NA |
| 332 | 795 | 332 | 795 | 129859.3296 | 56 | 0 | NA |
| 333 | 795 | 333 | 795 | 115246.6704 | 111 | 0 | 0 |
| 337 | 795 | 337 | 795 | 118446.1056 | 0 | 0 | NA |
| 338 | 795 | 338 | 795 | 250040.0016 | 0 | 0 | NA |
| 339 | 795 | 339 | 795 | 260222.4144 | 0 | 0 | NA |
| 340 | 795 | 340 | 795 | 167837.7024 | 0 | 0 | 0 |
| 341 | 795 | 341 | 795 | 153742.41 0 | 0 | NA | |
| 342 | 795 | 342 | 795 | 198737.64 0 | 0 | 0 | |
| 343 | 795 | 343 | 795 | 213222.2976 | 0 | 0 | NA |
| 344 | 795 | 344 | 795 | 132947.7444 | 2 | 0 | NA |
| 345 | 795 | 345 | 795 | 80951.6304 | 74 | 0 | NA |
| 346 | 795 | 346 | 795 | 54952.7364 | 24 | 0 | NA |
| 347 | 795 | 347 | 795 | 45701.8884 | 9 | 0 | NA |
| 348 | 795 | 348 | 795 | 90324.2916 | 1 | 0 | NA |
| 349 | 795 | 349 | 795 | 148656.5136 | 89 | 0 | 0 |
| 350 | 795 | 350 | 795 | 157815.5076 | 0 | 0 | NA |
| 351 | 795 | 351 | 795 | 137048.04 0 | 0 | 0 | |
| 352 | 795 | 352 | 795 | 130855.8276 | 0 | 0 | NA |
| 353 | 795 | 353 | 795 | 87355.7136 | 0 | 0 | NA |
| 329 | 794 | 329 | 794 | 175343.1876 | 17 | 0 | NA |
| 330 | 794 | 330 | 794 | 209910.5856 | 0 | 0 | 0 |
| 331 | 794 | 331 | 794 | 124298.5536 | 0 | 0 | 1 |
| 332 | 794 | 332 | 794 | 135188.5824 | 0 | 0 | NA |
| 333 | 794 | 333 | 794 | 127277.6976 | 0 | 0 | NA |
| 336 | 794 | 336 | 794 | 122990.49 0 | 0 | NA | |
| 337 | 794 | 337 | 794 | 212447.2464 | 0 | 0 | NA |
| 338 | 794 | 338 | 794 | 292183.4916 | 0 | 0 | NA |
| 339 | 794 | 339 | 794 | 267557.9076 | 0 | 0 | NA |
| 340 | 794 | 340 | 794 | 242536.5504 | 0 | 0 | NA |
| 341 | 794 | 341 | 794 | 245500.4304 | 0 | 0 | NA |
| 342 | 794 | 342 | 794 | 199326.5316 | 0 | 0 | NA |
| 343 | 794 | 343 | 794 | 123158.8836 | 147 | 0 | NA |
| 344 | 794 | 344 | 794 | 94888.6416 | 267 | 0 | 0 |
| 345 | 794 | 345 | 794 | 123833.61 95 | 0 | NA | |
| 346 | 794 | 346 | 794 | 64719.36 11 | 0 | NA | |
| 347 | 794 | 347 | 794 | 83116.89 0 | 0 | NA | |
| 348 | 794 | 348 | 794 | 121773.0816 | 34 | 0 | 0 |
| 349 | 794 | 349 | 794 | 190375.1424 | 53 | 0 | 0 |
| 350 | 794 | 350 | 794 | 190375.1424 | 0 | 0 | NA |
| 351 | 794 | 351 | 794 | 179945.64 28 | 0 | NA | |
| 352 | 794 | 352 | 794 | 177510.5424 | 0 | 0 | 0 |
| 353 | 794 | 353 | 794 | 82564.2756 | 0 | 0 | 0 |
| 356 | 794 | 356 | 794 | 51438.24 0 | 0 | NA | |
| 326 | 793 | 326 | 793 | 202716.0576 | 0 | 0 | 0 |
| 327 | 793 | 327 | 793 | 169068.9924 | 0 | 0 | NA |
| 330 | 793 | 330 | 793 | 170998.7904 | 0 | 0 | NA |
| 331 | 793 | 331 | 793 | 113703.84 0 | 0 | NA | |
| 332 | 793 | 332 | 793 | 204611.4756 | 0 | 0 | 0 |
| 335 | 793 | 335 | 793 | 93856.4496 | 0 | 0 | NA |
| 336 | 793 | 336 | 793 | 159153.1236 | 0 | 0 | NA |
| 337 | 793 | 337 | 793 | 287660.5956 | 0 | 0 | NA |
| 338 | 793 | 338 | 793 | 303512.8464 | 0 | 0 | NA |
| 339 | 793 | 339 | 793 | 304836.4944 | 0 | 0 | NA |
| 340 | 793 | 340 | 793 | 270545.6196 | 0 | 0 | 0 |
| 341 | 793 | 341 | 793 | 226556.9604 | 0 | 0 | NA |
| 342 | 793 | 342 | 793 | 144476.01 3 | 0 | NA | |
| 343 | 793 | 343 | 793 | 112855.6836 | 293 | 0 | NA |
| 344 | 793 | 344 | 793 | 178067.1204 | 175 | 0 | NA |
| 345 | 793 | 345 | 793 | 158531.3856 | 167 | 0 | NA |
| 346 | 793 | 346 | 793 | 88994.8224 | 298 | 0 | 0 |
| 347 | 793 | 347 | 793 | 86966.01 168 | 0 | NA | |
| 348 | 793 | 348 | 793 | 99717.0084 | 16 | 0 | NA |
| 349 | 793 | 349 | 793 | 138607.29 5 | 0 | NA | |
| 350 | 793 | 350 | 793 | 189433.8576 | 0 | 0 | 0 |
| 351 | 793 | 351 | 793 | 173838.9636 | 0 | 0 | NA |
| 352 | 793 | 352 | 793 | 162135.0756 | 0 | 0 | NA |
| 353 | 793 | 353 | 793 | 100895.1696 | 0 | 0 | NA |
| 354 | 793 | 354 | 793 | 106145.64 0 | 0 | NA | |
| 355 | 793 | 355 | 793 | 121898.7396 | 0 | 0 | NA |
| 356 | 793 | 356 | 793 | 97082.0964 | 0 | 0 | NA |
| 315 | 792 | 315 | 792 | 146260.3536 | 0 | 0 | NA |
| 322 | 792 | 322 | 792 | 128522.25 286 | 0 | NA | |
| 324 | 792 | 324 | 792 | 158292.5796 | 0 | 0 | NA |
| 325 | 792 | 325 | 792 | 175896.36 0 | 0 | NA | |
| 326 | 792 | 326 | 792 | 284771.6496 | 0 | 0 | NA |
| 327 | 792 | 327 | 792 | 233173.0944 | 0 | 0 | 0 |
| 328 | 792 | 328 | 792 | 179843.8464 | 0 | 0 | 0 |
| 329 | 792 | 329 | 792 | 135144.4644 | 0 | 0 | NA |
| 330 | 792 | 330 | 792 | 120520.0656 | 0 | 0 | NA |
| 331 | 792 | 331 | 792 | 141300.81 0 | 0 | NA | |
| 332 | 792 | 332 | 792 | 241768.89 0 | 0 | NA | |
| 335 | 792 | 335 | 792 | 123538.1904 | 0 | 0 | NA |
| 336 | 792 | 336 | 792 | 197509.1364 | 0 | 0 | NA |
| 337 | 792 | 337 | 792 | 331499.5776 | 0 | 0 | 1 |
| 338 | 792 | 338 | 792 | 356480.6436 | 0 | 0 | NA |
| 339 | 792 | 339 | 792 | 339842.3616 | 0 | 0 | NA |
| 340 | 792 | 340 | 792 | 218294.5284 | 0 | 0 | NA |
| 341 | 792 | 341 | 792 | 135851.2164 | 0 | 0 | NA |
| 342 | 792 | 342 | 792 | 161555.7636 | 0 | 0 | 0 |
| 343 | 792 | 343 | 792 | 207334.5156 | 6 | 0 | NA |
| 344 | 792 | 344 | 792 | 272296.1124 | 0 | 0 | NA |
| 345 | 792 | 345 | 792 | 150559.5204 | 57 | 0 | NA |
| 346 | 792 | 346 | 792 | 145389.69 96 | 0 | NA | |
| 347 | 792 | 347 | 792 | 148656.5136 | 322 | 0 | NA |
| 348 | 792 | 348 | 792 | 139502.25 173 | 0 | NA | |
| 349 | 792 | 349 | 792 | 226785.4884 | 70 | 0 | NA |
| 350 | 792 | 350 | 792 | 230227.2324 | 0 | 0 | NA |
| 351 | 792 | 351 | 792 | 151212.0996 | 0 | 0 | NA |
| 352 | 792 | 352 | 792 | 121856.8464 | 0 | 0 | NA |
| 353 | 792 | 353 | 792 | 110529.6516 | 0 | 0 | 0 |
| 354 | 792 | 354 | 792 | 96671.2464 | 0 | 0 | NA |
| 355 | 792 | 355 | 792 | 78142.6116 | 0 | 0 | NA |
| 356 | 792 | 356 | 792 | 82254.24 0 | 0 | NA | |
| 360 | 792 | 360 | 792 | 15470.3844 | 0 | 0 | NA |
| 361 | 792 | 361 | 792 | 13146.9156 | 0 | 0 | NA |
| 362 | 792 | 362 | 792 | 10514.4516 | 0 | 0 | NA |
| 363 | 792 | 363 | 792 | 10040.04 0 | 0 | 0 | |
| 315 | 791 | 315 | 791 | 151819.3296 | 0 | 0 | NA |
| 316 | 791 | 316 | 791 | 316676.3076 | 0 | 0 | 1 |
| 317 | 791 | 317 | 791 | 275499.0144 | 42 | 0 | NA |
| 321 | 791 | 321 | 791 | 200292.0516 | 378 | 0 | NA |
| 322 | 791 | 322 | 791 | 229134.5424 | 188 | 0 | 1 |
| 323 | 791 | 323 | 791 | 197775.8784 | 70 | 0 | NA |
| 324 | 791 | 324 | 791 | 193952.16 0 | 0 | NA | |
| 325 | 791 | 325 | 791 | 188442.81 0 | 19 | NA | |
| 326 | 791 | 326 | 791 | 252144.5796 | 0 | 0 | NA |
| 327 | 791 | 327 | 791 | 230112.09 0 | 0 | 0 | |
| 328 | 791 | 328 | 791 | 159776.0784 | 0 | 0 | NA |
| 329 | 791 | 329 | 791 | 146581.7796 | 0 | 0 | NA |
| 330 | 791 | 330 | 791 | 140040.6084 | 0 | 0 | NA |
| 331 | 791 | 331 | 791 | 174590.2656 | 0 | 0 | NA |
| 332 | 791 | 332 | 791 | 180098.3844 | 0 | 0 | NA |
| 335 | 791 | 335 | 791 | 172739.9844 | 0 | 0 | 0 |
| 336 | 791 | 336 | 791 | 311029.29 0 | 0 | 1 | |
| 337 | 791 | 337 | 791 | 424504.3716 | 0 | 0 | NA |
| 338 | 791 | 338 | 791 | 400764.9636 | 0 | 0 | 0 |
| 339 | 791 | 339 | 791 | 368400.4416 | 0 | 0 | NA |
| 340 | 791 | 340 | 791 | 228962.25 0 | 0 | NA | |
| 341 | 791 | 341 | 791 | 191371.2516 | 0 | 0 | 0 |
| 342 | 791 | 342 | 791 | 309491.9424 | 0 | 0 | NA |
| 343 | 791 | 343 | 791 | 308958.1056 | 0 | 0 | 1 |
| 344 | 791 | 344 | 791 | 239708.16 0 | 0 | NA | |
| 345 | 791 | 345 | 791 | 176248.8324 | 108 | 0 | NA |
| 346 | 791 | 346 | 791 | 218911.6944 | 6 | 0 | NA |
| 347 | 791 | 347 | 791 | 203527.2996 | 98 | 0 | NA |
| 348 | 791 | 348 | 791 | 215778.8304 | 105 | 0 | NA |
| 349 | 791 | 349 | 791 | 261018.81 0 | 0 | NA | |
| 350 | 791 | 350 | 791 | 166659.8976 | 0 | 0 | NA |
| 351 | 791 | 351 | 791 | 89676.2916 | 0 | 0 | NA |
| 352 | 791 | 352 | 791 | 64688.8356 | 0 | 0 | NA |
| 353 | 791 | 353 | 791 | 64171.0224 | 0 | 0 | NA |
| 354 | 791 | 354 | 791 | 68465.9556 | 0 | 0 | NA |
| 355 | 791 | 355 | 791 | 60959.61 0 | 0 | NA | |
| 356 | 791 | 356 | 791 | 51438.24 0 | 0 | NA | |
| 359 | 791 | 359 | 791 | 47698.56 0 | 0 | NA | |
| 360 | 791 | 360 | 791 | 31584.3984 | 0 | 0 | NA |
| 362 | 791 | 362 | 791 | 20857.1364 | 0 | 0 | NA |
| 363 | 791 | 363 | 791 | 19293.21 0 | 0 | 0 | |
| 315 | 790 | 315 | 790 | 160929.3456 | 0 | 0 | NA |
| 316 | 790 | 316 | 790 | 355621.3956 | 0 | 0 | NA |
| 317 | 790 | 317 | 790 | 407631.1716 | 0 | 0 | NA |
| 319 | 790 | 319 | 790 | 153836.5284 | 362 | 0 | NA |
| 320 | 790 | 320 | 790 | 187766.2224 | 400 | 0 | NA |
| 321 | 790 | 321 | 790 | 293027.3424 | 266 | 0 | 1 |
| 322 | 790 | 322 | 790 | 343067.9184 | 2 | 68 | NA |
| 323 | 790 | 323 | 790 | 311497.9344 | 0 | 21 | NA |
| 324 | 790 | 324 | 790 | 262000.6596 | 0 | 0 | NA |
| 325 | 790 | 325 | 790 | 220204.9476 | 0 | 95 | NA |
| 326 | 790 | 326 | 790 | 256258.6884 | 0 | 124 | NA |
| 327 | 790 | 327 | 790 | 244312.7184 | 0 | 0 | NA |
| 328 | 790 | 328 | 790 | 180149.3136 | 0 | 0 | 0 |
| 329 | 790 | 329 | 790 | 197349.1776 | 0 | 0 | NA |
| 330 | 790 | 330 | 790 | 169958.3076 | 0 | 0 | NA |
| 331 | 790 | 331 | 790 | 215834.5764 | 0 | 0 | NA |
| 332 | 790 | 332 | 790 | 150233.76 0 | 0 | NA | |
| 335 | 790 | 335 | 790 | 226499.8464 | 0 | 0 | NA |
| 336 | 790 | 336 | 790 | 381553.29 0 | 0 | 0 | |
| 337 | 790 | 337 | 790 | 379702.44 0 | 0 | NA | |
| 338 | 790 | 338 | 790 | 293612.2596 | 0 | 0 | NA |
| 339 | 790 | 339 | 790 | 261080.1216 | 0 | 0 | 0 |
| 340 | 790 | 340 | 790 | 183423.7584 | 0 | 0 | 0 |
| 341 | 790 | 341 | 790 | 227300.0976 | 0 | 0 | NA |
| 342 | 790 | 342 | 790 | 320129.64 0 | 0 | NA | |
| 343 | 790 | 343 | 790 | 265822.7364 | 0 | 0 | NA |
| 344 | 790 | 344 | 790 | 229479.3216 | 0 | 0 | NA |
| 345 | 790 | 345 | 790 | 299099.61 0 | 0 | NA | |
| 346 | 790 | 346 | 790 | 337398.3396 | 0 | 0 | NA |
| 347 | 790 | 347 | 790 | 204285.9204 | 0 | 0 | NA |
| 348 | 790 | 348 | 790 | 264895.5024 | 0 | 0 | NA |
| 349 | 790 | 349 | 790 | 277202.25 0 | 0 | NA | |
| 350 | 790 | 350 | 790 | 187194.6756 | 0 | 0 | 0 |
| 351 | 790 | 351 | 790 | 100933.29 0 | 0 | NA | |
| 352 | 790 | 352 | 790 | 96372.9936 | 0 | 0 | NA |
| 353 | 790 | 353 | 790 | 142611.9696 | 0 | 0 | NA |
| 354 | 790 | 354 | 790 | 143883.6624 | 0 | 14 | 0 |
| 355 | 790 | 355 | 790 | 184814.01 0 | 20 | NA | |
| 356 | 790 | 356 | 790 | 133254.2016 | 0 | 60 | NA |
| 357 | 790 | 357 | 790 | 186675.8436 | 0 | 0 | NA |
| 358 | 790 | 358 | 790 | 186935.1696 | 0 | 0 | NA |
| 359 | 790 | 359 | 790 | 127020.96 0 | 76 | NA | |
| 360 | 790 | 360 | 790 | 50625 0 | 0 | NA | |
| 315 | 789 | 315 | 789 | 159680.16 0 | 0 | 0 | |
| 316 | 789 | 316 | 789 | 303314.5476 | 0 | 0 | NA |
| 317 | 789 | 317 | 789 | 492635.5344 | 0 | 0 | NA |
| 318 | 789 | 318 | 789 | 335310.4836 | 141 | 0 | NA |
| 319 | 789 | 319 | 789 | 272484 193 | 44 | NA | |
| 320 | 789 | 320 | 789 | 294523.29 182 | 106 | NA | |
| 321 | 789 | 321 | 789 | 401448.96 47 | 120 | NA | |
| 322 | 789 | 322 | 789 | 568908.1476 | 0 | 119 | NA |
| 323 | 789 | 323 | 789 | 525161.1024 | 0 | 95 | NA |
| 324 | 789 | 324 | 789 | 482080.2624 | 0 | 3 | NA |
| 325 | 789 | 325 | 789 | 308558.0304 | 0 | 197 | NA |
| 326 | 789 | 326 | 789 | 289336.41 0 | 372 | NA | |
| 327 | 789 | 327 | 789 | 324968.4036 | 0 | 40 | 0 |
| 328 | 789 | 328 | 789 | 279417.96 0 | 0 | 0 | |
| 329 | 789 | 329 | 789 | 215333.1216 | 0 | 0 | NA |
| 330 | 789 | 330 | 789 | 207443.8116 | 0 | 0 | NA |
| 331 | 789 | 331 | 789 | 170899.56 0 | 0 | NA | |
| 333 | 789 | 333 | 789 | 206406.6624 | 0 | 0 | NA |
| 334 | 789 | 334 | 789 | 239766.9156 | 0 | 0 | NA |
| 335 | 789 | 335 | 789 | 305698.41 0 | 0 | NA | |
| 336 | 789 | 336 | 789 | 441400.7844 | 0 | 0 | NA |
| 337 | 789 | 337 | 789 | 359208.4356 | 0 | 0 | 0 |
| 338 | 789 | 338 | 789 | 313644.8016 | 0 | 0 | NA |
| 339 | 789 | 339 | 789 | 319518.8676 | 0 | 0 | NA |
| 340 | 789 | 340 | 789 | 282407.2164 | 0 | 0 | 0 |
| 341 | 789 | 341 | 789 | 386759.61 0 | 0 | NA | |
| 342 | 789 | 342 | 789 | 513486.8964 | 0 | 0 | NA |
| 343 | 789 | 343 | 789 | 328329 0 | 0 | NA | |
| 344 | 789 | 344 | 789 | 377192.5056 | 0 | 0 | NA |
| 345 | 789 | 345 | 789 | 449650.7136 | 0 | 0 | NA |
| 346 | 789 | 346 | 789 | 422084.1024 | 0 | 0 | 0 |
| 347 | 789 | 347 | 789 | 292962.3876 | 0 | 0 | NA |
| 348 | 789 | 348 | 789 | 331154.2116 | 0 | 0 | NA |
| 349 | 789 | 349 | 789 | 420526.3104 | 0 | 0 | NA |
| 350 | 789 | 350 | 789 | 271420.1604 | 0 | 0 | NA |
| 351 | 789 | 351 | 789 | 126252.3024 | 0 | 0 | NA |
| 352 | 789 | 352 | 789 | 179792.9604 | 0 | 0 | 0 |
| 353 | 789 | 353 | 789 | 286310.6064 | 0 | 40 | 0 |
| 354 | 789 | 354 | 789 | 261509.5044 | 0 | 259 | 0 |
| 355 | 789 | 355 | 789 | 277581.4596 | 0 | 222 | NA |
| 356 | 789 | 356 | 789 | 249380.3844 | 0 | 295 | NA |
| 357 | 789 | 357 | 789 | 285733.0116 | 0 | 13 | NA |
| 358 | 789 | 358 | 789 | 161941.8564 | 0 | 2 | NA |
| 359 | 789 | 359 | 789 | 90432.5184 | 0 | 46 | NA |
| 360 | 789 | 360 | 789 | 104122.3824 | 0 | 0 | NA |
| 361 | 789 | 361 | 789 | 96895.2384 | 0 | 0 | 0 |
| 362 | 789 | 362 | 789 | 59750.9136 | 0 | 0 | 0 |
| 363 | 789 | 363 | 789 | 84297.3156 | 0 | 0 | NA |
| 364 | 789 | 364 | 789 | 35359.0416 | 0 | 0 | NA |
| 315 | 788 | 315 | 788 | 206624.7936 | 0 | 0 | NA |
| 316 | 788 | 316 | 788 | 254016 0 | 0 | 1 | |
| 317 | 788 | 317 | 788 | 453117.4596 | 0 | 0 | NA |
| 318 | 788 | 318 | 788 | 347793.2676 | 20 | 0 | NA |
| 319 | 788 | 319 | 788 | 397656.36 0 | 172 | NA | |
| 320 | 788 | 320 | 788 | 367090.5744 | 0 | 254 | 1 |
| 321 | 788 | 321 | 788 | 443635.9236 | 0 | 147 | NA |
| 322 | 788 | 322 | 788 | 646995.0096 | 0 | 25 | 0 |
| 323 | 788 | 323 | 788 | 670662.7236 | 0 | 36 | NA |
| 324 | 788 | 324 | 788 | 678251.0736 | 0 | 4 | 0 |
| 325 | 788 | 325 | 788 | 449328.9024 | 0 | 24 | NA |
| 326 | 788 | 326 | 788 | 349422.8544 | 0 | 356 | 0 |
| 327 | 788 | 327 | 788 | 494490.24 0 | 142 | NA | |
| 328 | 788 | 328 | 788 | 541107.36 0 | 0 | NA | |
| 329 | 788 | 329 | 788 | 257718.6756 | 0 | 0 | 0 |
| 330 | 788 | 330 | 788 | 190689.4224 | 0 | 0 | NA |
| 331 | 788 | 331 | 788 | 166561.9344 | 0 | 0 | NA |
| 332 | 788 | 332 | 788 | 223199.5536 | 0 | 0 | NA |
| 333 | 788 | 333 | 788 | 321216.8976 | 0 | 0 | NA |
| 334 | 788 | 334 | 788 | 389151.3924 | 0 | 0 | NA |
| 335 | 788 | 335 | 788 | 385864.5924 | 0 | 0 | NA |
| 336 | 788 | 336 | 788 | 423410.49 0 | 0 | NA | |
| 337 | 788 | 337 | 788 | 412318.0944 | 0 | 0 | NA |
| 338 | 788 | 338 | 788 | 422630.01 0 | 0 | NA | |
| 339 | 788 | 339 | 788 | 484499.5236 | 0 | 0 | NA |
| 340 | 788 | 340 | 788 | 522382.0176 | 0 | 0 | NA |
| 341 | 788 | 341 | 788 | 523770.6384 | 0 | 0 | NA |
| 342 | 788 | 342 | 788 | 656488.8576 | 0 | 0 | NA |
| 343 | 788 | 343 | 788 | 485502.3684 | 0 | 0 | 0 |
| 344 | 788 | 344 | 788 | 506887.0416 | 0 | 0 | NA |
| 347 | 788 | 347 | 788 | 396522.09 0 | 0 | 0 | |
| 348 | 788 | 348 | 788 | 447882.1776 | 0 | 0 | NA |
| 349 | 788 | 349 | 788 | 479667.0564 | 0 | 1 | NA |
| 350 | 788 | 350 | 788 | 286053.8256 | 0 | 111 | 0 |
| 351 | 788 | 351 | 788 | 152193.6144 | 0 | 1 | NA |
| 352 | 788 | 352 | 788 | 267247.6416 | 0 | 0 | NA |
| 353 | 788 | 353 | 788 | 317081.61 0 | 78 | NA | |
| 354 | 788 | 354 | 788 | 255166.4196 | 0 | 384 | NA |
| 355 | 788 | 355 | 788 | 227872.5696 | 0 | 269 | NA |
| 356 | 788 | 356 | 788 | 182756.25 0 | 219 | NA | |
| 357 | 788 | 357 | 788 | 158435.8416 | 0 | 143 | NA |
| 358 | 788 | 358 | 788 | 154307.5524 | 0 | 291 | NA |
| 359 | 788 | 359 | 788 | 192106.89 0 | 160 | 0 | |
| 360 | 788 | 360 | 788 | 200990.8224 | 0 | 132 | NA |
| 361 | 788 | 361 | 788 | 144932.49 0 | 0 | NA | |
| 362 | 788 | 362 | 788 | 156388.6116 | 0 | 0 | 0 |
| 363 | 788 | 363 | 788 | 123791.3856 | 0 | 0 | NA |
| 315 | 787 | 315 | 787 | 291146.5764 | 0 | 0 | NA |
| 316 | 787 | 316 | 787 | 249020.9604 | 0 | 0 | NA |
| 317 | 787 | 317 | 787 | 474858.81 0 | 91 | NA | |
| 318 | 787 | 318 | 787 | 504668.16 0 | 0 | 0 | |
| 319 | 787 | 319 | 787 | 493478.1504 | 0 | 305 | NA |
| 320 | 787 | 320 | 787 | 466543.6416 | 0 | 275 | NA |
| 321 | 787 | 321 | 787 | 550504.6416 | 0 | 117 | NA |
| 322 | 787 | 322 | 787 | 639968.0004 | 0 | 112 | 0 |
| 323 | 787 | 323 | 787 | 820075.1364 | 0 | 8 | 0 |
| 324 | 787 | 324 | 787 | 795164.5584 | 0 | 0 | 0 |
| 325 | 787 | 325 | 787 | 610273.44 0 | 0 | 0 | |
| 326 | 787 | 326 | 787 | 471392.0964 | 0 | 160 | NA |
| 327 | 787 | 327 | 787 | 551662.7076 | 0 | 137 | NA |
| 328 | 787 | 328 | 787 | 500244.9984 | 0 | 0 | 0 |
| 329 | 787 | 329 | 787 | 251422.0164 | 0 | 0 | 0 |
| 330 | 787 | 330 | 787 | 168280.4484 | 0 | 0 | NA |
| 331 | 787 | 331 | 787 | 180098.3844 | 0 | 0 | 0 |
| 332 | 787 | 332 | 787 | 251602.56 | 0 | 0 | NA |
| 333 | 787 | 333 | 787 | 390125.16 | 0 | 0 | 0 |
| 334 | 787 | 334 | 787 | 412626.3696 | 0 | 0 | NA |
| 335 | 787 | 335 | 787 | 404954.0496 | 0 | 0 | NA |
| 336 | 787 | 336 | 787 | 416025 | 0 | 0 | NA |
| 337 | 787 | 337 | 787 | 405412.3584 | 0 | 0 | NA |
| 338 | 787 | 338 | 787 | 505776.9924 | 0 | 0 | NA |
| 339 | 787 | 339 | 787 | 530391.7584 | 0 | 0 | NA |
| 340 | 787 | 340 | 787 | 646319.5236 | 0 | 0 | NA |
| 341 | 787 | 341 | 787 | 611398.8864 | 0 | 0 | NA |
| 342 | 787 | 342 | 787 | 637474.4964 | 0 | 0 | NA |
| 349 | 787 | 349 | 787 | 441002.2464 | 0 | 0 | NA |
| 350 | 787 | 350 | 787 | 321557.0436 | 0 | 43 | NA |
| 351 | 787 | 351 | 787 | 295761.9456 | 0 | 58 | NA |
| 352 | 787 | 352 | 787 | 276003.1296 | 0 | 214 | NA |
| 353 | 787 | 353 | 787 | 271482.6816 | 0 | 274 | NA |
| 354 | 787 | 354 | 787 | 236837.9556 | 0 | 224 | NA |
| 355 | 787 | 355 | 787 | 245976.3216 | 0 | 281 | NA |
| 356 | 787 | 356 | 787 | 185175.3024 | 0 | 391 | NA |
| 357 | 787 | 357 | 787 | 241297.0884 | 0 | 296 | 0 |
| 358 | 787 | 358 | 787 | 281133.2484 | 0 | 378 | NA |
| 359 | 787 | 359 | 787 | 268987.4496 | 0 | 394 | NA |
| 360 | 787 | 360 | 787 | 265142.6064 | 0 | 400 | NA |
| 361 | 787 | 361 | 787 | 231264.81 | 0 | 62 | 0 |
| 362 | 787 | 362 | 787 | 119522.3184 | 0 | 0 | 0 |
| 363 | 787 | 363 | 787 | 52780.4676 | 0 | 0 | NA |
| 364 | 787 | 364 | 787 | 37558.44 | 0 | 0 | 0 |
| 315 | 786 | 315 | 786 | 396144.36 | 0 | 0 | NA |
| 316 | 786 | 316 | 786 | 260100 | 0 | 0 | 0 |
| 317 | 786 | 317 | 786 | 393129 | 0 | 56 | NA |
| 318 | 786 | 318 | 786 | 551662.7076 | 0 | 105 | NA |
| 319 | 786 | 319 | 786 | 524118.0816 | 0 | 336 | NA |
| 320 | 786 | 320 | 786 | 643236.0804 | 0 | 18 | 0 |
| 321 | 786 | 321 | 786 | 842393.5524 | 0 | 0 | NA |
| 322 | 786 | 322 | 786 | 838323.36 | 0 | 0 | NA |
| 323 | 786 | 323 | 786 | 913438.9476 | 0 | 0 | NA |
| 324 | 786 | 324 | 786 | 1045056.3984 | 0 | 0 | NA |
| 325 | 786 | 325 | 786 | 886008.0384 | 0 | 0 | 0 |
| 326 | 786 | 326 | 786 | 645162.3684 | 0 | 0 | NA |
| 327 | 786 | 327 | 786 | 453521.4336 | 0 | 0 | NA |
| 328 | 786 | 328 | 786 | 331914.2544 | 0 | 0 | 0 |
| 329 | 786 | 329 | 786 | 208319.2164 | 0 | 0 | NA |
| 330 | 786 | 330 | 786 | 160784.9604 | 0 | 0 | NA |
| 331 | 786 | 331 | 786 | 231322.5216 | 0 | 0 | NA |
| 332 | 786 | 332 | 786 | 292832.4996 | 0 | 0 | NA |
| 333 | 786 | 333 | 786 | 429994.9476 | 0 | 0 | NA |
| 334 | 786 | 334 | 786 | 475603.3296 | 0 | 0 | NA |
| 350 | 786 | 350 | 786 | 353049.8724 | 0 | 9 | 0 |
| 351 | 786 | 351 | 786 | 369712.6416 | 0 | 20 | NA |
| 352 | 786 | 352 | 786 | 320808.96 | 0 | 84 | 0 |
| 353 | 786 | 353 | 786 | 265328.01 | 0 | 160 | NA |
| 354 | 786 | 354 | 786 | 288626.8176 | 0 | 289 | NA |
| 355 | 786 | 355 | 786 | 327161.1204 | 0 | 134 | NA |
| 356 | 786 | 356 | 786 | 270795.3444 | 0 | 320 | NA |
| 357 | 786 | 357 | 786 | 285604.7364 | 0 | 178 | NA |
| 358 | 786 | 358 | 786 | 292767.5664 | 0 | 359 | 0 |
| 359 | 786 | 359 | 786 | 294067.5984 | 0 | 400 | NA |
| 360 | 786 | 360 | 786 | 258938.4996 | 0 | 396 | NA |
| 361 | 786 | 361 | 786 | 171694.2096 | 0 | 183 | NA |
| 362 | 786 | 362 | 786 | 110370.1284 | 0 | 0 | NA |
| 363 | 786 | 363 | 786 | 57168.81 | 0 | 0 | NA |
| 364 | 786 | 364 | 786 | 42131.6676 | 0 | 0 | NA |
| 314 | 785 | 314 | 785 | 225074.3364 | 0 | 0 | NA |
| 315 | 785 | 315 | 785 | 508597.1856 | 0 | 0 | NA |
| 316 | 785 | 316 | 785 | 362018.8224 | 0 | 0 | NA |
| 317 | 785 | 317 | 785 | 304704 | 0 | 8 | NA |
| 318 | 785 | 318 | 785 | 415638.09 | 0 | 63 | NA |
| 319 | 785 | 319 | 785 | 569904.2064 | 0 | 112 | NA |
| 320 | 785 | 320 | 785 | 909772.5924 | 0 | 0 | NA |
| 321 | 785 | 321 | 785 | 943967.6964 | 0 | 0 | NA |
| 322 | 785 | 322 | 785 | 1090562.49 | 0 | 0 | NA |
| 323 | 785 | 323 | 785 | 1044933.7284 | 0 | 0 | NA |
| 324 | 785 | 324 | 785 | 988791.5844 | 0 | 0 | NA |
| 325 | 785 | 325 | 785 | 889286.7204 | 0 | 0 | NA |
| 326 | 785 | 326 | 785 | 704995.3296 | 0 | 0 | NA |
| 327 | 785 | 327 | 785 | 482913.8064 | 0 | 0 | 0 |
| 328 | 785 | 328 | 785 | 419126.76 0 | 0 | 0 | |
| 329 | 785 | 329 | 785 | 195169.5684 | 0 | 0 | NA |
| 330 | 785 | 330 | 785 | 197349.1776 | 0 | 0 | NA |
| 331 | 785 | 331 | 785 | 320944.9104 | 0 | 0 | NA |
| 332 | 785 | 332 | 785 | 378003.6324 | 0 | 0 | NA |
| 333 | 785 | 333 | 785 | 448283.8116 | 0 | 0 | 0 |
| 354 | 785 | 354 | 785 | 459684 0 | 37 | NA | |
| 355 | 785 | 355 | 785 | 427872.9744 | 0 | 197 | NA |
| 356 | 785 | 356 | 785 | 311230.0944 | 0 | 353 | NA |
| 357 | 785 | 357 | 785 | 205970.7456 | 0 | 185 | NA |
| 358 | 785 | 358 | 785 | 169661.61 0 | 28 | NA | |
| 359 | 785 | 359 | 785 | 176148.09 0 | 108 | NA | |
| 360 | 785 | 360 | 785 | 175292.9424 | 0 | 336 | NA |
| 361 | 785 | 361 | 785 | 155867.04 0 | 208 | NA | |
| 362 | 785 | 362 | 785 | 97044.7104 | 0 | 0 | NA |
| 363 | 785 | 363 | 785 | 73777.4244 | 0 | 0 | NA |
| 364 | 785 | 364 | 785 | 51574.41 0 | 0 | NA | |
| 314 | 784 | 314 | 784 | 314855.6544 | 0 | 0 | NA |
| 315 | 784 | 315 | 784 | 438296.9616 | 0 | 18 | NA |
| 316 | 784 | 316 | 784 | 439091.7696 | 0 | 171 | NA |
| 317 | 784 | 317 | 784 | 365057.64 0 | 124 | 1 | |
| 318 | 784 | 318 | 784 | 309558.7044 | 0 | 39 | NA |
| 319 | 784 | 319 | 784 | 490364.0676 | 0 | 23 | 0 |
| 320 | 784 | 320 | 784 | 841842.9504 | 0 | 12 | NA |
| 321 | 784 | 321 | 784 | 846363.2004 | 0 | 0 | NA |
| 322 | 784 | 322 | 784 | 781738.9056 | 0 | 0 | 0 |
| 323 | 784 | 323 | 784 | 820944.7236 | 0 | 0 | NA |
| 324 | 784 | 324 | 784 | 812377.7424 | 0 | 0 | NA |
| 325 | 784 | 325 | 784 | 714802.6116 | 0 | 0 | NA |
| 326 | 784 | 326 | 784 | 567551.2896 | 0 | 0 | NA |
| 327 | 784 | 327 | 784 | 451987.29 0 | 0 | NA | |
| 328 | 784 | 328 | 784 | 276949.5876 | 0 | 0 | NA |
| 329 | 784 | 329 | 784 | 165583.8864 | 0 | 0 | NA |
| 330 | 784 | 330 | 784 | 295174.89 0 | 0 | NA | |
| 331 | 784 | 331 | 784 | 381924 0 | 0 | 1 | |
| 332 | 784 | 332 | 784 | 390575.0016 | 0 | 0 | NA |
| 355 | 784 | 355 | 784 | 348855.6096 | 0 | 243 | 0 |
| 356 | 784 | 356 | 784 | 225758.0196 | 0 | 324 | NA |
| 357 | 784 | 357 | 784 | 149073.21 0 | 20 | NA | |
| 358 | 784 | 358 | 784 | 121647.4884 | 0 | 12 | NA |
| 359 | 784 | 359 | 784 | 115613.6004 | 0 | 22 | NA |
| 360 | 784 | 360 | 784 | 140175.36 0 | 194 | NA | |
| 361 | 784 | 361 | 784 | 93709.4544 | 0 | 73 | 1 |
| 362 | 784 | 362 | 784 | 75548.0196 | 0 | 0 | NA |
| 363 | 784 | 363 | 784 | 69949.6704 | 0 | 78 | NA |
| 364 | 784 | 364 | 784 | 71385.1524 | 0 | 4 | NA |
| 314 | 783 | 314 | 783 | 376308.6336 | 0 | 0 | 0 |
| 315 | 783 | 315 | 783 | 510996.2256 | 0 | 19 | NA |
| 316 | 783 | 316 | 783 | 497702.0304 | 0 | 258 | NA |
| 317 | 783 | 317 | 783 | 538638.5664 | 0 | 350 | 1 |
| 318 | 783 | 318 | 783 | 483330.8484 | 0 | 77 | NA |
| 319 | 783 | 319 | 783 | 434175.5664 | 0 | 18 | NA |
| 320 | 783 | 320 | 783 | 766535.2704 | 0 | 79 | NA |
| 321 | 783 | 321 | 783 | 912521.6676 | 0 | 0 | 0 |
| 322 | 783 | 322 | 783 | 828573.2676 | 0 | 0 | NA |
| 323 | 783 | 323 | 783 | 531528.4836 | 0 | 0 | NA |
| 324 | 783 | 324 | 783 | 483164.01 0 | 0 | NA | |
| 325 | 783 | 325 | 783 | 398261.9664 | 0 | 0 | NA |
| 326 | 783 | 326 | 783 | 316068.84 0 | 0 | NA | |
| 327 | 783 | 327 | 783 | 199540.89 0 | 0 | NA | |
| 328 | 783 | 328 | 783 | 176349.6036 | 0 | 0 | NA |
| 329 | 783 | 329 | 783 | 311832.8964 | 0 | 0 | 0 |
| 330 | 783 | 330 | 783 | 441639.9936 | 0 | 0 | NA |
| 331 | 783 | 331 | 783 | 419826.2436 | 0 | 0 | 0 |
| 332 | 783 | 332 | 783 | 420370.6896 | 0 | 0 | 0 |
| 356 | 783 | 356 | 783 | 166855.9104 | 0 | 303 | 0 |
| 357 | 783 | 357 | 783 | 135762.7716 | 0 | 59 | NA |
| 358 | 783 | 358 | 783 | 154826.5104 | 0 | 96 | NA |
| 359 | 783 | 359 | 783 | 130812.4224 | 0 | 203 | NA |
| 360 | 783 | 360 | 783 | 89712.2304 | 0 | 30 | NA |
| 361 | 783 | 361 | 783 | 91518.3504 | 0 | 0 | NA |
| 362 | 783 | 362 | 783 | 160112.0196 | 0 | 5 | NA |
| 363 | 783 | 363 | 783 | 151305.4404 | 0 | 395 | NA |
| 364 | 783 | 364 | 783 | 128049.4656 | 0 | 92 | NA |
| 314 | 782 | 314 | 782 | 366727.1364 | 0 | 0 | NA |
| 315 | 782 | 315 | 782 | 591391.7604 | 0 | 0 | NA |
| 316 | 782 | 316 | 782 | 593793.5364 | 0 | 44 | NA |
| 317 | 782 | 317 | 782 | 729486.81 0 | 273 | 0 | |
| 318 | 782 | 318 | 782 | 682110.81 0 | 65 | NA | |
| 319 | 782 | 319 | 782 | 441799.5024 | 0 | 33 | NA |
| 320 | 782 | 320 | 782 | 554876.01 0 | 1 | NA | |
| 321 | 782 | 321 | 782 | 846804.8484 | 0 | 0 | NA |
| 322 | 782 | 322 | 782 | 850010.2416 | 0 | 0 | NA |
| 323 | 782 | 323 | 782 | 822794.1264 | 0 | 0 | 0 |
| 324 | 782 | 324 | 782 | 650861.6976 | 0 | 0 | NA |
| 325 | 782 | 325 | 782 | 450777.96 0 | 0 | NA | |
| 326 | 782 | 326 | 782 | 366727.1364 | 0 | 0 | NA |
| 327 | 782 | 327 | 782 | 332813.61 0 | 0 | 1 | |
| 328 | 782 | 328 | 782 | 366872.49 0 | 0 | NA | |
| 329 | 782 | 329 | 782 | 401220.8964 | 0 | 0 | 0 |
| 330 | 782 | 330 | 782 | 460090.89 0 | 0 | NA | |
| 331 | 782 | 331 | 782 | 445075.7796 | 0 | 0 | NA |
| 332 | 782 | 332 | 782 | 517191.1056 | 0 | 0 | NA |
| 356 | 782 | 356 | 782 | 192738.5604 | 0 | 148 | NA |
| 357 | 782 | 357 | 782 | 189277.2036 | 0 | 313 | NA |
| 358 | 782 | 358 | 782 | 184350.0096 | 0 | 383 | NA |
| 359 | 782 | 359 | 782 | 124976.3904 | 0 | 161 | 0 |
| 360 | 782 | 360 | 782 | 120395.1204 | 0 | 0 | NA |
| 361 | 782 | 361 | 782 | 110489.76 0 | 3 | NA | |
| 362 | 782 | 362 | 782 | 177106.3056 | 0 | 148 | NA |
| 363 | 782 | 363 | 782 | 185330.25 0 | 400 | NA | |
| 364 | 782 | 364 | 782 | 157006.1376 | 0 | 338 | NA |
| 314 | 781 | 314 | 781 | 362813.4756 | 0 | 4 | NA |
| 315 | 781 | 315 | 781 | 401905.2816 | 0 | 0 | NA |
| 316 | 781 | 316 | 781 | 536790.6756 | 0 | 0 | NA |
| 317 | 781 | 317 | 781 | 912521.6676 | 0 | 0 | NA |
| 318 | 781 | 318 | 781 | 825299.5716 | 0 | 0 | NA |
| 319 | 781 | 319 | 781 | 770638.1796 | 0 | 234 | NA |
| 320 | 781 | 320 | 781 | 583665.4404 | 0 | 83 | NA |
| 321 | 781 | 321 | 781 | 744838.0416 | 0 | 0 | 0 |
| 325 | 781 | 325 | 781 | 467773.9236 | 0 | 0 | NA |
| 326 | 781 | 326 | 781 | 500839.29 0 | 0 | NA | |
| 327 | 781 | 327 | 781 | 425364.84 0 | 0 | 1 | |
| 328 | 781 | 328 | 781 | 424113.5376 | 0 | 0 | NA |
| 329 | 781 | 329 | 781 | 452875.1616 | 0 | 0 | NA |
| 330 | 781 | 330 | 781 | 501348.9636 | 0 | 0 | NA |
| 331 | 781 | 331 | 781 | 481330.6884 | 0 | 0 | NA |
| 357 | 781 | 357 | 781 | 168773.0724 | 0 | 10 | NA |
| 358 | 781 | 358 | 781 | 194639.7924 | 0 | 279 | NA |
| 359 | 781 | 359 | 781 | 164511.36 0 | 360 | NA | |
| 360 | 781 | 360 | 781 | 146719.6416 | 0 | 150 | NA |
| 361 | 781 | 361 | 781 | 125996.6016 | 0 | 280 | NA |
| 362 | 781 | 362 | 781 | 167739.3936 | 0 | 400 | NA |
| 363 | 781 | 363 | 781 | 201313.7424 | 0 | 348 | NA |
| 364 | 781 | 364 | 781 | 166904.9316 | 0 | 212 | NA |
| 314 | 780 | 314 | 780 | 370515.69 0 | 0 | NA | |
| 315 | 780 | 315 | 780 | 378741.7764 | 0 | 0 | NA |
| 316 | 780 | 316 | 780 | 491288.8464 | 0 | 0 | NA |
| 317 | 780 | 317 | 780 | 732359.4084 | 0 | 0 | NA |
| 318 | 780 | 318 | 780 | 1012559.1876 | 0 | 0 | NA |
| 319 | 780 | 319 | 780 | 893818.9764 | 0 | 19 | NA |
| 320 | 780 | 320 | 780 | 779618.3616 | 0 | 10 | NA |
| 321 | 780 | 321 | 780 | 733900.6224 | 0 | 0 | NA |
| 327 | 780 | 327 | 780 | 235147.4064 | 0 | 0 | NA |
| 328 | 780 | 328 | 780 | 377266.2084 | 0 | 0 | NA |
| 329 | 780 | 329 | 780 | 461394.1476 | 0 | 0 | NA |
| 330 | 780 | 330 | 780 | 550771.7796 | 0 | 0 | NA |
| 331 | 780 | 331 | 780 | 496348.4304 | 0 | 0 | NA |
| 358 | 780 | 358 | 780 | 207826.5744 | 0 | 131 | NA |
| 359 | 780 | 359 | 780 | 188651.2356 | 0 | 400 | NA |
| 360 | 780 | 360 | 780 | 149073.21 0 | 400 | NA | |
| 361 | 780 | 361 | 780 | 208209.69 0 | 400 | NA | |
| 362 | 780 | 362 | 780 | 213554.8944 | 0 | 230 | NA |
| 363 | 780 | 363 | 780 | 175997.0304 | 0 | 12 | 0 |
| 364 | 780 | 364 | 780 | 120228.6276 | 0 | 0 | 0 |
| 315 | 779 | 315 | 779 | 515294.2656 | 0 | 0 | NA |
| 316 | 779 | 316 | 779 | 646416 0 | 0 | NA | |

| | | | | | | | | |
|---|---|---|---|---|---|---|---|---|
| 317 | 779 | 317 | 779 | 773696.16 | 0 | 0 | 0 | |
| 318 | 779 | 318 | 779 | 970146.2016 | | 0 | 0 | NA |
| 331 | 779 | 331 | 779 | 573988.0644 | | 0 | 0 | NA |
| 332 | 779 | 332 | 779 | 568998.6624 | | 0 | 0 | NA |
| 359 | 779 | 359 | 779 | 225644.0004 | | 0 | 373 | NA |
| 360 | 779 | 360 | 779 | 216727.4916 | | 0 | 386 | NA |
| 361 | 779 | 361 | 779 | 233926.9956 | | 0 | 285 | NA |
| 362 | 779 | 362 | 779 | 143474.2884 | | 0 | 3 | NA |
| 363 | 779 | 363 | 779 | 100704.6756 | | 0 | 0 | 0 |
| 364 | 779 | 364 | 779 | 82702.2564 | | 0 | 0 | NA |
| 365 | 779 | 365 | 779 | 54363.5856 | | 0 | 0 | 0 |
| 314 | 778 | 314 | 778 | 523944.3456 | | 0 | 0 | 0 |
| 315 | 778 | 315 | 778 | 672826.4676 | | 0 | 0 | 0 |
| 316 | 778 | 316 | 778 | 828573.2676 | | 0 | 0 | NA |
| 317 | 778 | 317 | 778 | 800201.8116 | | 0 | 0 | NA |
| 359 | 778 | 359 | 778 | 207771.8724 | | 0 | 50 | NA |
| 360 | 778 | 360 | 778 | 206897.6196 | | 0 | 77 | 0 |
| 361 | 778 | 361 | 778 | 161941.8564 | | 0 | 34 | NA |
| 362 | 778 | 362 | 778 | 97269.1344 | | 0 | 0 | NA |
| 363 | 778 | 363 | 778 | 78142.6116 | | 0 | 0 | NA |
| 364 | 778 | 364 | 778 | 53333.2836 | | 0 | 0 | NA |
| 365 | 778 | 365 | 778 | 41812.0704 | | 0 | 0 | NA |
| 366 | 778 | 366 | 778 | 22626.1764 | | 0 | 0 | 0 |
| 316 | 777 | 316 | 777 | 1024913.2644 | | 0 | 0 | 0 |
| 359 | 777 | 359 | 777 | 99300.6144 | | 0 | 0 | NA |
| 360 | 777 | 360 | 777 | 108227.8404 | | 0 | 0 | NA |
| 361 | 777 | 361 | 777 | 89712.2304 | | 0 | 0 | NA |
| 362 | 777 | 362 | 777 | 59049 | 0 | 0 | NA | |
| 363 | 777 | 363 | 777 | 50113.2996 | | 0 | 0 | NA |
| 364 | 777 | 364 | 777 | 28425.96 | 0 | 0 | NA | |
| 365 | 777 | 365 | 777 | 23097.9204 | | 0 | 0 | NA |
| 366 | 777 | 366 | 777 | 16998.9444 | | 0 | 0 | NA |
| 360 | 776 | 360 | 776 | 84715.9236 | | 0 | 0 | 0 |
| 361 | 776 | 361 | 776 | 58032.81 | 0 | 0 | NA | |
| 362 | 776 | 362 | 776 | 33160.41 | 0 | 0 | NA | |
| 363 | 776 | 363 | 776 | 20839.8096 | | 0 | 0 | NA |
| 364 | 776 | 364 | 776 | 15055.29 | 0 | 0 | NA | |
| 365 | 776 | 365 | 776 | 13298.7024 | | 0 | 0 | 0 |
| 360 | 775 | 360 | 775 | 50005.9044 | | 0 | 0 | 1 |
| 361 | 775 | 361 | 775 | 29652.84 | 0 | 0 | NA | |
| 362 | 775 | 362 | 775 | 17838.2736 | | 0 | 0 | NA |
| 363 | 775 | 363 | 775 | 11702.9124 | | 0 | 0 | NA |
| 364 | 775 | 364 | 775 | 8805.9456 | 0 | 0 | 0 | |
| 365 | 775 | 365 | 775 | 7475.3316 | 0 | 0 | NA | |
| 360 | 774 | 360 | 774 | 20632.4496 | | 0 | 0 | NA |
| 361 | 774 | 361 | 774 | 13312.5444 | | 0 | 0 | 0 |
| 362 | 774 | 362 | 774 | 11689.9344 | | 0 | 0 | NA |
| 363 | 774 | 363 | 774 | 10637.8596 | | 0 | 0 | 0 |
| 364 | 774 | 364 | 774 | 7066.0836 | 0 | 0 | NA | |
| 365 | 774 | 365 | 774 | 5340.6864 | 0 | 0 | NA | |
| 359 | 773 | 359 | 773 | 10004.0004 | | 0 | 0 | NA |
| 360 | 773 | 360 | 773 | 10172.7396 | | 0 | 0 | NA |
| 361 | 773 | 361 | 773 | 10811.8404 | | 0 | 0 | 0 |
| 362 | 773 | 362 | 773 | 7299.9936 | 0 | 0 | NA | |
| 359 | 772 | 359 | 772 | 8328.3876 | 0 | 0 | NA | |
| 360 | 772 | 360 | 772 | 7371.9396 | 0 | 0 | 0 | |
| 361 | 772 | 361 | 772 | 6493.1364 | 0 | 0 | NA | |
| 362 | 772 | 362 | 772 | 4483.6416 | 0 | 0 | 0 | |
| 364 | 772 | 364 | 772 | 3694.2084 | 0 | 0 | NA | |
| 360 | 771 | 360 | 771 | 4129.3476 | 0 | 0 | NA | |
| 361 | 771 | 361 | 771 | 4621.2804 | 0 | 0 | 0 | |
| 350 | 797 | 350 | 797 | 58786.8516 | | 0 | 0 | NA |
| 356 | 797 | 356 | 797 | 30653.0064 | | 0 | 0 | NA |
| 352 | 796 | 352 | 796 | 89030.6244 | | 0 | 0 | 0 |
| 355 | 796 | 355 | 796 | 59750.9136 | | 0 | 0 | NA |
| 356 | 796 | 356 | 796 | 35675.6544 | | 0 | 0 | NA |
| 355 | 795 | 355 | 795 | 67215.7476 | | 0 | 0 | NA |
| 361 | 791 | 361 | 791 | 31456.5696 | | 0 | 0 | NA |
| 362 | 790 | 362 | 790 | 30758.1444 | | 0 | 0 | NA |
| 363 | 790 | 363 | 790 | 25606.4004 | | 0 | 0 | NA |
| 366 | 776 | 366 | 776 | 9285.2496 | 0 | 0 | NA | |
| 363 | 773 | 363 | 773 | 4686.7716 | 0 | 0 | NA | |
| 364 | 773 | 364 | 773 | 4129.3476 | 0 | 0 | 0 | |
| 363 | 772 | 363 | 772 | 3464.4996 | 0 | 0 | 0 | |
| 321 | 812 | 321 | 812 | 344615.9616 | | 0 | 0 | 0 |

| | | | | | | | |
|---|---|---|---|---|---|---|---|
| 323 | 812 | 323 | 812 | 489356.2116 | 0 | 0 | NA |
| 324 | 812 | 324 | 812 | 307026.81 0 | 0 | NA | |
| 320 | 811 | 320 | 811 | 300808.3716 | 0 | 0 | NA |
| 321 | 811 | 321 | 811 | 274555.0404 | 0 | 0 | NA |
| 322 | 811 | 322 | 811 | 298574.8164 | 0 | 0 | NA |
| 323 | 811 | 323 | 811 | 408857.9364 | 0 | 0 | NA |
| 324 | 811 | 324 | 811 | 265142.6064 | 0 | 0 | 0 |
| 319 | 810 | 319 | 810 | 219361.0896 | 0 | 0 | NA |
| 320 | 810 | 320 | 810 | 239766.9156 | 0 | 0 | NA |
| 321 | 810 | 321 | 810 | 248961.0816 | 0 | 0 | NA |
| 322 | 810 | 322 | 810 | 217566.2736 | 0 | 0 | NA |
| 323 | 810 | 323 | 810 | 209031.84 0 | 0 | 1 | |
| 324 | 810 | 324 | 810 | 203527.2996 | 0 | 12 | NA |
| 318 | 809 | 318 | 809 | 235554.9156 | 0 | 0 | 0 |
| 319 | 809 | 319 | 809 | 245143.8144 | 0 | 0 | NA |
| 320 | 809 | 320 | 809 | 235846.2096 | 0 | 0 | NA |
| 321 | 809 | 321 | 809 | 209965.5684 | 0 | 0 | NA |
| 322 | 809 | 322 | 809 | 217174.6404 | 0 | 0 | 0 |
| 323 | 809 | 323 | 809 | 262861.29 0 | 0 | NA | |
| 324 | 809 | 324 | 809 | 240296.04 0 | 0 | NA | |
| 325 | 809 | 325 | 809 | 203310.81 0 | 0 | 0 | |
| 319 | 808 | 319 | 808 | 338863.6944 | 0 | 0 | NA |
| 320 | 808 | 320 | 808 | 395163.1044 | 0 | 58 | NA |
| 321 | 808 | 321 | 808 | 277265.4336 | 0 | 0 | NA |
| 322 | 808 | 322 | 808 | 274932.4356 | 0 | 0 | 0 |
| 323 | 808 | 323 | 808 | 396144.36 0 | 47 | NA | |
| 324 | 808 | 324 | 808 | 345885.1344 | 0 | 0 | NA |
| 325 | 808 | 325 | 808 | 223256.25 0 | 0 | NA | |
| 319 | 807 | 319 | 807 | 422786.0484 | 0 | 0 | NA |
| 320 | 807 | 320 | 807 | 387805.1076 | 0 | 9 | NA |
| 321 | 807 | 321 | 807 | 327847.8564 | 0 | 0 | NA |
| 322 | 807 | 322 | 807 | 382443.2964 | 0 | 0 | NA |
| 323 | 807 | 323 | 807 | 419049.0756 | 0 | 0 | 0 |
| 324 | 807 | 324 | 807 | 360576.2304 | 0 | 0 | NA |
| 325 | 807 | 325 | 807 | 256744.89 0 | 0 | 0 | |
| 326 | 807 | 326 | 807 | 317216.7684 | 0 | 0 | NA |
| 327 | 807 | 327 | 807 | 281769.8724 | 0 | 0 | NA |
| 328 | 807 | 328 | 807 | 202446.0036 | 0 | 0 | NA |
| 319 | 806 | 319 | 806 | 452471.4756 | 0 | 0 | NA |
| 320 | 806 | 320 | 806 | 498210.1056 | 0 | 0 | NA |
| 321 | 806 | 321 | 806 | 449409.3444 | 0 | 0 | 0 |
| 322 | 806 | 322 | 806 | 460742.2884 | 0 | 0 | 1 |
| 323 | 806 | 323 | 806 | 533542.5936 | 0 | 0 | NA |
| 324 | 806 | 324 | 806 | 432358.8516 | 0 | 0 | NA |
| 325 | 806 | 325 | 806 | 351767.61 0 | 0 | NA | |
| 326 | 806 | 326 | 806 | 408857.9364 | 0 | 0 | NA |
| 327 | 806 | 327 | 806 | 394710.6276 | 0 | 0 | NA |
| 328 | 806 | 328 | 806 | 320401.2816 | 0 | 0 | NA |
| 329 | 806 | 329 | 806 | 251662.7556 | 0 | 0 | NA |
| 320 | 805 | 320 | 805 | 504838.6704 | 0 | 0 | 0 |
| 321 | 805 | 321 | 805 | 630213.6996 | 0 | 0 | NA |
| 322 | 805 | 322 | 805 | 637570.3104 | 0 | 0 | NA |
| 323 | 805 | 323 | 805 | 562500 0 | 0 | NA | |
| 324 | 805 | 324 | 805 | 423410.49 0 | 0 | 0 | |
| 325 | 805 | 325 | 805 | 345250.2564 | 0 | 0 | NA |
| 326 | 805 | 326 | 805 | 365492.7936 | 0 | 0 | NA |
| 327 | 805 | 327 | 805 | 389825.4096 | 0 | 0 | NA |
| 328 | 805 | 328 | 805 | 388328.3856 | 0 | 22 | NA |
| 329 | 805 | 329 | 805 | 389675.5776 | 0 | 40 | NA |
| 320 | 804 | 320 | 804 | 383260.0464 | 0 | 0 | NA |
| 321 | 804 | 321 | 804 | 464987.61 0 | 0 | NA | |
| 322 | 804 | 322 | 804 | 462209.6196 | 0 | 0 | NA |
| 323 | 804 | 323 | 804 | 338514.5124 | 0 | 0 | NA |
| 324 | 804 | 324 | 804 | 269361 0 | 0 | NA | |
| 325 | 804 | 325 | 804 | 256866.5124 | 0 | 82 | NA |
| 326 | 804 | 326 | 804 | 247287.3984 | 0 | 77 | NA |
| 327 | 804 | 327 | 804 | 249080.8464 | 0 | 120 | NA |
| 328 | 804 | 328 | 804 | 334199.61 0 | 124 | 0 | |
| 329 | 804 | 329 | 804 | 399777.9984 | 0 | 1 | 0 |
| 330 | 804 | 330 | 804 | 266441.7924 | 0 | 0 | NA |
| 319 | 803 | 319 | 803 | 240178.4064 | 0 | 0 | 1 |
| 320 | 803 | 320 | 803 | 254318.49 0 | 0 | NA | |
| 321 | 803 | 321 | 803 | 259488.36 0 | 0 | NA | |
| 322 | 803 | 322 | 803 | 251662.7556 | 0 | 0 | NA |
| 323 | 803 | 323 | 803 | 252204.84 0 | 0 | NA | |
| 324 | 803 | 324 | 803 | 237831.7824 | 0 | 0 | NA |
| 325 | 803 | 325 | 803 | 206134.1604 | 0 | 81 | NA |
| 326 | 803 | 326 | 803 | 182294.8416 | 0 | 7 | NA |
| 327 | 803 | 327 | 803 | 207334.5156 | 0 | 37 | 0 |
| 328 | 803 | 328 | 803 | 335518.9776 | 0 | 0 | 0 |
| 329 | 803 | 329 | 803 | 238944.9924 | 0 | 62 | NA |
| 330 | 803 | 330 | 803 | 228732.6276 | 0 | 0 | NA |
| 317 | 802 | 317 | 802 | 382665.96 0 | 0 | 0 | |
| 318 | 802 | 318 | 802 | 278910.7344 | 0 | 0 | NA |
| 319 | 802 | 319 | 802 | 297199.4256 | 0 | 0 | 0 |
| 320 | 802 | 320 | 802 | 235904.49 0 | 0 | 0 | |
| 321 | 802 | 321 | 802 | 242241.1524 | 0 | 0 | NA |
| 322 | 802 | 322 | 802 | 242891.2656 | 0 | 0 | 0 |
| 323 | 802 | 323 | 802 | 211839.2676 | 0 | 22 | 1 |
| 324 | 802 | 324 | 802 | 196656.7716 | 0 | 0 | 1 |
| 325 | 802 | 325 | 802 | 169661.61 0 | 23 | 0 | |
| 326 | 802 | 326 | 802 | 162909.1044 | 0 | 67 | NA |
| 327 | 802 | 327 | 802 | 162231.7284 | 0 | 0 | NA |
| 328 | 802 | 328 | 802 | 207225.2484 | 0 | 0 | NA |
| 329 | 802 | 329 | 802 | 171445.6836 | 0 | 42 | NA |
| 330 | 802 | 330 | 802 | 154119.0564 | 0 | 0 | NA |
| 313 | 801 | 313 | 801 | 999800.01 0 | 0 | NA | |
| 314 | 801 | 314 | 801 | 813026.8224 | 0 | 0 | NA |
| 315 | 801 | 315 | 801 | 581375.7504 | 0 | 0 | NA |
| 316 | 801 | 316 | 801 | 400840.9344 | 0 | 0 | NA |
| 317 | 801 | 317 | 801 | 300150.5796 | 0 | 0 | NA |
| 318 | 801 | 318 | 801 | 375352.2756 | 0 | 0 | 0 |
| 319 | 801 | 319 | 801 | 426461.2416 | 0 | 0 | NA |
| 320 | 801 | 320 | 801 | 334338.3684 | 0 | 0 | NA |
| 321 | 801 | 321 | 801 | 334407.7584 | 0 | 0 | 0 |
| 322 | 801 | 322 | 801 | 341313.0084 | 0 | 119 | 0 |
| 323 | 801 | 323 | 801 | 224221.1904 | 0 | 223 | NA |
| 324 | 801 | 324 | 801 | 204991.6176 | 0 | 0 | NA |
| 325 | 801 | 325 | 801 | 205861.8384 | 0 | 0 | NA |
| 326 | 801 | 326 | 801 | 184401.5364 | 0 | 0 | NA |
| 327 | 801 | 327 | 801 | 159105.2544 | 0 | 0 | NA |
| 328 | 801 | 328 | 801 | 142747.9524 | 0 | 46 | NA |
| 329 | 801 | 329 | 801 | 164316.7296 | 0 | 4 | NA |
| 297 | 800 | 297 | 800 | 715005.5364 | 0 | 155 | NA |
| 298 | 800 | 298 | 800 | 1428072.8004 | 0 | 0 | NA |
| 302 | 800 | 302 | 800 | 640352.0484 | 0 | 0 | NA |
| 303 | 800 | 303 | 800 | 543404.8656 | 0 | 30 | 0 |
| 304 | 800 | 304 | 800 | 681219.1296 | 0 | 238 | NA |
| 306 | 800 | 306 | 800 | 680526.0036 | 0 | 317 | NA |
| 307 | 800 | 307 | 800 | 785775.8736 | 0 | 47 | NA |
| 309 | 800 | 309 | 800 | 1011472.7184 | 0 | 0 | NA |
| 310 | 800 | 310 | 800 | 830649.96 0 | 0 | NA | |
| 311 | 800 | 311 | 800 | 778453.29 0 | 0 | NA | |
| 312 | 800 | 312 | 800 | 1081558.4004 | 0 | 0 | NA |
| 313 | 800 | 313 | 800 | 862223.6736 | 0 | 0 | 0 |
| 314 | 800 | 314 | 800 | 521515.0656 | 0 | 3 | 0 |
| 315 | 800 | 315 | 800 | 339562.5984 | 0 | 142 | NA |
| 316 | 800 | 316 | 800 | 319858.1136 | 0 | 208 | 0 |
| 317 | 800 | 317 | 800 | 324763.2144 | 0 | 345 | NA |
| 318 | 800 | 318 | 800 | 476514.09 0 | 222 | NA | |
| 319 | 800 | 319 | 800 | 599261.7744 | 0 | 158 | NA |
| 320 | 800 | 320 | 800 | 413474.7204 | 0 | 129 | NA |
| 321 | 800 | 321 | 800 | 374911.29 0 | 6 | NA | |
| 322 | 800 | 322 | 800 | 393731.1504 | 0 | 3 | NA |
| 323 | 800 | 323 | 800 | 302786.0676 | 0 | 67 | 0 |
| 324 | 800 | 324 | 800 | 272484 0 | 0 | NA | |
| 325 | 800 | 325 | 800 | 228560.4864 | 0 | 13 | NA |
| 326 | 800 | 326 | 800 | 167985.2196 | 0 | 38 | 1 |
| 327 | 800 | 327 | 800 | 179843.8464 | 0 | 0 | 0 |
| 328 | 800 | 328 | 800 | 207662.49 0 | 0 | NA | |
| 329 | 800 | 329 | 800 | 323874.81 0 | 0 | NA | |
| 294 | 799 | 294 | 799 | 1204813.5696 | 0 | 0 | NA |
| 295 | 799 | 295 | 799 | 804070.89 0 | 0 | NA | |
| 296 | 799 | 296 | 799 | 573988.0644 | 0 | 31 | 0 |
| 297 | 799 | 297 | 799 | 455220.09 0 | 231 | NA | |
| 298 | 799 | 298 | 799 | 1069114.6404 | 0 | 0 | NA |
| 300 | 799 | 300 | 799 | 967311.5904 | 0 | 0 | NA |
| 301 | 799 | 301 | 799 | 1131670.44 | 0 | 0 | 0 |
| 302 | 799 | 302 | 799 | 653284.2276 | 0 | 0 | NA |
| 303 | 799 | 303 | 799 | 367745.2164 | 0 | 30 | 0 |
| 304 | 799 | 304 | 799 | 720597.2544 | 0 | 223 | NA |
| 305 | 799 | 305 | 799 | 699063.21 0 | 47 | NA | |
| 306 | 799 | 306 | 799 | 591022.6884 | 0 | 114 | 0 |
| 307 | 799 | 307 | 799 | 800631.2484 | 0 | 11 | NA |
| 308 | 799 | 308 | 799 | 1124914.7844 | 0 | 0 | NA |
| 309 | 799 | 309 | 799 | 1182134.3076 | 0 | 0 | NA |
| 310 | 799 | 310 | 799 | 692456.9796 | 0 | 69 | NA |
| 311 | 799 | 311 | 799 | 596478.1824 | 0 | 142 | NA |
| 312 | 799 | 312 | 799 | 842724 0 | 0 | NA | |
| 313 | 799 | 313 | 799 | 736369.9344 | 0 | 0 | 0 |
| 314 | 799 | 314 | 799 | 407937.69 0 | 74 | NA | |
| 315 | 799 | 315 | 799 | 559982.8224 | 0 | 9 | NA |
| 316 | 799 | 316 | 799 | 504412.4484 | 0 | 45 | NA |
| 317 | 799 | 317 | 799 | 433701.2736 | 0 | 148 | NA |
| 318 | 799 | 318 | 799 | 488265.5376 | 0 | 113 | NA |
| 319 | 799 | 319 | 799 | 611774.2656 | 0 | 0 | NA |
| 320 | 799 | 320 | 799 | 461231.1396 | 0 | 35 | NA |
| 321 | 799 | 321 | 799 | 384300.8064 | 0 | 235 | 0 |
| 322 | 799 | 322 | 799 | 341944.2576 | 0 | 6 | NA |
| 323 | 799 | 323 | 799 | 295957.7604 | 0 | 230 | NA |
| 324 | 799 | 324 | 799 | 231668.9424 | 0 | 100 | 1 |
| 325 | 799 | 325 | 799 | 196922.9376 | 0 | 72 | NA |
| 326 | 799 | 326 | 799 | 224278.0164 | 0 | 14 | NA |
| 327 | 799 | 327 | 799 | 255469.5936 | 0 | 0 | NA |
| 294 | 798 | 294 | 798 | 1109820.1104 | 0 | 0 | NA |
| 295 | 798 | 295 | 798 | 1254848.04 | 0 | 0 | 0 |
| 296 | 798 | 296 | 798 | 1071846.09 | 0 | 0 | NA |
| 297 | 798 | 297 | 798 | 454653.5184 | 0 | 177 | 0 |
| 298 | 798 | 298 | 798 | 549614.6496 | 0 | 15 | NA |
| 299 | 798 | 299 | 798 | 858105.7956 | 0 | 0 | NA |
| 300 | 798 | 300 | 798 | 748674.8676 | 0 | 95 | NA |
| 301 | 798 | 301 | 798 | 900980.64 0 | 0 | NA | |
| 302 | 798 | 302 | 798 | 704693.0916 | 0 | 0 | NA |
| 303 | 798 | 303 | 798 | 345461.8176 | 4 | 19 | 1 |
| 304 | 798 | 304 | 798 | 645258.7584 | 0 | 8 | NA |
| 305 | 798 | 305 | 798 | 748674.8676 | 0 | 12 | NA |
| 306 | 798 | 306 | 798 | 425599.6644 | 0 | 0 | NA |
| 307 | 798 | 307 | 798 | 630880.7184 | 0 | 0 | NA |
| 308 | 798 | 308 | 798 | 968256 0 | 0 | NA | |
| 309 | 798 | 309 | 798 | 943035.21 0 | 0 | NA | |
| 310 | 798 | 310 | 798 | 532491.2784 | 0 | 117 | 0 |
| 311 | 798 | 311 | 798 | 413397.5616 | 0 | 280 | 0 |
| 312 | 798 | 312 | 798 | 566286.3504 | 0 | 27 | NA |
| 313 | 798 | 313 | 798 | 445155.84 0 | 185 | NA | |
| 314 | 798 | 314 | 798 | 433306.2276 | 0 | 136 | NA |
| 315 | 798 | 315 | 798 | 576172.0836 | 0 | 0 | NA |
| 316 | 798 | 316 | 798 | 635464.0656 | 0 | 0 | NA |
| 317 | 798 | 317 | 798 | 465888.1536 | 0 | 23 | NA |
| 318 | 798 | 318 | 798 | 351625.2804 | 0 | 380 | NA |
| 319 | 798 | 319 | 798 | 345744 0 | 149 | 0 | |
| 320 | 798 | 320 | 798 | 404496 0 | 175 | NA | |
| 321 | 798 | 321 | 798 | 454815.36 0 | 39 | NA | |
| 322 | 798 | 322 | 798 | 350274.5856 | 0 | 61 | NA |
| 323 | 798 | 323 | 798 | 276823.2996 | 0 | 0 | NA |
| 324 | 798 | 324 | 798 | 217790.2224 | 0 | 53 | NA |
| 325 | 798 | 325 | 798 | 174540.1284 | 0 | 40 | NA |
| 326 | 798 | 326 | 798 | 234623.9844 | 0 | 0 | NA |
| 294 | 797 | 294 | 797 | 970146.2016 | 0 | 0 | NA |
| 295 | 797 | 295 | 797 | 1145242.4256 | 0 | 0 | NA |
| 296 | 797 | 296 | 797 | 1193774.76 | 0 | 0 | NA |
| 297 | 797 | 297 | 797 | 551840.9796 | 0 | 42 | 0 |
| 298 | 797 | 298 | 797 | 473867.0244 | 0 | 153 | NA |
| 299 | 797 | 299 | 797 | 635368.41 0 | 55 | NA | |
| 300 | 797 | 300 | 797 | 594718.5924 | 0 | 132 | 0 |
| 301 | 797 | 301 | 797 | 649507.0464 | 0 | 66 | NA |
| 302 | 797 | 302 | 797 | 655128.36 0 | 0 | NA | |
| 303 | 797 | 303 | 797 | 314249.9364 | 63 | 91 | NA |
| 304 | 797 | 304 | 797 | 438614.7984 | 18 | 0 | 0 |
| 305 | 797 | 305 | 797 | 628404.9984 | 0 | 0 | NA |
| 306 | 797 | 306 | 797 | 421616.4624 | 20 | 0 | NA |
| 307 | 797 | 307 | 797 | 450214.1604 | 2 | 0 | NA |
| 308 | 797 | 308 | 797 | 552197.61 0 | 0 | NA | |
| 309 | 797 | 309 | 797 | 597127.1076 | 0 | 56 | NA |
| 310 | 797 | 310 | 797 | 403809.4116 | 0 | 199 | 0 |
| 311 | 797 | 311 | 797 | 370588.7376 | 0 | 267 | NA |
| 312 | 797 | 312 | 797 | 481580.4816 | 0 | 136 | NA |
| 313 | 797 | 313 | 797 | 417729.5424 | 0 | 73 | NA |
| 314 | 797 | 314 | 797 | 398413.44 0 | 14 | NA | |
| 315 | 797 | 315 | 797 | 409702.4064 | 0 | 0 | 0 |
| 316 | 797 | 316 | 797 | 522555.4944 | 0 | 0 | NA |
| 317 | 797 | 317 | 797 | 345109.2516 | 0 | 19 | NA |
| 318 | 797 | 318 | 797 | 295957.7604 | 0 | 137 | NA |
| 319 | 797 | 319 | 797 | 276507.7056 | 0 | 32 | NA |
| 320 | 797 | 320 | 797 | 364477.8384 | 0 | 0 | NA |
| 321 | 797 | 321 | 797 | 351482.9796 | 0 | 0 | NA |
| 322 | 797 | 322 | 797 | 312973.1136 | 0 | 0 | NA |
| 323 | 797 | 323 | 797 | 233810.9316 | 0 | 0 | NA |
| 324 | 797 | 324 | 797 | 203256.7056 | 0 | 0 | NA |
| 325 | 797 | 325 | 797 | 154873.7316 | 0 | 0 | NA |
| 294 | 796 | 294 | 796 | 701372.7504 | 0 | 52 | NA |
| 295 | 796 | 295 | 796 | 871384.9104 | 0 | 0 | 0 |
| 296 | 796 | 296 | 796 | 904515.1236 | 0 | 0 | NA |
| 297 | 796 | 297 | 796 | 546061.8816 | 0 | 0 | 0 |
| 298 | 796 | 298 | 796 | 392978.5344 | 0 | 196 | NA |
| 299 | 796 | 299 | 796 | 759826.0224 | 0 | 10 | NA |
| 300 | 796 | 300 | 796 | 472051.4436 | 0 | 191 | NA |
| 301 | 796 | 301 | 796 | 481913.64 0 | 214 | NA | |
| 302 | 796 | 302 | 796 | 616476.2256 | 0 | 71 | NA |
| 303 | 796 | 303 | 796 | 316271.2644 | 34 | 101 | NA |
| 304 | 796 | 304 | 796 | 307758.6576 | 222 | 116 | NA |
| 305 | 796 | 305 | 796 | 440046.4896 | 0 | 110 | NA |
| 306 | 796 | 306 | 796 | 320808.96 52 | 0 | NA | |
| 307 | 796 | 307 | 796 | 274492.1664 | 204 | 0 | 1 |
| 308 | 796 | 308 | 796 | 280878.8004 | 135 | 0 | NA |
| 309 | 796 | 309 | 796 | 303843.4884 | 244 | 16 | 0 |
| 310 | 796 | 310 | 796 | 331016.1156 | 207 | 3 | NA |
| 311 | 796 | 311 | 796 | 364695.21 0 | 118 | NA | |
| 312 | 796 | 312 | 796 | 423488.5776 | 0 | 74 | NA |
| 313 | 796 | 313 | 796 | 388926.8496 | 0 | 108 | NA |
| 314 | 796 | 314 | 796 | 352836 0 | 0 | NA | |
| 315 | 796 | 315 | 796 | 387506.25 0 | 0 | NA | |
| 316 | 796 | 316 | 796 | 408321 0 | 0 | NA | |
| 317 | 796 | 317 | 796 | 334754.8164 | 45 | 0 | NA |
| 318 | 796 | 318 | 796 | 201529.1664 | 13 | 0 | NA |
| 319 | 796 | 319 | 796 | 187038.9504 | 0 | 0 | NA |
| 320 | 796 | 320 | 796 | 207061.4016 | 0 | 43 | NA |
| 321 | 796 | 321 | 796 | 230457.6036 | 0 | 144 | NA |
| 322 | 796 | 322 | 796 | 244550.0304 | 0 | 0 | NA |
| 323 | 796 | 323 | 796 | 157672.5264 | 0 | 0 | 0 |
| 324 | 796 | 324 | 796 | 130292.1216 | 10 | 0 | 0 |
| 325 | 796 | 325 | 796 | 105092.6724 | 0 | 0 | NA |
| 294 | 795 | 294 | 795 | 685087.29 0 | 0 | NA | |
| 295 | 795 | 295 | 795 | 479334.6756 | 0 | 43 | NA |
| 296 | 795 | 296 | 795 | 504071.6004 | 0 | 80 | NA |
| 297 | 795 | 297 | 795 | 396824.4036 | 0 | 134 | 0 |
| 298 | 795 | 298 | 795 | 349706.6496 | 0 | 292 | NA |
| 299 | 795 | 299 | 795 | 691957.7856 | 0 | 4 | 0 |
| 300 | 795 | 300 | 795 | 496094.8356 | 0 | 0 | NA |
| 301 | 795 | 301 | 795 | 407554.56 0 | 107 | NA | |
| 302 | 795 | 302 | 795 | 432753.4656 | 0 | 0 | 0 |
| 303 | 795 | 303 | 795 | 337328.64 60 | 0 | NA | |
| 304 | 795 | 304 | 795 | 272859.9696 | 253 | 29 | NA |
| 305 | 795 | 305 | 795 | 471969 0 | 146 | NA | |
| 306 | 795 | 306 | 795 | 463189.1364 | 0 | 62 | 0 |
| 307 | 795 | 307 | 795 | 253411.56 8 | 9 | NA | |
| 308 | 795 | 308 | 795 | 229306.8996 | 212 | 0 | NA |
| 309 | 795 | 309 | 795 | 334199.61 53 | 0 | NA | |
| 310 | 795 | 310 | 795 | 420448.4964 | 2 | 0 | NA |
| 311 | 795 | 311 | 795 | 463025.8116 | 0 | 35 | NA |
| 312 | 795 | 312 | 795 | 395163.1044 | 0 | 8 | 0 |
| 313 | 795 | 313 | 795 | 246393.1044 | 0 | 0 | NA |
| 314 | 795 | 314 | 795 | 223937.1684 | 0 | 0 | NA |
| 315 | 795 | 315 | 795 | 255894.3396 | 3 | 0 | NA |
| 316 | 795 | 316 | 795 | 295370.5104 | 47 | 0 | 1 |
| 317 | 795 | 317 | 795 | 247347.0756 | 0 | 0 | NA |
| 318 | 795 | 318 | 795 | 288240.1344 | 13 | 0 | NA |
| 319 | 795 | 319 | 795 | 256076.4816 | 58 | 0 | NA |
| 320 | 795 | 320 | 795 | 153977.76 13 | 65 | NA | |
| 321 | 795 | 321 | 795 | 142566.6564 | 0 | 0 | NA |
| 322 | 795 | 322 | 795 | 163005.9876 | 0 | 0 | NA |
| 323 | 795 | 323 | 795 | 139457.4336 | 0 | 0 | NA |
| 324 | 795 | 324 | 795 | 90216.1296 | 0 | 0 | 0 |
| 294 | 794 | 294 | 794 | 824427.6804 | 0 | 0 | 0 |
| 295 | 794 | 295 | 794 | 521255.1204 | 0 | 0 | NA |
| 296 | 794 | 296 | 794 | 607183.8084 | 0 | 0 | 0 |
| 297 | 794 | 297 | 794 | 438614.7984 | 0 | 21 | NA |
| 298 | 794 | 298 | 794 | 303909.6384 | 0 | 339 | NA |
| 299 | 794 | 299 | 794 | 336423.2004 | 0 | 338 | 0 |
| 300 | 794 | 300 | 794 | 327367.0656 | 0 | 289 | NA |
| 301 | 794 | 301 | 794 | 266689.6164 | 17 | 316 | NA |
| 302 | 794 | 302 | 794 | 226100.25 112 | 96 | NA | |
| 303 | 794 | 303 | 794 | 215778.8304 | 198 | 0 | 0 |
| 304 | 794 | 304 | 794 | 239414.49 165 | 0 | 0 | |
| 305 | 794 | 305 | 794 | 354191.6196 | 0 | 110 | NA |
| 306 | 794 | 306 | 794 | 349777.6164 | 0 | 147 | NA |
| 307 | 794 | 307 | 794 | 234391.5396 | 1 | 0 | 0 |
| 308 | 794 | 308 | 794 | 235147.4064 | 76 | 0 | NA |
| 309 | 794 | 309 | 794 | 306030.24 0 | 0 | 0 | |
| 310 | 794 | 310 | 794 | 332398.3716 | 0 | 0 | NA |
| 311 | 794 | 311 | 794 | 413706.24 0 | 0 | 0 | |
| 312 | 794 | 312 | 794 | 553268.1924 | 0 | 0 | NA |
| 313 | 794 | 313 | 794 | 402894.8676 | 0 | 0 | NA |
| 314 | 794 | 314 | 794 | 278593.9524 | 31 | 0 | 0 |
| 315 | 794 | 315 | 794 | 156673.4724 | 0 | 0 | NA |
| 316 | 794 | 316 | 794 | 164025 11 | 0 | NA | |
| 317 | 794 | 317 | 794 | 189695.0916 | 12 | 0 | NA |
| 318 | 794 | 318 | 794 | 269921.8116 | 23 | 0 | NA |
| 319 | 794 | 319 | 794 | 196497.1584 | 0 | 0 | NA |
| 320 | 794 | 320 | 794 | 138607.29 2 | 0 | NA | |
| 321 | 794 | 321 | 794 | 136161 0 | 0 | 0 | |
| 322 | 794 | 322 | 794 | 95703.6096 | 6 | 0 | NA |
| 323 | 794 | 323 | 794 | 91409.4756 | 169 | 0 | NA |
| 294 | 793 | 294 | 793 | 942219.6624 | 0 | 0 | NA |
| 295 | 793 | 295 | 793 | 1113742.5156 | 0 | 0 | NA |
| 296 | 793 | 296 | 793 | 1043584.8336 | 0 | 0 | NA |
| 297 | 793 | 297 | 793 | 695055.69 0 | 0 | NA | |
| 298 | 793 | 298 | 793 | 305698.41 0 | 153 | NA | |
| 299 | 793 | 299 | 793 | 419748.4944 | 0 | 151 | NA |
| 300 | 793 | 300 | 793 | 366654.4704 | 0 | 400 | NA |
| 301 | 793 | 301 | 793 | 336980.25 0 | 302 | NA | |
| 302 | 793 | 302 | 793 | 352978.5744 | 0 | 219 | NA |
| 303 | 793 | 303 | 793 | 385491.9744 | 15 | 0 | NA |
| 304 | 793 | 304 | 793 | 235263.8016 | 0 | 0 | NA |
| 305 | 793 | 305 | 793 | 236196 0 | 0 | NA | |
| 306 | 793 | 306 | 793 | 291405.6324 | 8 | 17 | NA |
| 307 | 793 | 307 | 793 | 290822.9184 | 0 | 0 | NA |
| 308 | 793 | 308 | 793 | 207279.8784 | 216 | 0 | NA |
| 309 | 793 | 309 | 793 | 199326.5316 | 173 | 0 | NA |
| 310 | 793 | 310 | 793 | 206352.1476 | 14 | 0 | NA |
| 311 | 793 | 311 | 793 | 281706.1776 | 0 | 0 | NA |
| 312 | 793 | 312 | 793 | 316338.7536 | 14 | 0 | NA |
| 313 | 793 | 313 | 793 | 253471.9716 | 2 | 0 | 0 |
| 314 | 793 | 314 | 793 | 171396 30 | 0 | NA | |
| 315 | 793 | 315 | 793 | 118942.2144 | 0 | 0 | NA |
| 316 | 793 | 316 | 793 | 120062.25 0 | 0 | NA | |
| 317 | 793 | 317 | 793 | 150233.76 5 | 0 | NA | |
| 318 | 793 | 318 | 793 | 230457.6036 | 70 | 0 | NA |
| 319 | 793 | 319 | 793 | 130812.4224 | 0 | 0 | NA |
| 320 | 793 | 320 | 793 | 153225.2736 | 0 | 0 | 0 |
| 321 | 793 | 321 | 793 | 121605.6384 | 117 | 0 | NA |
| 322 | 793 | 322 | 793 | 105443.0784 | 339 | 0 | NA |
| 294 | 792 | 294 | 792 | 801383.04 0 | 0 | NA | |
| 295 | 792 | 295 | 792 | 990224.01 0 | 0 | 0 | |
| 296 | 792 | 296 | 792 | 880707.1716 | 0 | 0 | NA |
| 297 | 792 | 297 | 792 | 559803.24 0 | 0 | NA | |
| 298 | 792 | 298 | 792 | 302720.04 0 | 141 | 0 | |
| 299 | 792 | 299 | 792 | 389675.5776 | 0 | 116 | NA |
| 300 | 792 | 300 | 792 | 502879.5396 | 0 | 41 | NA |
| 301 | 792 | 301 | 792 | 487176.0804 | 0 | 0 | NA |
| 302 | 792 | 302 | 792 | 448042.8096 | 0 | 0 | NA |
| 303 | 792 | 303 | 792 | 450939.1104 | 0 | 0 | NA |
| 304 | 792 | 304 | 792 | 340402.2336 | 0 | 0 | 0 |
| 305 | 792 | 305 | 792 | 229306.8996 | 2 | 0 | NA |
| 306 | 792 | 306 | 792 | 220036.0464 | 221 | 0 | NA |
| 307 | 792 | 307 | 792 | 297853.9776 | 201 | 0 | NA |
| 308 | 792 | 308 | 792 | 314721 22 | 0 | 1 | |
| 309 | 792 | 309 | 792 | 284259.5856 | 31 | 0 | NA |
| 310 | 792 | 310 | 792 | 216615.7764 | 132 | 0 | NA |
| 311 | 792 | 311 | 792 | 153742.41 50 | 0 | NA | |
| 312 | 792 | 312 | 792 | 135056.25 19 | 0 | NA | |
| 313 | 792 | 313 | 792 | 120770.1504 | 0 | 0 | NA |
| 314 | 792 | 314 | 792 | 119771.3664 | 0 | 0 | NA |
| 316 | 792 | 316 | 792 | 181322.6724 | 47 | 0 | NA |
| 317 | 792 | 317 | 792 | 142113.9204 | 10 | 0 | NA |
| 318 | 792 | 318 | 792 | 133429.4784 | 0 | 0 | NA |
| 319 | 792 | 319 | 792 | 152755.9056 | 0 | 0 | NA |
| 320 | 792 | 320 | 792 | 189486.09 18 | 0 | NA | |
| 321 | 792 | 321 | 792 | 120020.6736 | 280 | 0 | NA |
| 294 | 791 | 294 | 791 | 572080.4496 | 0 | 90 | 0 |
| 295 | 791 | 295 | 791 | 651345.8436 | 0 | 71 | NA |
| 296 | 791 | 296 | 791 | 726926.76 0 | 11 | NA | |
| 297 | 791 | 297 | 791 | 503730.8676 | 0 | 0 | 0 |
| 298 | 791 | 298 | 791 | 354906.1476 | 0 | 26 | NA |
| 299 | 791 | 299 | 791 | 316271.2644 | 0 | 33 | NA |
| 300 | 791 | 300 | 791 | 407018.4804 | 0 | 0 | NA |
| 301 | 791 | 301 | 791 | 384524.01 0 | 0 | NA | |
| 302 | 791 | 302 | 791 | 363753.7344 | 0 | 0 | NA |
| 303 | 791 | 303 | 791 | 348501.3156 | 0 | 0 | NA |
| 304 | 791 | 304 | 791 | 347085.9396 | 0 | 0 | NA |
| 305 | 791 | 305 | 791 | 229824.36 72 | 0 | NA | |
| 306 | 791 | 306 | 791 | 162715.4244 | 200 | 0 | NA |
| 307 | 791 | 307 | 791 | 195222.5856 | 103 | 0 | NA |
| 308 | 791 | 308 | 791 | 253592.8164 | 138 | 0 | 0 |
| 309 | 791 | 309 | 791 | 272045.6964 | 59 | 0 | 0 |
| 310 | 791 | 310 | 791 | 231437.9664 | 24 | 0 | NA |
| 311 | 791 | 311 | 791 | 128436.2244 | 2 | 0 | NA |
| 312 | 791 | 312 | 791 | 150094.2564 | 0 | 0 | NA |
| 313 | 791 | 313 | 791 | 166317.1524 | 0 | 14 | NA |
| 314 | 791 | 314 | 791 | 151585.6356 | 0 | 0 | NA |
| 318 | 791 | 318 | 791 | 141661.9044 | 105 | 0 | NA |
| 319 | 791 | 319 | 791 | 107794.0224 | 240 | 0 | NA |
| 320 | 791 | 320 | 791 | 117456.9984 | 369 | 0 | 0 |
| 294 | 790 | 294 | 790 | 460009.4976 | 0 | 324 | NA |
| 295 | 790 | 295 | 790 | 394635.24 0 | 370 | NA | |
| 296 | 790 | 296 | 790 | 422474.0004 | 0 | 195 | NA |
| 297 | 790 | 297 | 790 | 381479.1696 | 0 | 327 | NA |
| 298 | 790 | 298 | 790 | 334615.9716 | 0 | 257 | NA |
| 299 | 790 | 299 | 790 | 247048.7616 | 0 | 5 | NA |
| 300 | 790 | 300 | 790 | 249620.1444 | 0 | 0 | NA |
| 301 | 790 | 301 | 790 | 284899.7376 | 0 | 0 | NA |
| 302 | 790 | 302 | 790 | 230861.0304 | 0 | 0 | 0 |
| 303 | 790 | 303 | 790 | 199219.3956 | 0 | 0 | NA |
| 304 | 790 | 304 | 790 | 181476 32 | 0 | NA | |
| 305 | 790 | 305 | 790 | 170701.1856 | 85 | 0 | NA |
| 306 | 790 | 306 | 790 | 164608.7184 | 0 | 0 | NA |
| 307 | 790 | 307 | 790 | 138786.0516 | 13 | 0 | NA |
| 308 | 790 | 308 | 790 | 126081.8064 | 9 | 0 | NA |
| 310 | 790 | 310 | 790 | 145664.3556 | 0 | 0 | NA |
| 311 | 790 | 311 | 790 | 157958.5536 | 0 | 0 | NA |
| 312 | 790 | 312 | 790 | 288949.2516 | 48 | 0 | NA |
| 313 | 790 | 313 | 790 | 415638.09 22 | 0 | 1 | |
| 314 | 790 | 314 | 790 | 261386.7876 | 0 | 0 | 1 |
| 293 | 789 | 293 | 789 | 407401.3584 | 0 | 354 | 1 |
| 294 | 789 | 294 | 789 | 373883.3316 | 0 | 400 | NA |
| 295 | 789 | 295 | 789 | 352194.7716 | 0 | 400 | 1 |
| 296 | 789 | 296 | 789 | 343911.8736 | 0 | 398 | NA |
| 297 | 789 | 297 | 789 | 346803.21 0 | 400 | 0 | |
| 298 | 789 | 298 | 789 | 328672.89 0 | 385 | NA | |
| 299 | 789 | 299 | 789 | 322033.5504 | 0 | 188 | NA |
| 300 | 789 | 300 | 789 | 216615.7764 | 0 | 0 | 0 |
| 301 | 789 | 301 | 789 | 193635.2016 | 0 | 0 | 0 |
| 302 | 789 | 302 | 789 | 188182.44 0 | 2 | NA | |
| 303 | 789 | 303 | 789 | 239297.0724 | 0 | 17 | NA |
| 304 | 789 | 304 | 789 | 255166.4196 | 0 | 40 | NA |
| 305 | 789 | 305 | 789 | 314990.3376 | 0 | 40 | NA |
| 306 | 789 | 306 | 789 | 328672.89 0 | 43 | NA | |
| 307 | 789 | 307 | 789 | 268738.56 0 | 44 | NA | |
| 308 | 789 | 308 | 789 | 160544.4624 | 0 | 0 | 1 |
| 309 | 789 | 309 | 789 | 302984.1936 | 0 | 0 | 0 |
| 310 | 789 | 310 | 789 | 242300.2176 | 0 | 0 | NA |
| 311 | 789 | 311 | 789 | 294197.76 | 0 | 0 | NA |
| 312 | 789 | 312 | 789 | 476017.2036 | 0 | 0 | NA |
| 313 | 789 | 313 | 789 | 545441.3316 | 0 | 0 | NA |
| 314 | 789 | 314 | 789 | 270545.6196 | 0 | 0 | 1 |
| 293 | 788 | 293 | 788 | 306827.3664 | 0 | 304 | 1 |
| 294 | 788 | 294 | 788 | 301796.4096 | 0 | 184 | NA |
| 295 | 788 | 295 | 788 | 297526.6116 | 0 | 166 | NA |
| 296 | 788 | 296 | 788 | 264957.2676 | 0 | 119 | 0 |
| 297 | 788 | 297 | 788 | 262430.7984 | 0 | 215 | NA |
| 298 | 788 | 298 | 788 | 263538.4896 | 0 | 159 | NA |
| 299 | 788 | 299 | 788 | 267682.0644 | 0 | 6 | NA |
| 300 | 788 | 300 | 788 | 213887.7504 | 0 | 0 | NA |
| 301 | 788 | 301 | 788 | 233231.0436 | 0 | 0 | NA |
| 302 | 788 | 302 | 788 | 339982.2864 | 0 | 17 | NA |
| 303 | 788 | 303 | 788 | 446918.9904 | 0 | 32 | 0 |
| 304 | 788 | 304 | 788 | 429365.6676 | 0 | 2 | NA |
| 305 | 788 | 305 | 788 | 472876.2756 | 0 | 4 | NA |
| 306 | 788 | 306 | 788 | 406023.84 | 0 | 0 | NA |
| 307 | 788 | 307 | 788 | 252687.1824 | 0 | 0 | 1 |
| 308 | 788 | 308 | 788 | 184143.9744 | 0 | 0 | NA |
| 309 | 788 | 309 | 788 | 324763.2144 | 0 | 0 | 0 |
| 310 | 788 | 310 | 788 | 334824.2496 | 0 | 0 | NA |
| 311 | 788 | 311 | 788 | 360288.0576 | 0 | 90 | NA |
| 312 | 788 | 312 | 788 | 551751.84 | 0 | 174 | NA |
| 313 | 788 | 313 | 788 | 472381.29 | 0 | 26 | 1 |
| 314 | 788 | 314 | 788 | 237539.2644 | 0 | 0 | NA |
| 292 | 787 | 292 | 787 | 318299.0724 | 0 | 161 | NA |
| 293 | 787 | 293 | 787 | 324353.0304 | 0 | 55 | NA |
| 294 | 787 | 294 | 787 | 316878.9264 | 0 | 329 | NA |
| 295 | 787 | 295 | 787 | 304637.7636 | 0 | 131 | NA |
| 296 | 787 | 296 | 787 | 323943.1056 | 0 | 230 | NA |
| 297 | 787 | 297 | 787 | 342225 | 0 | 46 | 0 |
| 298 | 787 | 298 | 787 | 296807.04 | 0 | 66 | NA |
| 299 | 787 | 299 | 787 | 233810.9316 | 0 | 0 | NA |
| 300 | 787 | 300 | 787 | 222067.1376 | 0 | 0 | NA |
| 301 | 787 | 301 | 787 | 329154.6384 | 0 | 55 | NA |
| 302 | 787 | 302 | 787 | 435916.8576 | 0 | 9 | NA |
| 303 | 787 | 303 | 787 | 565383.6864 | 0 | 87 | NA |
| 304 | 787 | 304 | 787 | 437582.25 | 0 | 0 | NA |
| 305 | 787 | 305 | 787 | 364912.6464 | 0 | 0 | NA |
| 306 | 787 | 306 | 787 | 303711.21 | 0 | 0 | NA |
| 307 | 787 | 307 | 787 | 240413.7024 | 0 | 0 | NA |
| 308 | 787 | 308 | 787 | 214387.5204 | 0 | 0 | NA |
| 309 | 787 | 309 | 787 | 326063.8404 | 0 | 0 | 0 |
| 310 | 787 | 310 | 787 | 406559.2644 | 0 | 0 | 0 |
| 311 | 787 | 311 | 787 | 446678.3556 | 0 | 242 | 1 |
| 312 | 787 | 312 | 787 | 558277.9524 | 0 | 172 | NA |
| 313 | 787 | 313 | 787 | 463842.7236 | 0 | 0 | NA |
| 314 | 787 | 314 | 787 | 203581.44 | 0 | 0 | NA |
| 291 | 786 | 291 | 786 | 414632.9664 | 0 | 346 | NA |
| 292 | 786 | 292 | 786 | 406406.25 | 0 | 241 | 0 |
| 293 | 786 | 293 | 786 | 474528.0996 | 0 | 22 | NA |
| 294 | 786 | 294 | 786 | 391775.8464 | 0 | 274 | NA |
| 295 | 786 | 295 | 786 | 404724.9924 | 0 | 0 | NA |
| 296 | 786 | 296 | 786 | 423020.16 | 0 | 5 | NA |
| 297 | 786 | 297 | 786 | 357269.1984 | 0 | 190 | 0 |
| 298 | 786 | 298 | 786 | 313846.4484 | 0 | 74 | 0 |
| 299 | 786 | 299 | 786 | 251843.3856 | 0 | 0 | 0 |
| 300 | 786 | 300 | 786 | 248422.4964 | 0 | 0 | NA |
| 301 | 786 | 301 | 786 | 398943.8244 | 0 | 0 | NA |
| 302 | 786 | 302 | 786 | 447882.1776 | 0 | 11 | NA |
| 303 | 786 | 303 | 786 | 553446.7236 | 0 | 315 | NA |
| 304 | 786 | 304 | 786 | 442437.8256 | 0 | 69 | NA |
| 305 | 786 | 305 | 786 | 461964.9024 | 0 | 51 | NA |
| 306 | 786 | 306 | 786 | 353763.2484 | 0 | 0 | NA |
| 307 | 786 | 307 | 786 | 334754.8164 | 0 | 3 | NA |
| 308 | 786 | 308 | 786 | 297199.4256 | 0 | 0 | NA |
| 309 | 786 | 309 | 786 | 225758.0196 | 0 | 0 | NA |
| 310 | 786 | 310 | 786 | 312704.64 | 0 | 7 | 0 |
| 311 | 786 | 311 | 786 | 450939.1104 | 0 | 46 | NA |
| 312 | 786 | 312 | 786 | 502964.64 | 0 | 7 | NA |
| 313 | 786 | 313 | 786 | 325995.3216 | 0 | 0 | NA |
| 314 | 786 | 314 | 786 | 191528.7696 | 0 | 0 | NA |
| 291 | 785 | 291 | 785 | 488852.6724 | 0 | 345 | NA |
| 292 | 785 | 292 | 785 | 518400 | 0 | 274 | NA |
| 293 | 785 | 293 | 785 | 663899.04 0 | 95 | NA | |
| 294 | 785 | 294 | 785 | 560791.2996 | 0 | 16 | NA |
| 295 | 785 | 295 | 785 | 504156.8016 | 0 | 0 | NA |
| 296 | 785 | 296 | 785 | 426382.8804 | 0 | 66 | NA |
| 297 | 785 | 297 | 785 | 369931.5684 | 0 | 317 | NA |
| 298 | 785 | 298 | 785 | 384821.7156 | 0 | 107 | NA |
| 299 | 785 | 299 | 785 | 280306.7136 | 0 | 78 | NA |
| 300 | 785 | 300 | 785 | 273298.9284 | 0 | 18 | NA |
| 301 | 785 | 301 | 785 | 383185.7604 | 0 | 69 | NA |
| 302 | 785 | 302 | 785 | 425756.25 0 | 45 | NA | |
| 303 | 785 | 303 | 785 | 551930.1264 | 0 | 222 | NA |
| 304 | 785 | 304 | 785 | 447882.1776 | 0 | 165 | NA |
| 305 | 785 | 305 | 785 | 369712.6416 | 0 | 68 | NA |
| 306 | 785 | 306 | 785 | 380146.2336 | 0 | 26 | NA |
| 307 | 785 | 307 | 785 | 469910.25 0 | 206 | NA | |
| 308 | 785 | 308 | 785 | 533893.2624 | 0 | 0 | NA |
| 309 | 785 | 309 | 785 | 313712.01 0 | 0 | NA | |
| 310 | 785 | 310 | 785 | 332744.3856 | 0 | 22 | NA |
| 311 | 785 | 311 | 785 | 519869.8404 | 0 | 0 | 0 |
| 312 | 785 | 312 | 785 | 507228.84 0 | 0 | NA | |
| 313 | 785 | 313 | 785 | 284259.5856 | 0 | 0 | NA |
| 291 | 784 | 291 | 784 | 568365.21 0 | 185 | NA | |
| 292 | 784 | 292 | 784 | 740288.16 0 | 73 | 0 | |
| 293 | 784 | 293 | 784 | 874973.16 0 | 0 | NA | |
| 294 | 784 | 294 | 784 | 796234.9824 | 0 | 0 | NA |
| 295 | 784 | 295 | 784 | 615534.3936 | 0 | 9 | NA |
| 296 | 784 | 296 | 784 | 393731.1504 | 0 | 170 | 0 |
| 297 | 784 | 297 | 784 | 421460.64 0 | 236 | NA | |
| 298 | 784 | 298 | 784 | 424582.56 0 | 22 | 0 | |
| 299 | 784 | 299 | 784 | 272233.4976 | 0 | 86 | NA |
| 300 | 784 | 300 | 784 | 303314.5476 | 0 | 81 | NA |
| 301 | 784 | 301 | 784 | 431806.6944 | 0 | 234 | NA |
| 302 | 784 | 302 | 784 | 538990.9056 | 0 | 103 | 0 |
| 303 | 784 | 303 | 784 | 631929.6036 | 0 | 78 | 1 |
| 304 | 784 | 304 | 784 | 582382.6596 | 0 | 117 | NA |
| 305 | 784 | 305 | 784 | 605128.41 0 | 93 | 1 | |
| 306 | 784 | 306 | 784 | 554161.1364 | 0 | 228 | 0 |
| 307 | 784 | 307 | 784 | 678547.5876 | 0 | 64 | NA |
| 308 | 784 | 308 | 784 | 460497.96 0 | 0 | NA | |
| 309 | 784 | 309 | 784 | 404267.0724 | 0 | 0 | 0 |
| 310 | 784 | 310 | 784 | 426853.1556 | 0 | 0 | NA |
| 311 | 784 | 311 | 784 | 613840.9104 | 0 | 0 | NA |
| 312 | 784 | 312 | 784 | 394484.4864 | 0 | 0 | NA |
| 313 | 784 | 313 | 784 | 207498.4704 | 0 | 0 | NA |
| 291 | 783 | 291 | 783 | 586694.7216 | 0 | 9 | NA |
| 292 | 783 | 292 | 783 | 641985.5376 | 0 | 0 | NA |
| 294 | 783 | 294 | 783 | 679041.9216 | 0 | 4 | NA |
| 295 | 783 | 295 | 783 | 637474.4964 | 0 | 152 | NA |
| 296 | 783 | 296 | 783 | 457652.25 0 | 207 | NA | |
| 297 | 783 | 297 | 783 | 473701.8276 | 0 | 95 | NA |
| 298 | 783 | 298 | 783 | 341733.7764 | 0 | 0 | NA |
| 299 | 783 | 299 | 783 | 253592.8164 | 0 | 8 | NA |
| 300 | 783 | 300 | 783 | 330602.0004 | 0 | 34 | NA |
| 301 | 783 | 301 | 783 | 386237.3904 | 0 | 33 | NA |
| 303 | 783 | 303 | 783 | 692856.4644 | 0 | 0 | NA |
| 304 | 783 | 304 | 783 | 698060.25 0 | 139 | 0 | |
| 305 | 783 | 305 | 783 | 596014.8804 | 0 | 196 | NA |
| 306 | 783 | 306 | 783 | 575898.8544 | 0 | 167 | NA |
| 307 | 783 | 307 | 783 | 463352.49 0 | 34 | NA | |
| 308 | 783 | 308 | 783 | 340542.2736 | 0 | 0 | NA |
| 309 | 783 | 309 | 783 | 530566.56 0 | 11 | NA | |
| 310 | 783 | 310 | 783 | 548547.6096 | 0 | 0 | NA |
| 311 | 783 | 311 | 783 | 646030.1376 | 0 | 0 | NA |
| 312 | 783 | 312 | 783 | 294458.1696 | 0 | 0 | 0 |
| 313 | 783 | 313 | 783 | 204991.6176 | 0 | 0 | NA |
| 295 | 782 | 295 | 782 | 540842.5764 | 0 | 0 | NA |
| 296 | 782 | 296 | 782 | 556217.64 0 | 0 | NA | |
| 297 | 782 | 297 | 782 | 504242.01 0 | 0 | NA | |
| 304 | 782 | 304 | 782 | 586327.1184 | 0 | 167 | NA |
| 305 | 782 | 305 | 782 | 524204.9604 | 0 | 282 | NA |
| 306 | 782 | 306 | 782 | 390725.0064 | 0 | 251 | NA |
| 307 | 782 | 307 | 782 | 347722.5024 | 0 | 71 | 0 |
| 308 | 782 | 308 | 782 | 461964.9024 | 0 | 35 | NA |
| 309 | 782 | 309 | 782 | 613746.8964 | 0 | 0 | NA |
| 310 | 782 | 310 | 782 | 547304.04 0 | 0 | NA | |

| | | | | | | | | |
|---|---|---|---|---|---|---|---|---|
| 311 | 782 | 311 | 782 | 388253.61 | 0 | 0 | NA | |
| 312 | 782 | 312 | 782 | 234449.64 | 0 | 0 | 1 | |
| 313 | 782 | 313 | 782 | 283045.2804 | | 0 | 0 | NA |
| 304 | 781 | 304 | 781 | 620345.2644 | | 0 | 45 | NA |
| 305 | 781 | 305 | 781 | 633170.3184 | | 0 | 151 | NA |
| 306 | 781 | 306 | 781 | 445476.1536 | | 0 | 52 | NA |
| 307 | 781 | 307 | 781 | 539695.9296 | | 0 | 18 | 0 |
| 308 | 781 | 308 | 781 | 654157.44 | 0 | 34 | NA | |
| 309 | 781 | 309 | 781 | 753042.1284 | | 0 | 0 | NA |
| 310 | 781 | 310 | 781 | 512283.7476 | | 0 | 0 | NA |
| 311 | 781 | 311 | 781 | 348572.16 | 0 | 0 | NA | |
| 312 | 781 | 312 | 781 | 360072.0036 | | 0 | 0 | NA |
| 313 | 781 | 313 | 781 | 440285.3316 | | 0 | 48 | NA |
| 305 | 780 | 305 | 780 | 755334.81 | 0 | 41 | NA | |
| 306 | 780 | 306 | 780 | 647574.2784 | | 0 | 5 | NA |
| 307 | 780 | 307 | 780 | 609055.3764 | | 0 | 129 | NA |
| 308 | 780 | 308 | 780 | 665170.7364 | | 0 | 50 | NA |
| 309 | 780 | 309 | 780 | 492130.3104 | | 0 | 106 | 0 |
| 310 | 780 | 310 | 780 | 357915.0276 | | 0 | 147 | NA |
| 311 | 780 | 311 | 780 | 372051.2016 | | 0 | 21 | NA |
| 312 | 780 | 312 | 780 | 564752.25 | 0 | 54 | NA | |
| 313 | 780 | 313 | 780 | 657266.9184 | | 0 | 53 | NA |
| 305 | 779 | 305 | 779 | 726722.1504 | | 0 | 97 | NA |
| 306 | 779 | 306 | 779 | 673318.7136 | | 0 | 80 | NA |
| 307 | 779 | 307 | 779 | 635081.4864 | | 0 | 116 | NA |
| 308 | 779 | 308 | 779 | 553089.69 | 0 | 199 | NA | |
| 309 | 779 | 309 | 779 | 391250.25 | 0 | 369 | NA | |
| 310 | 779 | 310 | 779 | 399398.7204 | | 0 | 214 | NA |
| 311 | 779 | 311 | 779 | 498040.7184 | | 0 | 282 | NA |
| 312 | 779 | 312 | 779 | 612149.76 | 0 | 79 | 0 | |
| 313 | 779 | 313 | 779 | 546594.0624 | | 0 | 0 | NA |
| 314 | 779 | 314 | 779 | 441879.2676 | | 0 | 0 | NA |
| 308 | 778 | 308 | 778 | 505521 | 0 | 32 | NA | |
| 309 | 778 | 309 | 778 | 562410.0036 | | 0 | 252 | NA |
| 310 | 778 | 310 | 778 | 657558.81 | 0 | 15 | NA | |
| 311 | 778 | 311 | 778 | 675684 | 0 | 124 | NA | |
| 312 | 778 | 312 | 778 | 640256.0256 | | 0 | 3 | NA |
| 313 | 778 | 313 | 778 | 617513.0724 | | 0 | 0 | NA |
| 309 | 777 | 309 | 777 | 690062.49 | 0 | 0 | NA | |
| 310 | 777 | 310 | 777 | 745252.3584 | | 0 | 0 | NA |
| 311 | 777 | 311 | 777 | 687871.1844 | | 0 | 0 | NA |
| 309 | 790 | 309 | 790 | 147133.6164 | | 0 | 0 | NA |